\documentclass{aa}
\usepackage{graphicx}
\usepackage[varg]{txfonts}

\usepackage{natbib}
\bibpunct{(}{)}{;}{a}{}{,}
\usepackage{subcaption}
\usepackage{tablefootnote}
\usepackage{threeparttable}
\usepackage{float}
\usepackage{url}
\makeatletter
\renewcommand*\aa@pageof{, page \thepage{} of \pageref*{LastPage}}
\makeatother

\newcommand \Cp[1]{[\ion{\element[][#1]{C}}{II}]}
\newcommand \kms[1]{km/s #1}

\usepackage{hyperref}
\hypersetup{
    colorlinks=true,
    linkcolor=blue,
    citecolor=blue,
    urlcolor = blue,
    unicode=true
}
\usepackage{amsmath}

\title{\Cp{} 158 $\mu$m self-absorption and optical depth effects}
\author{C. Guevara \inst{1} \and J. Stutzki \inst{1}  \and V. Ossenkopf-Okada \inst{1} \and R. Simon \inst{1} \and J. P. P\'{e}rez-Beaupuits \inst{2,3} \and H. Beuther \inst{4} \and S. Bihr \inst{4} \and R. Higgins \inst{1} \and U. Graf \inst{1} \and R. G\"{u}sten \inst{2}  }

\institute{I. Physikalisches Institut, Universit\"{a}t zu K\"{o}ln, Z\"{u}lpicher Str. 77, 50937 K\"{o}ln, Germany \\
              \email{guevara@ph1.uni-koeln.de} \and
             Max-Planck-Institut f\"{u}r Radioastronomie, Auf dem H\"{u}gel 69, D-53121 Bonn, Germany \and
             European Southern Observatory, Santiago, Chile \and
             Max Planck Institute for Astronomy, K\"{o}nigstuhl 17, 69117 Heidelberg, Germany
             }

\date{Received date /
Accepted date }

\begin{document}

 \abstract
 {
The \Cp{} 158~$\mu$m far-infrared (FIR) fine-structure line is one of the most important cooling lines of the star-forming interstellar medium (ISM). It is used as a tracer of star formation efficiency in external galaxies and to study feedback effects in parental clouds. High spectral resolution observations have shown complex structures in the line profiles of the \Cp{}  emission.
}
{
Our aim is to determine whether the complex profiles observed in \Cp{12} are due to individual velocity components along the line-of-sight or to self-absorption based on a comparison of the \Cp{12} and isotopic \Cp{13} line profiles.
}
{
Deep integrations with the SOFIA/upGREAT 7-pixel array receiver in the sources of M43, Horsehead~PDR, Monoceros~R2, and M17~SW allow for the detection  of optically thin \Cp{13} emission lines, along with the \Cp{12} emission
lines, with a high signal-to-noise ratio  (S/N).  We first derived the \Cp{12} optical depth and the \Cp{} column density from a single component model. However, the complex line profiles observed require a double layer model with an emitting background and an absorbing foreground. A multi-component velocity fit allows us to derive the physical conditions of the \Cp{} gas: column density and excitation temperature.
}
{
We find moderate to high \Cp{12} optical depths in all four sources and self-absorption of \Cp{12} in Mon~R2 and M17~SW. The high column density of the warm background emission corresponds to an equivalent A$_{\mathrm v}$ of up to 41~mag. The foreground absorption requires substantial column densities of cold and dense \Cp{} gas, with an equivalent A$_{\mathrm v}$ ranging up to about 13~mag.
}
{
The column density of the warm background material requires multiple photon-dominated region (PDR) surfaces stacked along the line of sight and in velocity. The substantial column density of dense and cold foreground \Cp{} gas detected in absorption cannot be explained with any known scenario and we can only speculate on its origins. }


\maketitle

\section{Introduction} \label{Introduction}

The far-infrared (FIR) fine-structure line of the singly ionized carbon \Cp{} at 158~$\mu$m is, along with the ground-state fine structure line of neutral oxygen [\ion{O}{I}] at 63~$\mu$m, one of the strongest cooling lines of the interstellar medium. Its ionization potential of 11.2~eV is below that of hydrogen. Hence, ionized carbon, C$^{+}$, is abundant in \ion{H}{II} regions and the UV illuminated surface of atomic and molecular clouds. The UV penetration results in a layered structure where from the inside to outside, the main carbon carrier (namely, the carbon monoxide molecule) is photo-dissociated, and the resulting neutral atomic carbon is photo-ionized. The layered structure where the hydrogen changes from ionized to neutral atomic and to its molecular form and where, in parallel but at different geometrical depths, the carbon changes from atomic ionized to atomic neutral and to carbon monoxide is commonly known as a photon-dominated region (PDR). It is understood as resulting from a detailed (photo-) chemical network that includes the radiative transfer and the energy balance between UV-heating and dust- and line-emission cooling. 
The turbulent, fractal structure of the star-forming interstellar medium (ISM) implies that a considerable fraction of the ISM is actually in surface regions and, hence, can be described as a PDR \citep{1997ARA&A..35..179H}. The prominent sources of UV-radiation are young, massive stars; the \Cp{} emission can thus be used as a tracer of star formation activity. \par

Already in the first detection paper of the \Cp{} fine structure line, \citet{1980ApJ...240L..99R} 
noted that "optical depth effects in
the 157~$\mu$m\footnote{At that time, the spectroscopic data were less accurate
and the wavelength of the transition was assumed to fall at 157~$\mu$m instead
of 158~$\mu$m.} line may have been significant" but the authors did not take them into account because the database was too restricted. The only observational method for checking the optical depth of the \Cp{} emission consists of using the rarer isotope, 
$^{13}$C$^{+}$. The spectral signature of the \Cp{13} transitions (see below) requires a high spectral resolution to separate it from the fine structure line of the main isotope. In addition, the isotopic lines are weak due to the lower abundance of the isotopic species. Hence, attempts to measure the optical depth were dependent on the future availability of instrumentation of high sensitivity and 
high spectral resolution. \par

The \Cp{12} fine structure transition is one single line between the two energy levels, $^{2}$P$_{3/2} \rightarrow \, ^{2}$P$_{1/2}$, at a frequency of 1900.5369~GHz \citep{1986ApJ...305L..89C}. 
The \Cp{13} transition, instead, splits into three hyperfine components due to the presence of the additional spin of the unpaired neutron in the nucleus. These components are labeled by the total angular momentum change F=2$\rightarrow$1, F=1$\rightarrow$0, and F=1$\rightarrow$1. The frequencies of the fine structure transitions of both isotopes were determined by \citet{1986ApJ...305L..89C}. The astronomical observations are fully consistent with these frequencies, as was discussed by \citet{2013A&A...550A..57O}, who also noted that the relative strengths of the \Cp{13} hyperfine satellites ($s_{\mathrm F\rightarrow \mathrm F'}$, see Table~\ref{table:13CII}) given by \cite{1986ApJ...305L..89C} are incorrect. We summarize all the relevant \Cp{12} and \Cp{13} spectroscopic parameters in Table~\ref{table:13CII}, including the velocity offsets of the \Cp{13} hyperfine components relative to \Cp{12}. 
The frequency separation of the hyperfine lines is small enough that all lines can be observed simultaneously with the bandwidth available in current, state-of-the-art high resolution heterodyne receivers (the 130~km/s separation of the outer hfs-satellites corresponds to slightly below 1~GHz frequency separation). \par

\begin{table*}
  \centering
    \caption{\Cp{12} and \Cp{13} Spectroscopic parameters}
  \begin{tabular}{l c c r r r }
      \hline
      \hline
\multicolumn{1}{c}{Line} & \multicolumn{1}{c}{Statistical} & \multicolumn{1}{c}{Weight}  & \multicolumn{1}{c}{Frequency} &\multicolumn{1}{c}{Vel.\ offset} &  \multicolumn{1}{c}{Relative } \\
                         & \multicolumn{1}{c}{g$_{u}$}     & \multicolumn{1}{c}{g$_{l}$} & \multicolumn{1}{c}{$\nu$}     &\multicolumn{1}{c}{$\delta$ v$_{\mathrm F\rightarrow \mathrm F'}$} &  \multicolumn{1}{c}{intensity} \\
     &             &         & \multicolumn{1}{c}{(GHz)}     & \multicolumn{1}{c}{(km/s)}                                        &  \multicolumn{1}{c}{ $s_{\mathrm F\rightarrow \mathrm F'}$} \\
\hline
 $\Cp{12}$ $^{2}$P$_{3/2} - ^{2}$P$_{1/2}$ & 4 & 2 & 1900.5369 &   0   & $-$    \\
 $\Cp{13}$ F=2$\rightarrow$1               & 5 & 3 & 1900.4661 & +11.2 & 0.625  \\
 $\Cp{13}$ F=1$\rightarrow$0               & 3 & 1 & 1900.9500 & $-$65.2 & 0.250  \\
 $\Cp{13}$ F=1$\rightarrow$1               & 3 & 3 & 1900.1360 & +63.2 & 0.125  \\
\hline
  \end{tabular}
         \label{table:13CII}
\end{table*}

Detection of the \Cp{13} lines, aimed at getting a handle of the optical depth of the \Cp{12} emission, requires high spectral resolution. The outer satellite lines, which are very weak, require a spectral resolution higher than at least half of their velocity separation from the main \Cp{} line, which is around 30~km/s. Only in the case of intrinsically very narrow lines, the bright F=2$\rightarrow$1 satellite can be used, with a required spectral resolution below 5~km/s. 
It is for these reasons that \Cp{13} has been observed in only a few cases up till now: a marginal detection of \Cp{13} F=2$\rightarrow$1 was reported by \citet{1988ApJ...325L..47B} in M42 Orion with the KAO using their pioneering FIR heterodyne receiver; \citet{1991ApJ...382L..37S} independently reported the detection of  \Cp{13} F=2$\rightarrow$1 in M42 with the ultra-high-resolution central pixel of the Fabry-Perot spectrometer instrument FIFI on board the KAO.
\citet{2013A&A...550A..57O} reported the detection of the \Cp{13} emission in the Orion Bar and several 
PDRs with Herschel/HIFI, followed by an extended analysis of the Orion Bar by \citet{2015ApJ...812...75G}. Finally, \citet{2012A&A...542L..16G}, with the improved sensitivity and broader bandwidth of the GREAT receiver \citep{2012A&A...542L...1H} on board SOFIA, detected, for the first time, all three \Cp{13} satellites and deep self-absorption by cold foreground gas in \Cp{12} as a clear indication of high optical depth in NGC 2024. They were even able to map the extended \Cp{13} emission. \par

The standard PDR model scenario \citep{1985ApJ...291..722T}, which is very successful in explaining the observed \Cp{} line intensities, as well as the line ratios relative to other tracers, predicts an optical depth around unity in the \Cp{12} line for a single PDR layer over a wide range of physical parameters, suggesting, hence, that optical depth effects are not relevant for the analysis of \Cp{} observations. \par

In order to use the isotopic emission to derive the optical depth, the isotopic abundance ratio must be known. The \element[][12]{C}/\element[][13]{C} elemental isotopic ratio (from hereon named $\alpha$) has been studied for several decades \citep[e.g.,][]{1985A&A...149..195G,1985A&A...143..148H,1990ApJ...357..477L,1993ApJ...408..539L,1994ARA&A..32..191W,1996A&AS..119..439W,2002ApJ...578..211S,2005ApJ...634.1126M,2014A&A...570A..65G}. 
The common approach is to derive the ratio from the comparison of the line intensities of the \element[][12]{C} and \element[][13]{C} isotopic species of common molecules containing carbon, such as CO, H$_{2}$CO, or CN. All the studies show that $\alpha$ increases with the Galactic radius. \citet{2005ApJ...634.1126M}, by compiling CO and H$_{2}$CO and CN data, derived a common galacto-centric gradient according to $\alpha = (6.21\pm1.00)\times D_{ \mathrm G \mathrm C} \, [\mathrm{kpc}] + (18.71\pm7.37)$, where D$_{ \mathrm G \mathrm C}$ is the galacto-centric radius. \par

Going beyond the elemental abundances, the isotopic ratio of ionized carbon, \element[][12]{C}$^{+}$/\element[][13]{C}$^{+}$ (in the following denoted as $\alpha^{+}$) is also affected by fractionation. The slightly endothermic reaction $\element[][13]{C}$$^{+} + \element[][12]{CO} \rightleftharpoons \element[][12]{C}$$^{+} +  \element[][13]{CO} + 34.8$~K favors \element[][12]{C}$^{+}$ over \element[][13]{C}$^{+}$ at low temperatures, thus increasing the ratio over the elemental one: $\alpha^{+} > \alpha$. In parallel, the self-shielding of CO against photo-dissociation predicts a larger fraction of \element[][12]{C} to be bound in CO and, hence, would lower the ratio: $\alpha^{+} < \alpha$. With these two competing processes, model calculations are necessary for predicting $\alpha^{+}$. Model predictions by \citet{2013A&A...550A..56R} using the KOSMA-$\tau$ PDR model predict $\alpha^{+}$ to be slightly higher or equal to the elemental ratio $\alpha$. \par

In the optically thin case, the observed \Cp{12}/\Cp{13} line integrated intensity ratio is equal to the abundance ratio $\alpha^{+}$ when taking the summed-up intensities of the three \Cp{13} hyperfine lines. Higher optical depth in \Cp{12} gives a lower intensity ratio than $\alpha^+$ (and, hence, runs opposite to the fractionation effect). The HIFI \Cp{13} measurements by \citet{2013A&A...550A..57O} have shown that most line intensity ratios are
below the isotopic ratio, which has been interpreted as evidence of higher than unity optical depth of the \Cp{12} line. Given the uncertainty in the fractionation effect derived through simulations \citep{2013A&A...550A..56R}, where warmer C$^+$ ($>$100~K) shows no fractionation effects, and given the uncertainty in the temperature of the gas, we take in the following the value of the elemental abundance ratio also for the abundance ratio of the ions, $\alpha^{+}=\alpha$, so that the values derived below for the \Cp{12} optical depth can be regarded as the lower limits. \par 

This paper presents new \Cp{12} and \Cp{13} observations with the upGREAT\footnote{GREAT is a development by the MPI für Radioastronomie and KOSMA/Universität zu Köln, in cooperation with the MPI für Sonnensystemforschung and the DLR Institut für Optische Sensorsysteme.} instrument on board SOFIA. The increased sensitivity and the multi-pixel capability now allows for a more systematic study of several sources and positions within each source. From an observing program covering six sources, here we present the observational results and the analysis of four of the six sources. We focus on a detailed line profile analysis of the observed \Cp{12} and \Cp{13} spectra to study the optical depth of \Cp{12} and to derive physical properties of the gas traced by C$^{+}$, such as excitation temperature and column density. In Section~\ref{Observed Sources}, we describe the sources. In Sect.~\ref{Observations},
we describe the observations and data reduction. In Sect.~\ref{Results}, we present the observational results and present, as a first approximation, a single layer model to derive the \Cp{13} column density, the optical depth of \Cp{12} and the \Cp{12}/\Cp{13} line intensity ratio. As the complex line profiles indicate evidence that this first-order analysis is insufficient, we also perform an analysis through a multi-component fit to the \Cp{12} and 
\Cp{13} lines simultaneously to derive the physical properties of the \Cp{} emission. In Sect.~\ref{Discussion}, we discuss the details and implications of the multi-component analysis by itself. Finally, in Sect.~\ref{Conclusions}, we summarize the study and we discuss the implications of the derived optical depths and the column densities with regard to the physical properties of the different components. \par 


\section{Observed Sources} \label{Observed Sources}

To study the \Cp{12} and \Cp{13} emission in detail over a range of physical conditions and different astrophysical environments over the past two years, we conducted an observational program using SOFIA with the upGREAT heterodyne instrument to study six sources: DR21, S106, M43, the Horsehead PDR, Monoceros R2, and M17~SW. These sources were selected 
to cover a wide range of the main parameters that affect the \Cp{} intensity according to PDR models, such as UV intensities, densities, as well as the source intrinsic velocity distribution. Here we report on the observational results for four of these sources, M43, Horsehead, Mon~R2, and M17~SW, where the observations have been completed. DR21 and
S106 have only been partially observed, and the achieved S/N is not sufficient for a detailed analysis of the \Cp{13} emission. \par

\textbf{M43} is a close-by spherical \ion{H}{II} region, located northeast of the Orion Nebula \citep{1982ASSL...90.....G}. It is part of the Orion complex, located at a distance of 389~pc \citep{2018AJ....156...84K}. The region has one single ionizing source, namely, an early B-type star HD37061 \citep{2011A&A...530A..57S}. 
Due to its close distance, its simple spherical geometry and a single ionization source, M43 is well-suited as a simple, properly characterized test case for the present study. For M43, we use the abundance ratio of $\alpha^{+}=\alpha=67$ for solar galactocentric radii. \par

The \textbf{Horsehead nebula} is a dark cloud filament protruding out of the Orion Molecular Cloud complex \citep{2003A&A...410..577A} and it is visible in the optical against the prominent H$\alpha$ emission of the large-scale ionized surface of the Orion Molecular Cloud complex. The region is located at a distance of 360~pc \citep{2016A&A...595A...2G}. The cloud features a PDR with an edge-on geometry that corresponds  to the illuminated edge of the molecular cloud L1630 on the near side of the \ion{H}{II} region IC 434. The Horsehead PDR, with its simple, edge-on geometry is an excellent source for studying the PDR structure resulting from the penetration of the UV field into the molecular cloud. We use an abundance ratio $\alpha=\alpha^{+}=67$. \par

\textbf{Monoceros R2} (Mon R2) is an ultra-compact \ion{H}{II} region located at 830~pc \citep{1976AJ.....81..840H}. The region contains a reflection nebula and the UC\ion{H}{II} region is surrounded by several PDRs with different physical conditions \citep{2014A&A...561A..69P,2014A&A...569A..19T}. For Mon R2, we also use an abundance ratio  
of $\alpha^{+}=\alpha=67$ due to its close distance, similar to Orion. Mon R2 has a complex source morphology with different components along the line-of-sight and shows a correspondingly more complex \Cp{} profile (see Section~\ref{sub:lineprofile}) than M43 and the Horsehead Nebula. \par

\textbf{M17} is one of the brightest and most massive star-forming regions in the Galaxy. The \ion{H}{II} region is ionized by a highly obscured (Av > 10) cluster of many (>100) OB stars \citep{2008ApJ...686..310H}. The  M17 complex is located at a distance of 1.98~kpc \citep{2011ApJ...733...25X}. The \ion{H}{II} region, together with its 
associated giant molecular cloud located to the southwest has been considered as a prototype of an edge-on interface. M17~SW corresponds to the southwestern part of the GMC. The high column densities involved across this complex source make it an ideal testbed for optical depth studies \citep[eg.][]{1988ApJ...332.1049G,1993ApJ...405..249G}.
For M17~SW, we assume an abundance ratio $\alpha^{+}=\alpha=40$, although one can argue that the value should be higher, around 57, according to the Galactocentric gradient relation refered in Section~\ref{Introduction}, taking M17~SW's distance to the center of the Galaxy and also observational constraints \citep[eg.,][]{1976ApJ...206L..63M,1982A&A...109..344H}. We use a conservative lower value because any increase in the ratio would lead to an increase in the derived optical depth and column densities. Hence, the derived values have to be considered as lower limits to the actual values. We additionally note that fractionation, as discussed above, would also result in an increase in the abundance ratio, compared to the elemental abundance ratio. \par

\section{Observations} \label{Observations}

The observations reported here were all performed with the SOFIA airborne observatory \citep{2012ApJ...749L..17Y}. 
As the observations were performed over several observing campaigns, during which the receiver evolved and its configuration changed, the \Cp{} observations were done either with the single-pixel GREAT receiver \citep{2012A&A...542L...1H}, configured to the GREAT L2 single-pixel channel at 1900 GHz; or the upGREAT array receiver \citep{2016A&A...595A..34R}, with the 7 pixel/2 polarization configuration: LFA (Low Frequency Array) polarizations H and V at 1900 GHz. The \Cp{} channel was combined with different receivers in the other GREAT frequency channel: partly, we used the L1 single-pixel channel (frequency between 1200-1500 GHz)  tuned to [\ion{N}{II}] 205~$\mu$m, and for Mon~R2 we used the newly available upGREAT 7 pixel high frequency array HFA (High Frequency Array), tuned to [\ion{O}{I}] 63~$\mu$m. The observational setup is summarized for all observations and the positions observed in Table~\ref{Obstable}. Where available, data from both LFA subarrays (H and V polarization) were averaged together. As spectrometers, we used the FFTS backends with an intrinsic spectral resolution of 142~kHz, and after a resampling described below depending on the source, it is more than sufficient for even the narrow line Horsehead Nebula observations. All observations were done in total power mode. \par

   \begin{table*}
   \begin{threeparttable}
  \centering
  
  \caption{Observational parameters for the sources.}
  \begin{tabular}{l l l l l l l r c c}
      \hline
\hline
\multicolumn{1}{c}{Sources} & \multicolumn{1}{c}{Observing} & \multicolumn{1}{c}{Band}          & \multicolumn{1}{c}{RA}      & \multicolumn{1}{c}{DEC}    & \multicolumn{1}{c}{T$_{ \mathrm{sys}}$}  & \multicolumn{1}{c}{Total}  & \multicolumn{1}{c}{pwv\tnote{e}} & $\Delta$v & \multicolumn{1}{c}{rms} \\
                            & \multicolumn{1}{c}{Date}      & \multicolumn{1}{c}{Config.} & \multicolumn{1}{c}{(J2000)} & \multicolumn{1}{c}{(J2000)} &                               & \multicolumn{1}{c}{Obs. time} & \multicolumn{1}{c}{(zenith)} & & \multicolumn{1}{c}{(noise)}  \\
              &            &                        & \multicolumn{1}{c}{(h:m:s)}  & \multicolumn{1}{c}{(\degr:\arcmin:\arcsec)} & \multicolumn{1}{c}{(K)}        & \multicolumn{1}{c}{(min)}     & \multicolumn{1}{c}{($\mu$m)} & (km/s) & \multicolumn{1}{c}{(K)}  \\
          \hline
M43           & 12-08-2015 & L1\tnote{a} / LFA\tnote{c} & 05:35:31.36 & -05:16:02.6             & 2800       & \multicolumn{1}{c}{60}        & \multicolumn{1}{c}{12} & 0.1 & \multicolumn{1}{c}{0.15-0.20}  \\      
Horsehead PDR     & 12-17-2016 & L1\tnote{a} / LFA\tnote{c} & 05:40:54.27  & -02:28:00.0        & 2200       & \multicolumn{1}{c}{80}        & \multicolumn{1}{c}{8} & 0.3 & \multicolumn{1}{c}{0.08-0.10} \\ 
              & 02-10-2017 & L1\tnote{a} / LFA\tnote{c} & 05:40:54.27  & -02:28:00.0             & 2100       & \multicolumn{1}{c}{74} & \multicolumn{1}{c}{11} & 0.3 &\multicolumn{1}{c}{0.08-0.10} \\      
Monoceros R2  & 11-04-2016 & L2\tnote{b} / HFA\tnote{d} & 06:07:46.20  & -06:23:08.0             & 2430       & \multicolumn{1}{c}{32}       & \multicolumn{1}{c}{13} & 0.3 & \multicolumn{1}{c}{0.17-0.30}  \\      
M17SW          & 06-09-2016 & L1\tnote{a} / LFA\tnote{c} & 18:20:27.60 & -16:12:00.9             & 2800       & \multicolumn{1}{c}{84}        & \multicolumn{1}{c}{9} & 0.3 & \multicolumn{1}{c}{0.18-0.33} \\      
         \label{Obstable}
\end{tabular}
\begin{tablenotes}\footnotesize
\item[a] L1 corresponds to the old GREAT L1 band between 1200-1500 GHz.
\item[b] L2 to the old GREAT L2 band at 1900 GHz. 
\item[c] LFA to the upGREAT Low Frequency Array operating between 1810-2070 GHz.
\item[d] HFA to the upGREAT High Frequency Array operating at 4744 GHz.
\item[e] Precipitable water vapor.
\end{tablenotes}
\end{threeparttable}
 \end{table*}

The data were calibrated to the main beam brightness temperature intensity scale, $T_{ \mathrm m \mathrm b}$, with the {\it kalibrate} task \citep{2012A&A...542L...4G}, including bandpass gain calibration from counts into intensities and fitting an atmospheric model to the observed sky-hot scans to correct for the frequency dependent atmospheric transmission from the signal and image sideband. The main beam efficiencies of the individual pixels were derived through the observation of planets such as Jupiter and Saturn for each observing epoch. On average,  the main beam efficiencies are close to $\sim$0.65, consistent with the optical layout of the receiver and telescope. We use the main beam temperature scale because SOFIA's main beam pattern is clean, with low side lobes. We then further processed the data with the CLASS 90 package, part of the GILDAS\footnote{\url{https://www.iram.fr/IRAMFR/GILDAS/}} software.
In the following, we describe the specifics of the observations for each source. \par

\subsection{M43} 

\begin{figure}
\centering
\begin{subfigure}{\hsize}
  \centering
  \includegraphics[width=0.8\hsize]{{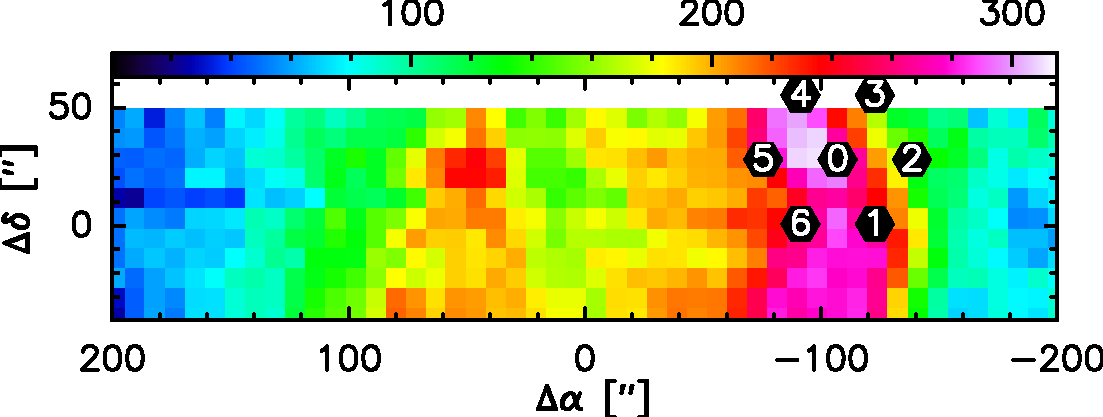}}
  \caption{}
  \label{fig:sub1}
\end{subfigure}%
\par
\begin{subfigure}{\hsize}
  \centering
   \includegraphics[width=0.60\hsize]{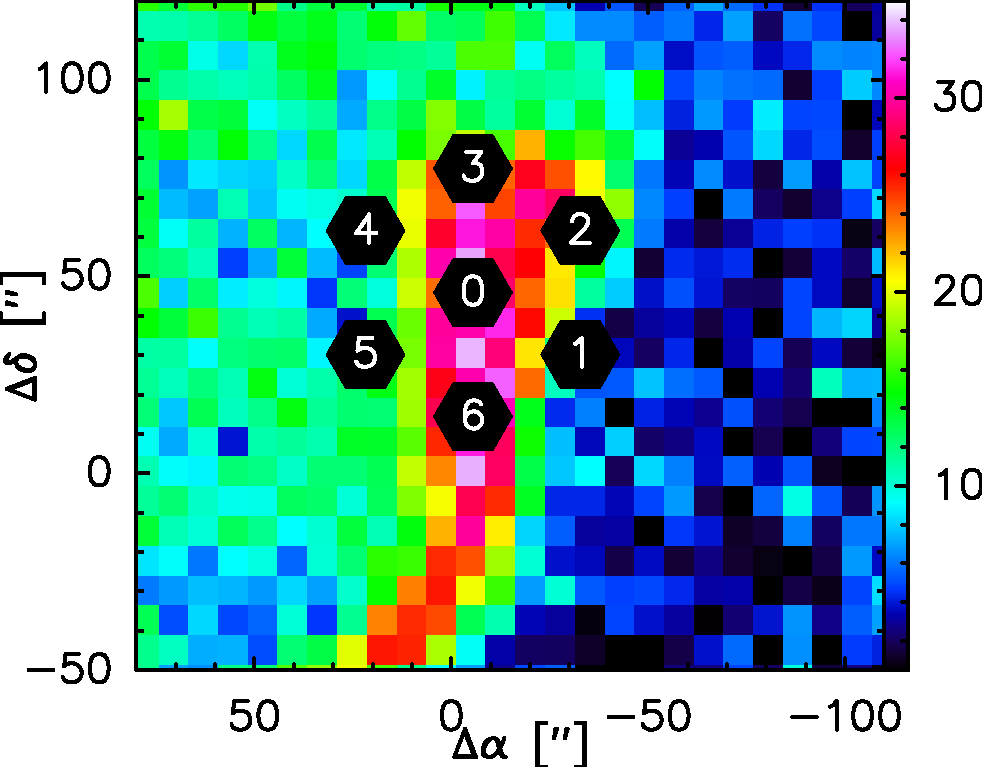}
  \caption{}
  \label{fig:sub3}
\end{subfigure}
\par
\begin{subfigure}{\hsize}
  \centering
   \includegraphics[width=0.9\hsize]{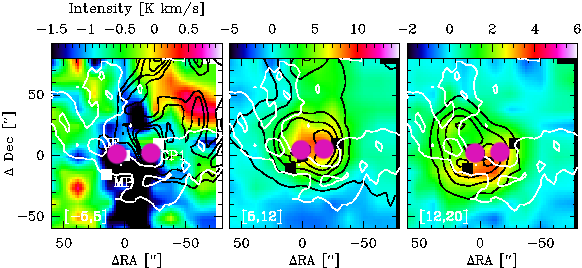}
  \caption{}
  \label{fig:sub4}
\end{subfigure}
\par
\begin{subfigure}{\hsize}
  \centering
  \includegraphics[width=0.70\hsize]{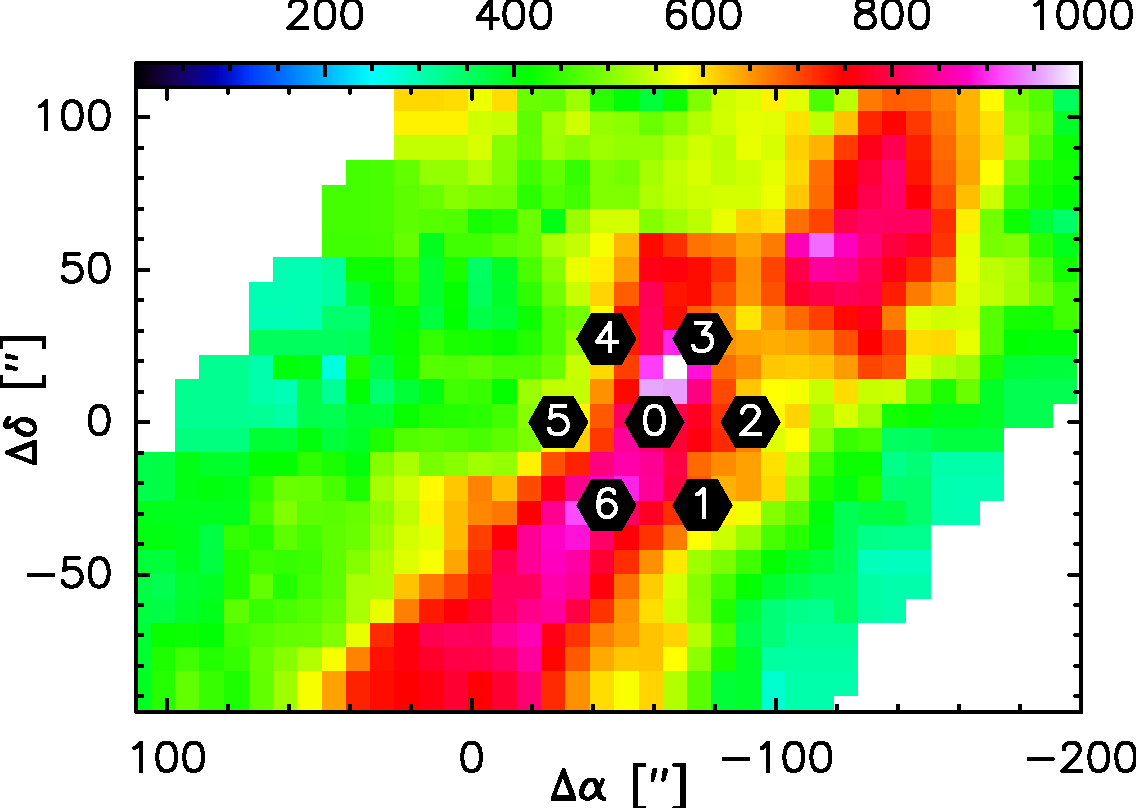}
  \caption{}
  \label{fig:sub2}
\end{subfigure}
\caption{(\subref{fig:sub1}) - M43 \Cp{} integrated intensity map between 5 and 15~km/s with the position of the upGREAT array for the deep integration marked as black hexagons. (\subref{fig:sub3}) - The Horsehead PDR \Cp{} integrated intensity map between 9 and 13~km/s with the position of the upGREAT array rotated at 30$^{\circ}$. (\subref{fig:sub4}) - Mon~R2 \Cp{} integrated map intensity in black contours in overlay with other species, see \citet{2014A&A...561A..69P}. We pointed the single L2 pixel of GREAT at the two positions marked by purple circles (The squares represent OH$^+$ positions). (\subref{fig:sub2}) - M17~SW \Cp{} integrated intensity map  \citep{2012A&A...542L..13P} between 15 to 25~km/s with the position of the upGREAT array at 0$^{\circ}$.}
\label{CIImap} 
\end{figure}

For M43, we first took a quick map of 600\arcsec $\times$ 140\arcsec extent, shown in Fig.~\ref{fig:sub1}, in total power on-the-fly mode for identifying the \Cp{} peak. We used an off-source reference position with an offset relative to the center of the map of (603\arcsec,76\arcsec). We selected this off position from CO (2-1) observations without emission, relatively far from the central emission. For the deep integration to detect the \Cp{13} line, we selected 
the position of peak emission at offsets relative to the center of ($-$107.6\arcsec,28.5\arcsec) for pointing the array with an orientation angle of 0$^{\circ}$  in total power mode. We found weak contamination in the off position for \Cp{} at a level of about 2~K. A multi-Gaussian profile fit to the OFF emission extracted from the sky-hot spectra was then applied as a correction to the contaminated observations (see Appendix~\ref{Offcp}). The velocity resolution after resampling is 0.3~km/s. \par

\subsection{The Horsehead PDR} 

The Horsehead PDR observations were performed in two separate flight legs. We selected the positions for the deep \Cp{12} integration from the previous \Cp{} Horsehead map observed within SOFIA Director's Discretionary Time. We pointed the LFA \ion{C}{II} array to the map coordinate offsets $(-5.5\arcsec,45.9\arcsec)$ with an array orientation angle of 30$^{\circ}$ (so that three pixels are aligned N-S) in total power mode, thus covering three positions along the bright \Cp{} ridge and the other 4 positions of the array being pointed slightly off, but parallel to the main ridge on both sides (see Fig.~\ref{fig:sub3}). We used an off-source position at $(-733\arcsec,-27.5\arcsec)$.  The velocity resolution after resampling is 0.1~km/s. \par

\subsection{Monoceros R2} 

For the Monoceros R2 (MonR2) observations, two positions were observed at map offsets (0\arcsec,05\arcsec) and 
($-$20\arcsec,05\arcsec) with the single-pixel L2 GREAT channel. The positions were selected for being the two main peaks of \Cp{} emission \citep[][see Fig.~\ref{fig:sub4}]{2014A&A...561A..69P}. The off-source position was at ($-$200\arcsec,0\arcsec). We found weak contamination for \Cp{}, at a level of around 2.5~K (it was corrected following the procedure described in Appendix~\ref{Offcp}) that was detected through the observation of the off against a further out off-source position at ($-$400\arcsec,0\arcsec). The observations were performed in total power mode. The velocity resolution after resampling is 0.3~km/s. \par

\subsection{M17~SW} 

The M17~SW observations use the position of the SAO star 161357 as the map center position. Based on the previous observations \citep{2012A&A...542L..13P} of M17~SW in \Cp{} with GREAT, we pointed the array to follow the main emission ridge, centering it at the map coordinates of ($-$60\arcsec,0\arcsec) with an angle of 0$^{\circ}$ (see Fig.~\ref{fig:sub2}). We used a close-by offset position at (537\arcsec,$-$67\arcsec) selected from a Spitzer 8~$\mu$m map. We observed this off-source position against a second far distant reference position at (1040\arcsec,$-$535\arcsec). Due to the broad extent of the \Cp{} emission around M17~SW, we find weak contamination for \Cp{} at the level of 3.5~K peak brightness temperature at the nearby OFF position, that was corrected according to Appendix~\ref{Offcp}. The velocity resolution after resampling is 0.3~km/s. \par


\section{\Cp{12}/\Cp{13} Results} \label{Results} 

In this section, we focus on the \Cp{12} and \Cp{13} observations and their respective analyses. The [\ion{N}{II}] observations are discussed in Section~\ref{NIIsec}. \par

Figures \ref{M43mosaic} to \ref{M17mosaic} show the observed \Cp{12} spectra for all four sources and observed positions. The top panel always shows the high S/N spectrum and a scaled-up version of the spectrum that makes the weak \Cp{13} satellites visible. The bottom panel shows an overlay of the averaged emission of the  \Cp{13} hyperfine structure satellites (red, see equation \ref{eq:tmbavg}), scaled-up with the nominal abundance values $\alpha^+$, as given above, with the \Cp{12} spectrum, combined with a plot of the \Cp{12}/\Cp{13} intensity ratio and the derived optical depth. \par

\subsection{Line Profiles} \label{sub:lineprofile}

In M43, all seven positions observed with the upGREAT-array show a narrow line profile with the main emission peak located at $v_{\mathrm{LSR}}\sim$10 km/s. The four positions with the strongest emission also show a secondary peak in velocity, located at $\sim$5~km/s, see Fig.~\ref{M43CIIbase}. The peak brightness temperature of $T_{ \mathrm m \mathrm b}\approx 50~K$ implies a minimum excitation temperature for C$^{+}$ of about 90~K due to the Rayleigh-Jeans correction (the high \Cp{} frequencies in the THz are well beyond the Rayleigh-Jeans regime). All three \Cp{13} satellites are clearly visible (although only barely for the case of the weak $\mathrm{F}=1\rightarrow$ 1 satellite) and separated from the main \Cp{12} line emission. The \Cp{13} line profile, scaled-up by $\alpha^{+}/s_{\mathrm F \rightarrow \mathrm F'}$ for each hfs satellite, averaged over all three hfs satellites (see Section~\ref{subsec:zero} for a detailed explanation of the averaging process), shows, within its higher noise, a shape consistent with the main isotope line. It is, however, consistently higher in intensity than the observed main isotopic line in four of the seven positions. This indicates that the emission is optically thick, as discussed in the next subsections. \par
   
\begin{figure}
   \centering
   \includegraphics[width=0.95\hsize]{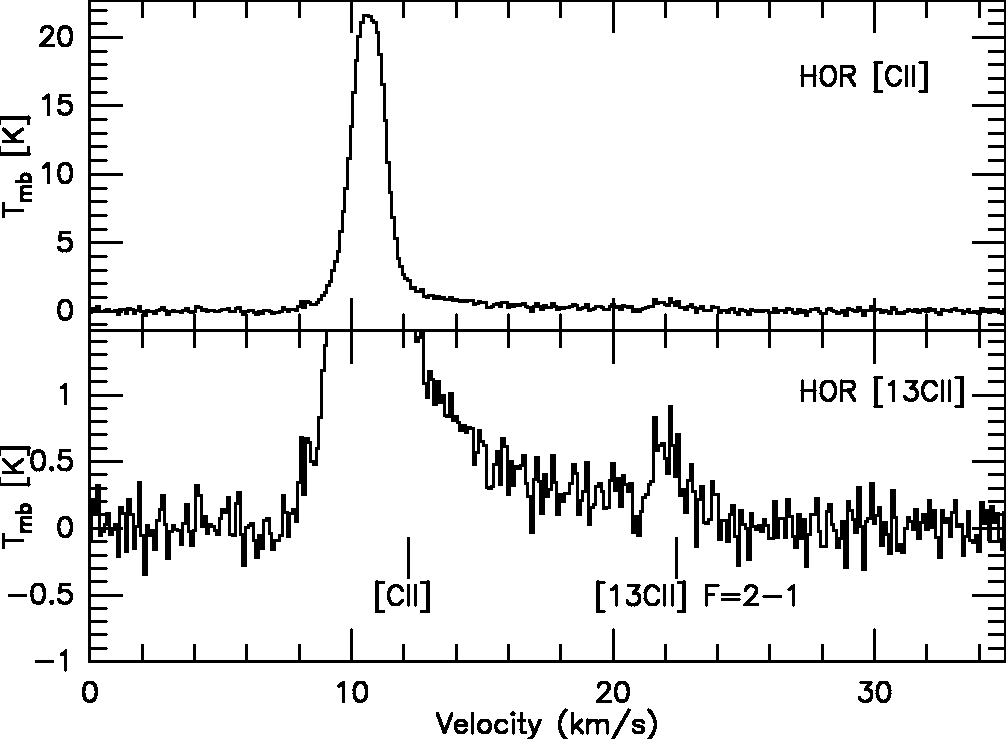}
      \caption{Horsehead \Cp{12} and \Cp{13} F=2-1 emission at 22~km/s (source $v_\mathrm{LSR}$ plus hyperfine velocity offset $\Delta$v$_{1-1}$) for position 0. The line profile shows a broad wing extending from 16 to 30~km/s. 
              }
         \label{hor_wing}
   \end{figure}

In the Horsehead PDR, the \Cp{12} line profile is also narrow, with a single peak at $\sim$10.5~km/s, see 
Fig.~\ref{HorCIIbase}. In addition, it shows an extended wing toward higher velocities from the 
\Cp{12} peak, from 16 to 30~km/s, see Fig.~\ref{hor_wing}, a feature that can only be detected thanks to the long integration time and correspondingly high S/N required for the \Cp{13} detection. The wing is visible at all seven positions as shown in Fig.~\ref{HORwingall} in the Appendix. In order to separate the \Cp{13} line profile, the wing emission has been fitted with a third-order polynomial in the velocity range between 12 and 30~km/s with a window between 19 and 25~km/s, which has been subtracted from the observed spectra shown in Fig.~\ref{HorCIIbase}; this is necessary to not confuse the derivation of the line ratios and the optical depths as below. Only the strongest \Cp{13} satellite (\Cp{13}$_{\mathrm{F=2-1}}$) is well-detected in the positions that trace the ridge emission: 
pixels 0, 2, 3 and 6 (Fig.~\ref{fig:sub3}). Similar to the case of M43, the \Cp{13} line profile, scaled-up by $\alpha^{+}/s_{\mathrm F \rightarrow \mathrm F'}$ (after subtraction of the wing emission, see Fig.~\ref{Horall}), shows a similar shape to \Cp{12} within its higher noise, but the intensity is higher than \Cp{12} in positions 0, 2 and 6, indicating an optical depth above 1. \par

For Mon~R2, the two positions observed show a broad emission from 0 to 35~km/s. The \Cp{12} line profile is very different between both positions but shares a strong dip at 12~km/s. This situation was already noted by \citet{2013A&A...550A..57O} from Herschel \Cp{} observations towards Mon~R2. In contrast, \Cp{13} shows a single peak, at both positions, filling the dip visible in the \Cp{12} emission, see Fig.~\ref{Monall}. All \Cp{13} satellites are strong enough for being detected, but \Cp{13}~${\mathrm{F=2-1}}$ is blended with the \Cp{12} line due to the width of the \Cp{12} emission line. After being scaled-up by $\alpha^{+}/s_{\mathrm F \rightarrow \mathrm F'}$ and averaged over the two outer satellites (see Fig.~\ref{Monall}), the \Cp{13} line profile has a much higher intensity than the main isotopic line. Thus, the \Cp{12} line is clearly optically thick with a significant opacity and the emission dip suggests self-absorption in \Cp{12}, as has been discussed by \citet{2013A&A...550A..57O}. \par

For M17~SW, the \Cp{12} emission is also broad, ranging from 0 to 40~km/s and the line profiles at the seven positions of the upGREAT array pixels, separated by 30\arcsec, show large differences among each other. The \Cp{12} profiles show several narrow spikes and dips as discussed already by \citet{2015A&A...575A...9P} (see Fig.~\ref{M17CIIall}). Only the two outer \Cp{13} satellites can be separated, F=$1\rightarrow0$ and F=$1\rightarrow1$. F=$2\rightarrow1$ is blended due to the width of the \Cp{12} emission line. The \Cp{13} profile, unlike \Cp{12}, shows a simple, close to Gaussian, profile with only one peak at $\sim$~20~km/s. As for the other sources, the scaled-up \Cp{13} shows a much higher intensity than the \Cp{12} one. Thus, also for M17~SW, the \Cp{12} emission is optically thick and the emission dips in the \Cp{12} profile are probably due to self-absorption. \par

In summary, we detect all \Cp{13} hfs satellites (unless the F=$2\rightarrow1$-satellite is blended with the \Cp{12} line), except for the case of the Horsehead PDR, where only the strongest satellite is detected. In all cases, the scaled-up \Cp{13} emission exceeds the \Cp{12} main isotopic emission in the central velocities of the sources.
The match is closer in the line wings, but typically the line center emission in \Cp{13} substantially overshoots.  This indicates that the low optical depth, implicitly assumed for this scaling, does not apply. \par

\subsection{Zeroth Order Analysis: Homogeneous single layer} \label{subsec:zero}

Although the narrow dips in the \Cp{12} profiles in two of the sources observed, Mon~R2, and M17, clearly indicate self-absorption effects and high optical depths--and, hence, the need for several, physically different source components along the line-of-sight to properly interpret the observed spectra--we first ignore these issues and analyze the observed spectra in terms of a single component, homogeneous source model. This is relevant because low spectral resolution observations, which are incapable of resolving the detailed emission profiles and, hence, only capable of obtaining line integrated intensities for \Cp{12} and the outer \Cp{13} hyperfine components, would be restricted to such an analysis and would have to quote the resulting source parameters as their observational results.\par

The optical depth is proportional to the line-of-sight integral of the population difference between the upper and lower states. Hence, the ratio of the optical depths of the \Cp{13} transitions and the \Cp{12} transition can be directly estimated as long as two conditions are met. First, the abundance ratio between the two isotopic species of \Cp{}, named $\alpha^+$ above in Section~\ref{Introduction}, is constant across the source; in other words, we ignore isotope selective fractionation. Second, the excitation temperature of the main isotopic line and all three \Cp{13} hyperfine satellites are identical at each position in the source, meaning that no hyperfine selective trapping effects (e.g.,\ in the optical ground state absorption transitions) result in different excitation. Thus, referring to the same Doppler-shift that is corrected by the hyperfine frequency shift, the optical depth ratio is given by: 

\begin{equation} \label{eq:odratio}
\tau_{13}(\nu-\Delta \nu_{13,\mathrm F\rightarrow \mathrm F'}) =
\beta_{\mathrm F\rightarrow \mathrm F'}\,\tau_{12}(\nu)
,\end{equation}

where $\Delta \nu_{13,\mathrm F\rightarrow \mathrm F'}= \nu_{\mathrm{rest},12} \, \times \Delta \mathrm v_{\mathrm F\rightarrow \mathrm F'}/ \mathrm c$, and 
$\beta_{\mathrm F\rightarrow \mathrm F'}$ corresponds to a single hyper-fine component and it is defined as:

\begin{equation} \label{eq:beta}
\beta_{\mathrm F\rightarrow \mathrm F'} = \frac{s_{\mathrm F \rightarrow \mathrm F'}}{\alpha^{+}}
.\end{equation}

The assumption from above is well justified: with regard to the assumed constant abundance ratio across the source,
detailed modeling within the context of a photo-dissociation region \citep{2013A&A...550A..57O} shows that fractionation between \Cp{13} and \Cp{12} at maximum can reach up to a factor of about two. With regard to the assumed identical $T_{\mathrm{ex}}$ for \Cp{12} and \Cp{13}, hyperfine selective trapping in the optical and UV ground state absorption transition can be ruled out as very unlikely because the frequency splitting of the \Cp{13} hyperfine states 
corresponds to a velocity splitting of below 0.3~km/s at optical and UV wavelength so that the hyperfine lines are fully blended in the optical transition. \footnote{We note that hyperfine selective trapping in the rotational FIR transitions of ammonia, NH$_3$, cause anomalous intensity ratios of the hyperfine satellites of the cm-wave inversion transitions; 
\citep{1977ApJ...214L..67M,1985A&A...144...13S}.} The observable intensities, according to the formal solution of the radiative transfer equation for a homogeneous source, $\mathrm T_{\mathrm{ex}}={\rm const.}$, are given by: 

\begin{equation} 
\label{eq:detequ}
\mathrm T_{\mathrm{mb}}(\nu) = \eta_{\phi} \left[ \cal{J}_{\nu}(\mathrm T_{\mathrm{ex}}) - \cal{J}_{\nu}(\mathrm T_\mathrm{bg})\right] \left( 1-e^{-\tau(\nu)}\right)  
.\end{equation}

Here $\mathcal{J}_{\nu}(T)=\dfrac{h\nu}{k} \left(e^{h\nu/kT}-1\right)^{-1}$ is the equivalent brightness temperature of a blackbody emission at temperature $T$ and $\eta_{\phi}$ is the beam filling factor of the layer. Also, as there is no bright continuum background emission in any of the sources observed, except for M17~SW, the only background is the cosmic microwave background with $T_{\mathrm{bg}}=2.7~{\rm K}$, corresponding to 70~mK at 1.9~THz, so that we can neglect the $\cal{J}_{\nu}(\mathrm T_\mathrm{bg})$ term. In the case of M17~SW, \citet{1992ApJ...390..499M} detected dust continuum emission, with temperature between 75 and 40~K and a dust optical depth between 0.021 and 0.106. At 1.9~THz, and taken into account its distribution, the background emission ranges between 1 and 4~K. We have ignored this contribution because for the optically thin emission, it would only result in an increase in the continuum that was already removed due to the baseline subtraction. For the optically thick \Cp{12} emission, it could affect the derivation of the excitation temperature as done in the analysis, but only by a few K, not significantly changing the estimated parameters. \par

\subsubsection{\Cp{12} optical depth} \label{C12od}

Combining the assumption of a homogeneous source, and of equal $T_{\mathrm{ex}}$ for all \Cp{12} and \Cp{13} transitions, a zeroth-order estimate of the optical depth of \Cp{12} can be derived from the intensity ratios of the \Cp{13} and \Cp{12} intensities at correspondingly shifted Doppler velocities. Following Eqs.~\ref{eq:odratio} and \ref{eq:detequ}, we can write :

\begin{equation} \label{eq:12tau}
\frac{\mathrm T_{\mathrm{mb},12}(\nu)}{\mathrm T_{\mathrm{mb},13}(\nu-\Delta \nu_{\mathrm F\rightarrow \mathrm F'})}
=
\frac{1-e^{-\tau_{12}(\nu)}}{1-e^{-\beta_{\mathrm F\rightarrow \mathrm F'}\,\tau_{12}(\mathrm \nu)}}
\simeq
\frac{1-e^{-\tau_{12}(\nu)}}{\beta_{\mathrm F\rightarrow \mathrm F'}\,\tau_{12}(\nu)}
,\end{equation} 

where the last step assumes that \Cp{13} is optically thin, an assumption well justified by the value of $\alpha^+$ in the range of 40 to 80 \citep{2012ApJS..203...13G,2013A&A...550A..57O}. \par

Instead of calculating the \Cp{12} optical depth for each \Cp{13} hyperfine satellite separately, we use the noise weighted average ( weighting factors $w_{\mathrm{ F\rightarrow F'}}$ defined below) of the appropriately velocity shifted and scaled-up three hyperfine satellites:

\begin{align} \label{eq:tmbavg}
 \begin{split}
\mathrm T_{\mathrm{mb},13, \mathrm{tot}}(v) &= \dfrac{\sum_{\mathrm{\mathrm{ F\rightarrow F'}}}  w_{\mathrm{ F\rightarrow F'}} \dfrac{\mathrm T_{\mathrm{mb},13}( v-\Delta v_{\mathrm{ F\rightarrow F'}})}{s_{\mathrm{ F\rightarrow F'}}}}{\sum_{\mathrm{ \mathrm{ F\rightarrow F'}}} w_{\mathrm{ F\rightarrow F'}}}, \\ 
w_{\mathrm{ F\rightarrow F'}} &= \left(\frac{s_{\mathrm F\rightarrow \mathrm F'}}{\sigma}\right)^{2},
\end{split}
\end{align}

with s$_{\mathrm{ F\rightarrow F'}}$ the relative intensities from Table~\ref{table:13CII} and $\sigma$ is the rms noise level of the observation. The $F,F'$-sum in Eq.~\ref{eq:tmbavg} runs overall satellites that are not blended with the main \Cp{12} line. Therefore, each satellite is scaled-up to the total \Cp{13} intensity and then averaged, independent of the number of satellites used. Using the \Cp{13} average spectrum, Eq.~\ref{eq:12tau} reads:

\begin{equation} \label{eq:avgtau}
\frac{\mathrm T_{\mathrm{mb},12}(v)}{\mathrm T_{\mathrm{mb},13,tot}(v)}
\simeq
\frac{1-e^{-\tau(v_{12})} \, \alpha^+}{\tau(v_{12})}
.\end{equation}

\begin{figure*}
\centering
\begin{subfigure}{\hsize}
   \centering
   \includegraphics[width=0.70\hsize]{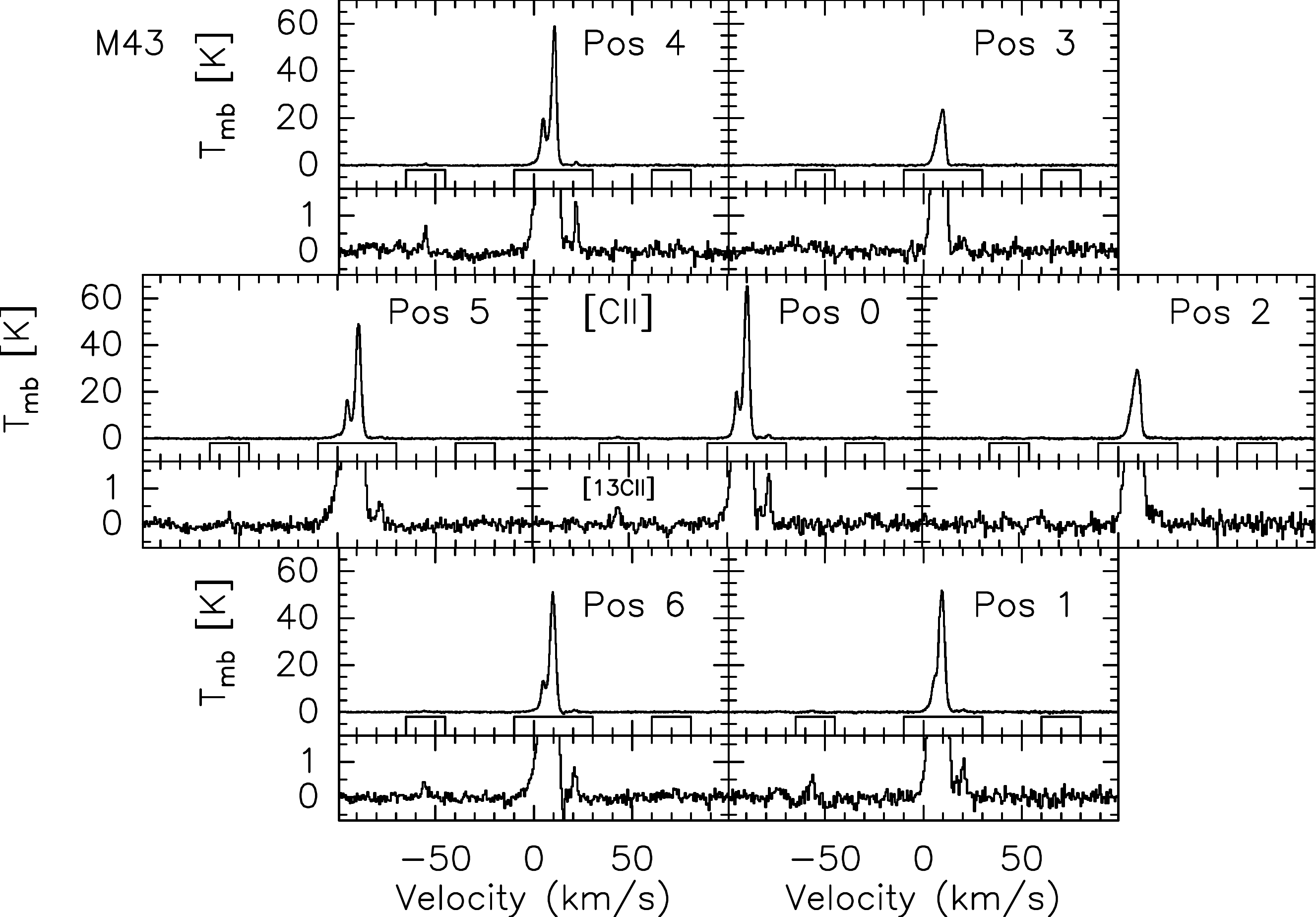}
     \caption{}
         \label{M43CIIbase}
\end{subfigure}%
\vspace{5mm}
\begin{subfigure}{\hsize}
   \centering
   \includegraphics[width=0.70\hsize]{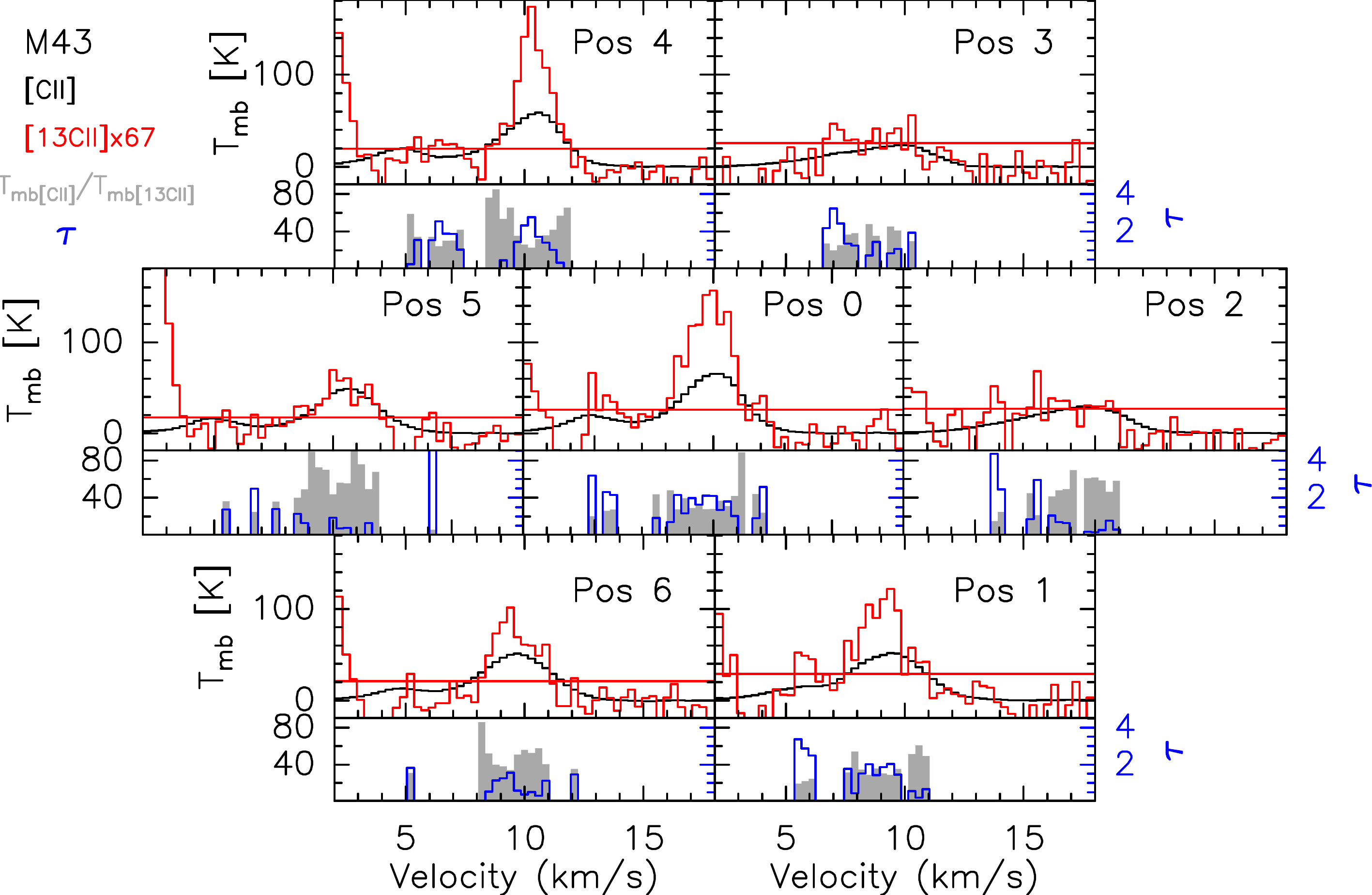}
   \caption{}
         \label{M43all}   
\end{subfigure}
\caption{(\subref{M43CIIbase}) - Mosaic observed in M43. For each position, the \Cp{12} line profile is shown in the top box. Below the spectra, we show the windows for the base line subtraction (-65,-45) (-10,30) (60,80)~km/s.
      The bottom box shows a zoom to the \Cp{13} satellites. (\subref{M43all}) - M43 mosaic of the seven positions observed by upGREAT. For each position, we show in the top panel a comparison between \Cp{12} (in black) and \Cp{13} (in red), the latter averaged over the hyperfine satellites and  scaled-up by the assumed value of $\alpha^+$ = 67. The red line corresponds to 1.5~$\sigma$ scaled-up by $\alpha^+$. The bottom panels show for all observation above 1.5 $\sigma$, the 
\Cp{12}/\Cp{13} intensity ratio per velocity bin (in gray) and the optical depth from the zeroth-order analysis (blue).  }
\label{M43mosaic}
\end{figure*}

\begin{figure*}
\centering
\begin{subfigure}{\hsize}
   \centering
   \includegraphics[width=0.70\hsize]{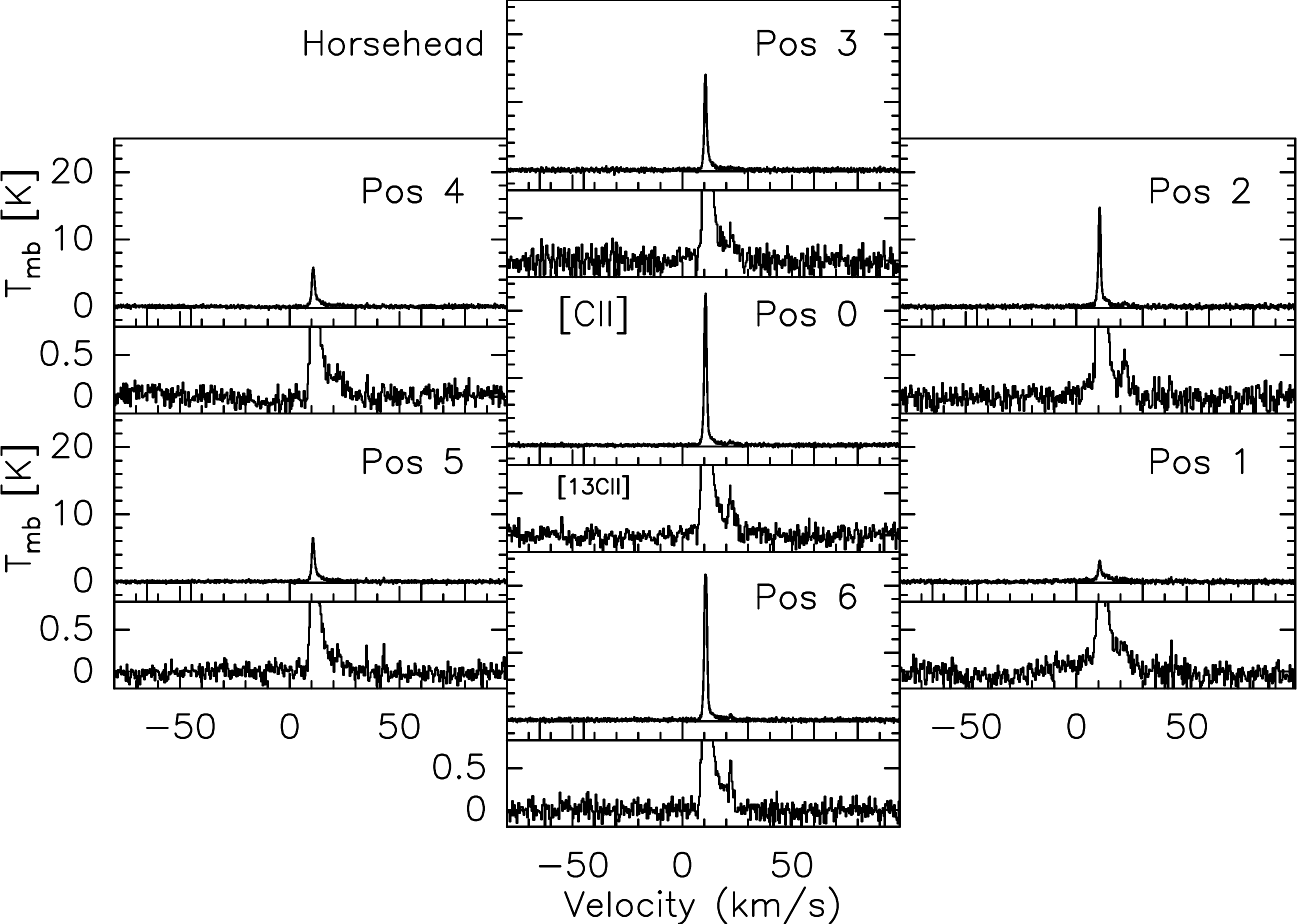}
      \caption{}
         \label{HorCIIbase}
\end{subfigure}%
\vspace{5mm}
\begin{subfigure}{\hsize}
   \centering
   \includegraphics[width=0.70\hsize]{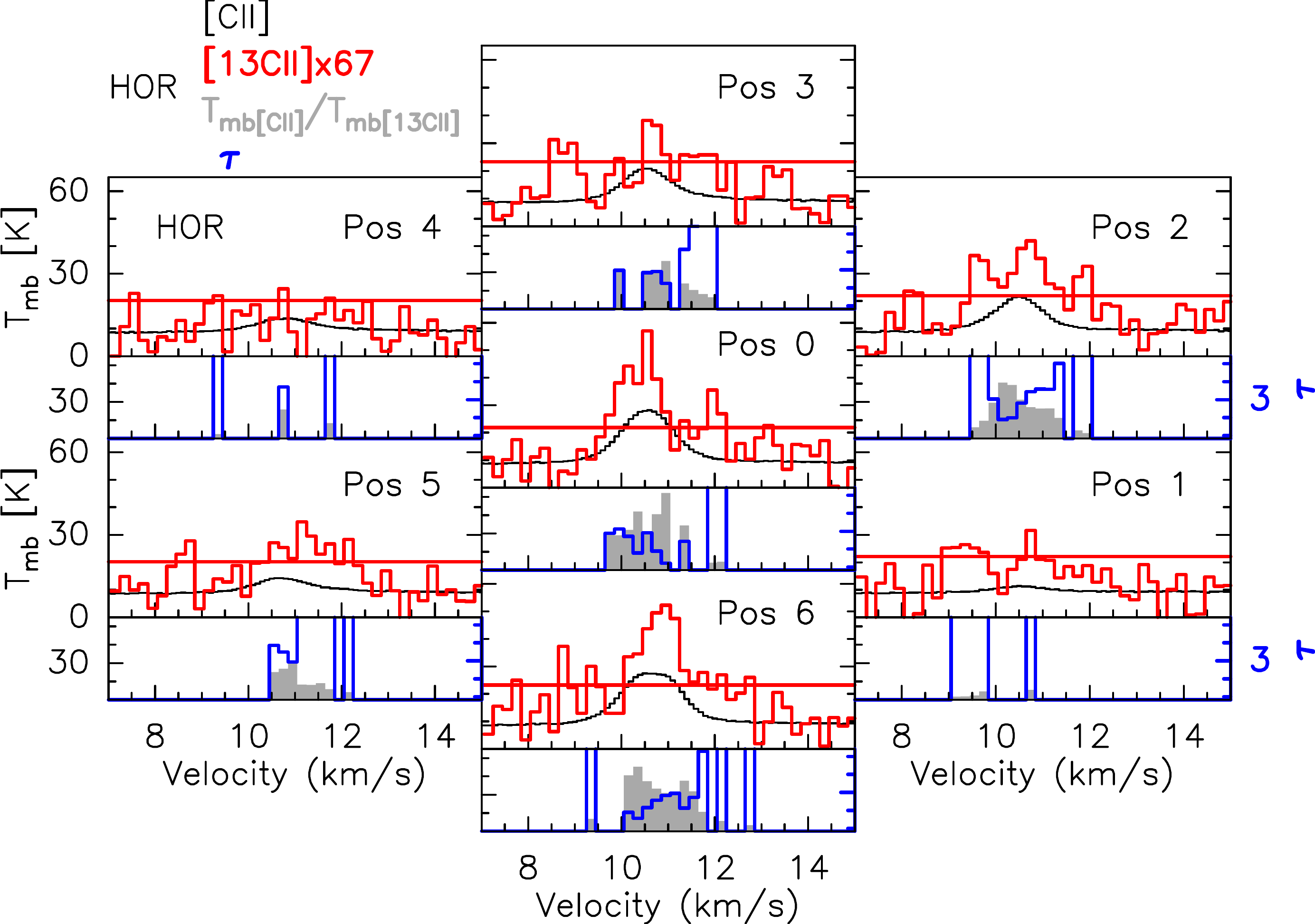}
      \caption{}
         \label{Horall} 
\end{subfigure}
\caption{(\subref{HorCIIbase}) - Same as Fig.~\ref{M43CIIbase}, but for the Horsehead PDR observations, with the windows for the base line subtraction at (-65,-45) (0,30) (60,80)~km/s. (\subref{Horall}) - Same as Fig.~\ref{M43all}, but for the Horsehead PDR observations and an assumed $\alpha^+$ = 67.}
\label{HORmosaic}
\end{figure*}

\begin{figure*}
\centering
\begin{subfigure}{\hsize}
   \centering
   \includegraphics[width=0.60\hsize]{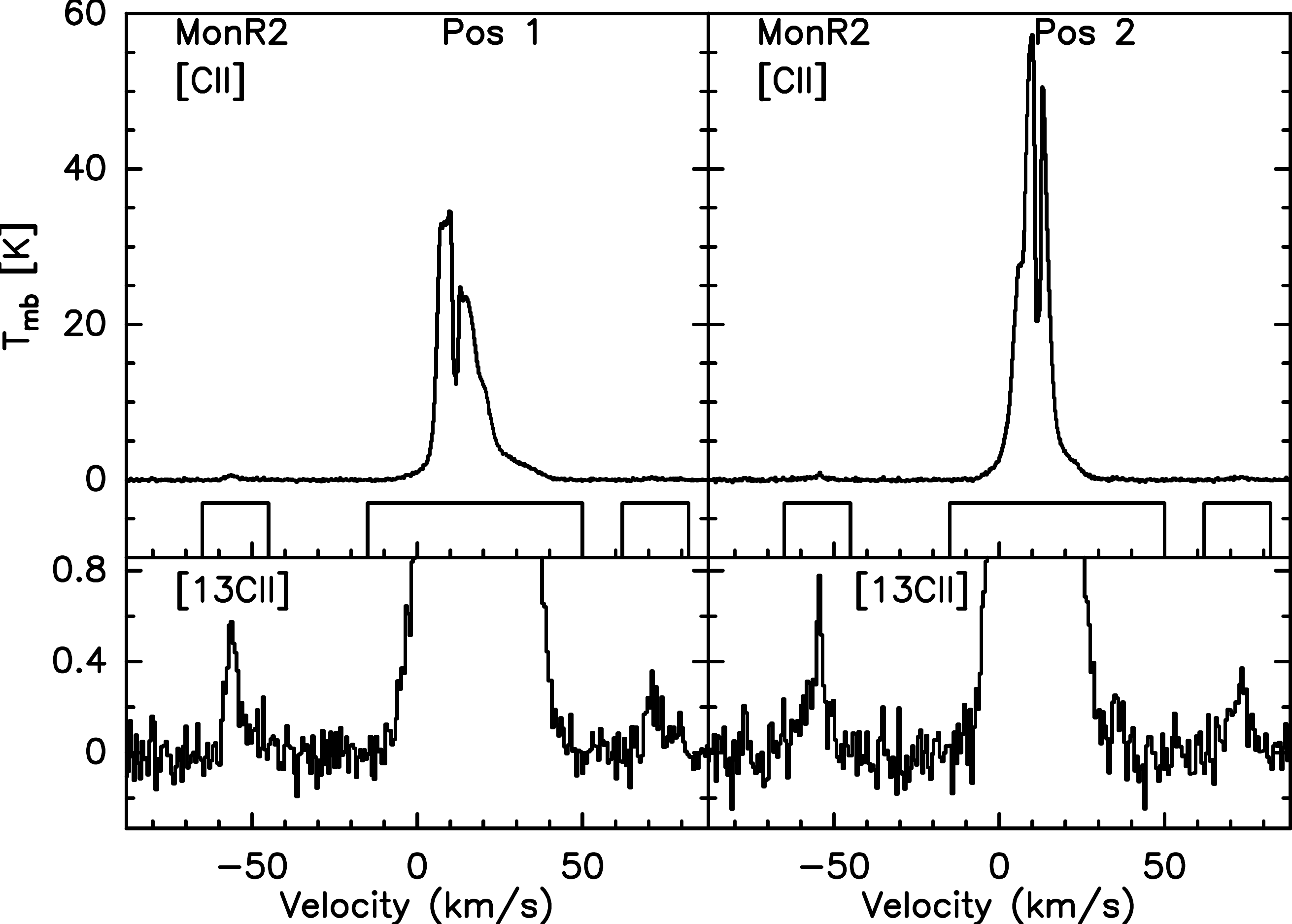}
      \caption{}
         \label{MonR2CIIbase}
\end{subfigure}%
\vspace{5mm}
\begin{subfigure}{\hsize}
   \centering
  \includegraphics[width=0.60\hsize]{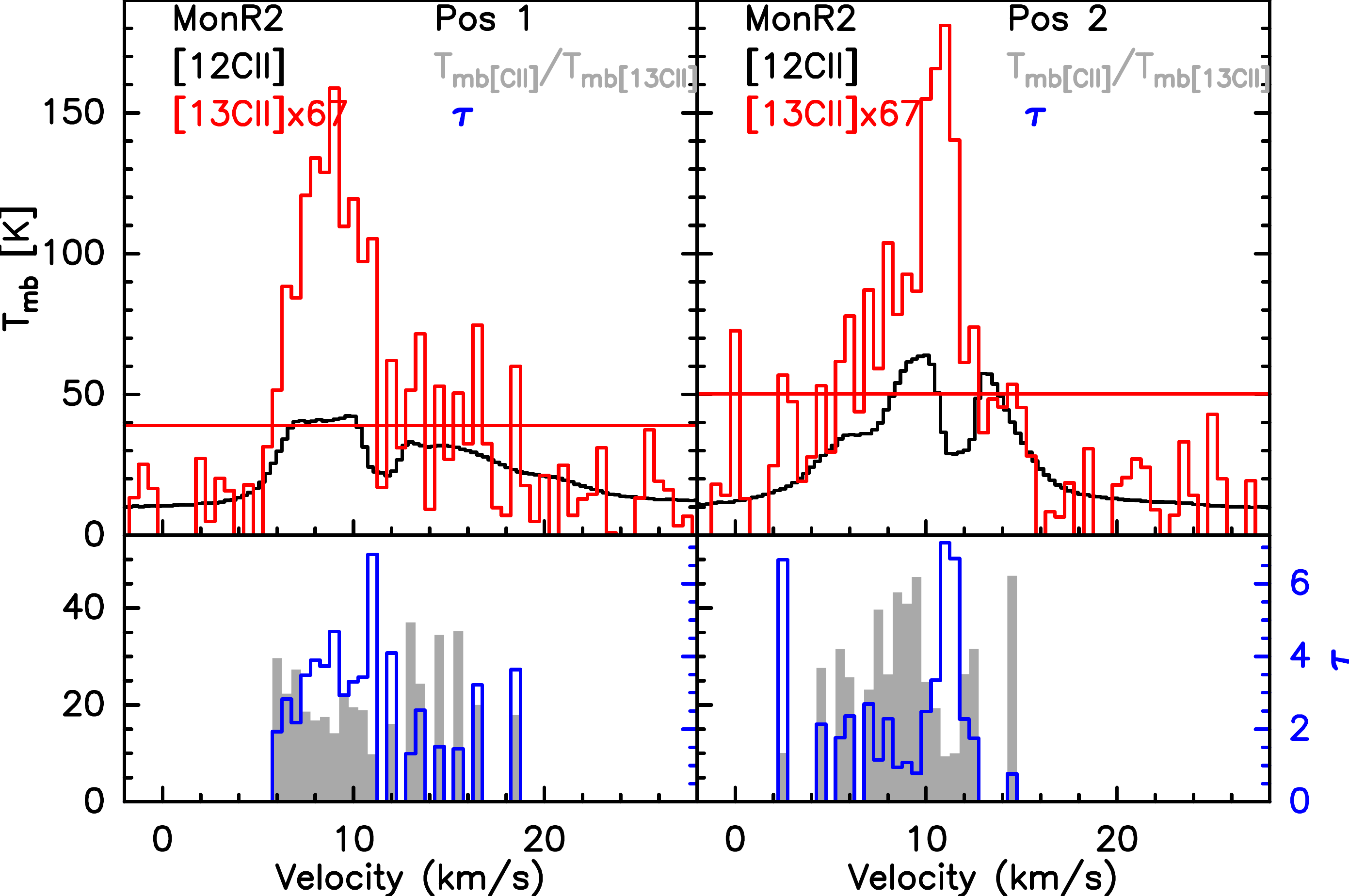}
      \caption{ }
         \label{Monall}
\end{subfigure}
\caption{(\subref{MonR2CIIbase}) - Same as Fig.~\ref{M43CIIbase}, but for the Mon~R2 observations, with the windows for the base line subtraction at (-65,-45) (0,30) (60,80)~km/s. (\subref{Monall}) - Same as Fig.~\ref{M43all}, but for the 2 positions in Mon~R2 observed with GREAT~L2 and an assumed $\alpha^+$ = 67.}
\label{MonR2mosaic}
\end{figure*}

\begin{figure*}
\centering
\begin{subfigure}{\hsize}
   \centering
   \includegraphics[width=0.70\hsize]{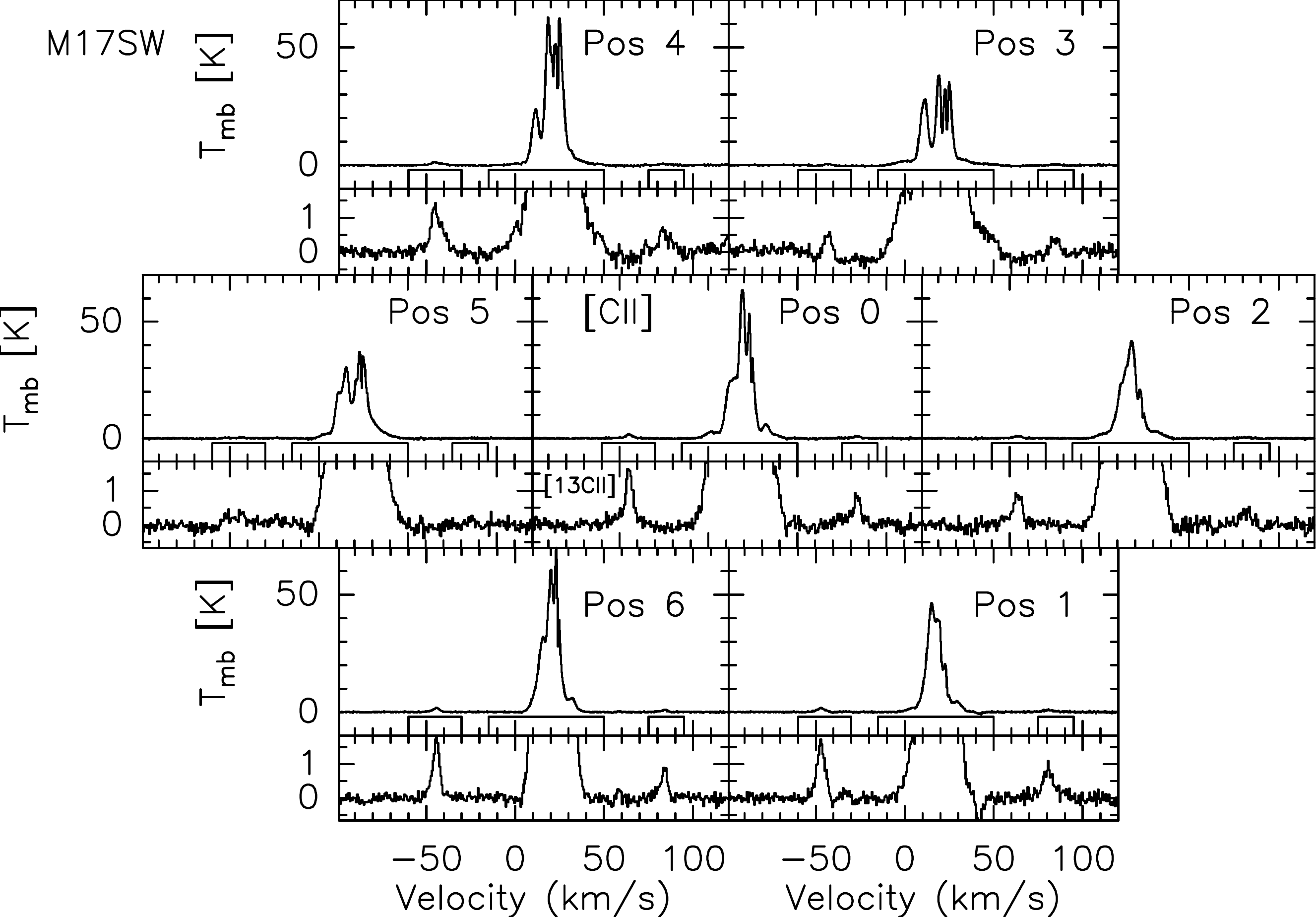}
      \caption{ }
         \label{M17CIIbase}
\end{subfigure}%
\vspace{5mm}
\begin{subfigure}{\hsize}
   \centering
   \includegraphics[width=0.70\hsize]{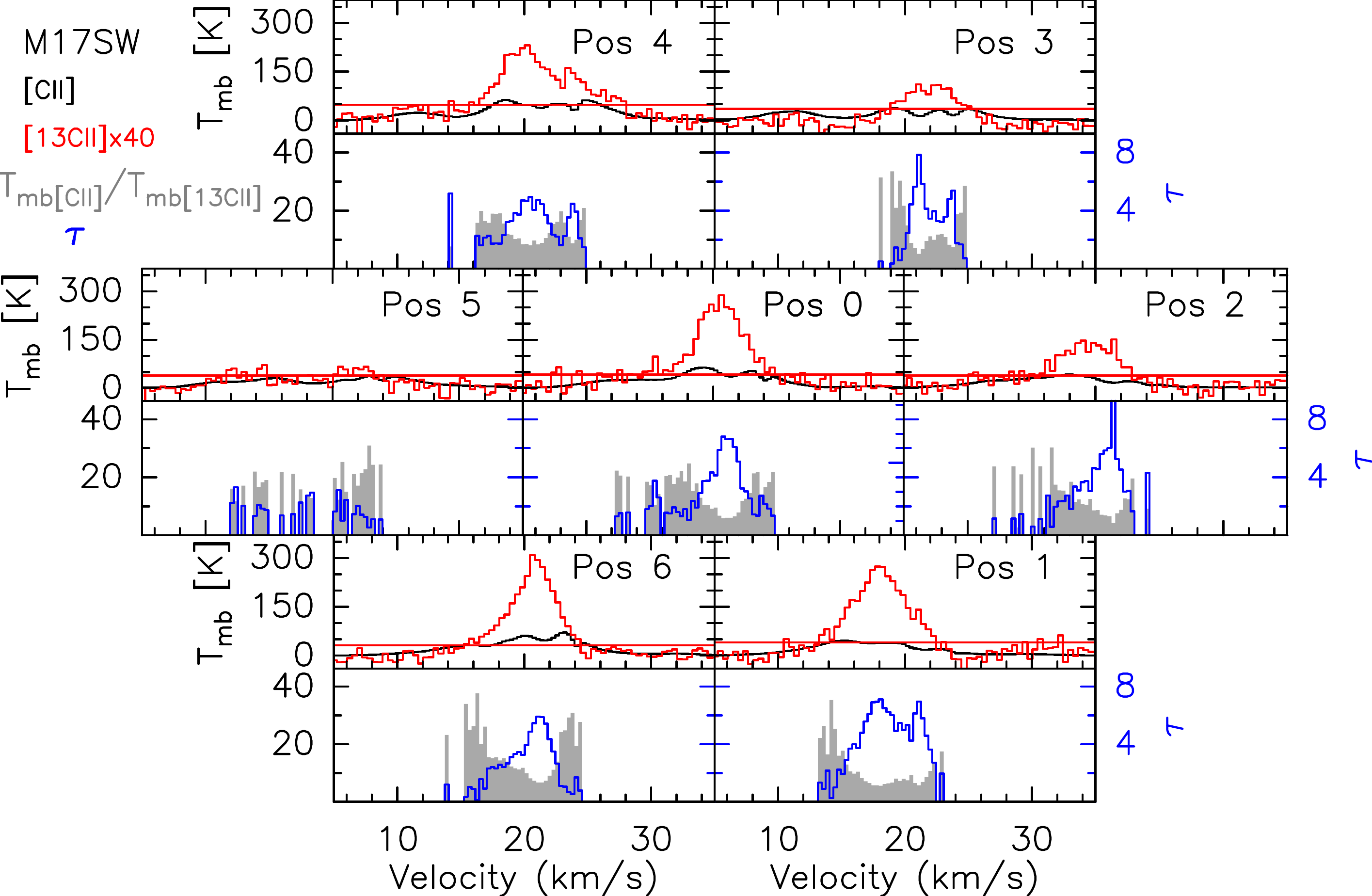}
      \caption{ }
         \label{M17CIIall}
\end{subfigure}
\caption{(\subref{M17CIIbase}) - Same as Fig.~\ref{M43CIIbase}, but for the M17~SW observations, with the windows for the base line subtraction at ($-$60,$-$30) ($-$15,50) (75,95)~km/s. (\subref{M17CIIall}) - Same as Fig.~\ref{M43all}, but for the M17~SW observations and an assumed $\alpha^+$ = 40.}
\label{M17mosaic}
\end{figure*}

These averaged \Cp{13} spectra, scaled-up by the factor $\alpha^+$, are plotted as the red histograms in Figures \ref{M43all} - \ref{M17CIIall}. The gray bar histograms in the lower panel give the \Cp{12}/\Cp{13} intensity ratio calculated from Eq.~\ref{eq:avgtau} for each velocity bin; the velocity range is restricted to where the \Cp{13} profiles show an intensity above $1.5~\sigma$ (see a discussion about the threshold in Section~\ref{abundanceratio} and the Appendix~\ref{app:dpeac}). For each spectrum observed, we numerically solve Eq.~\ref{eq:avgtau} 
for $\tau_{12}(v)$ for each velocity bin in this range. The thus derived opacity spectra are shown as the blue histograms in the lower panels of Figures \ref{M43all} to \ref{M17CIIall}. The velocity ranges above the \Cp{13} thresholds for each source are given below. \par

For M43, the $1.5~\sigma$ \Cp{13} threshold of 0.35 K results in a useful velocity range from 3~km/s to 17~km/s. In Fig.~\ref{M43all}, we see that the \Cp{12}/\Cp{13} line ratio ranges typically around 40 in those spectral regimes, where the \Cp{13} intensity is weak. Thus, it is close to the threshold emission level. The high S/N in these deep integrations with upGREAT/SOFIA thus results in the possibility of measuring directly the \Cp{12}/\Cp{13} abundance ratio from these regions of weak \Cp{13} emission (see Sect.~\ref{abundanceratio}). The opacity derived from the observed \Cp{12}/\Cp{13} line ratio shows the line to be optically thick along the main emission region for positions 0 and 4, with an optical depth of 2 around the \Cp{13} peak. \par

For the Horsehead PDR, we have subtracted the red wing emission visible in \Cp{12}. The \Cp{12} and \Cp{13} emission profiles are similar and peak at the same LSR-velocity, see Fig.~\ref{Horall}. The velocity range used above a \Cp{13} threshold of 0.22~K is 3~km/s to 17~km/s. The \Cp{13} S/N is not sufficient for a proper estimation of the line ratio outside the line center. The emission is optically thick along the main ridge for positions 0, 2, 3, and 6, with an optical depth of $\sim$ 2. For the outer positions, the low S/N does not allow for a good estimate of the optical depth. \par 

For Mon~R2, the useful velocity range above a \Cp{13} threshold of 0.45~K ranges from 6- 15~km/s (see Fig.~\ref{Monall}). The line profiles match only in the red line wing emission. The \Cp{12}/\Cp{13} T$_{\mathrm{mb}}$ ratio varies outside the peak and reaches average values of about 29, lower than the value of $\alpha^+=67$ we assumed for the source. Correspondingly, we also note that the derived \Cp{12} opacity is still well above unity in the wing region. The value of the optical depth, derived following the zeroth-order analysis, is very high in the line centers of both spectra, reaching up to a value of 7 around the \Cp{13} peak for both positions. The LSR-velocities of the peak emission in \Cp{13} and \Cp{12} are slightly shifted. The \Cp{12} at position~1 is flat-topped and both positions show strong dips in the emission profile. This indicates that the assumption of the zeroth-order analysis, namely that of a single component, is insufficient for explaining the line profiles of Mon~R2. \par

The situation is similar for M17~SW. At all seven positions observed the \Cp{13} spectra, scaled-up with the assumed abundance ratio, overshoot in the line centers, and match in the line wings. In addition, the \Cp{12} spectra show several emission peaks or absorption dips, whereas the \Cp{13} spectra exhibit smooth line profiles. With a \Cp{13} threshold of 0.5~K, the useful velocity range is from 12~km/s to 28~km/s (see Fig.~\ref{M17CIIall}). The \Cp{12}/\Cp{13} T$_{\mathrm{mb}}$ ratio in the line wings tends to be between 15 and 30, with considerable variations but on average well below 40, the value for $\alpha^+$ assumed for the source. Correspondingly, the optical depth is still around or 
above unity, even in the wings. We find that the emission is optically thick in the line centers in all positions, with an optical depth between 4 and 7, except at position 5, where it is located outside the main ridge of emission with an optical depth closer to unity. \par  

For all four sources, the zeroth-order analysis shows that the \Cp{12}  emission is optically thick, reaching high values for Mon~R2 and M17~SW, in particular. For both these sources, the \Cp{12} spectra are partially flat-topped, and they show a complex velocity structure with several emission components or absorption dips, clearly indicating that the simplified assumptions of the zeroth-order analysis are not met. Thus, the derived high optical depth values for these last sources are true only under the single layer assumption, and no further conclusions can be derived. A more sophisticated, multi-component analysis (see Sect.~\ref{Analysis}) must be used to analyze the line and to derive proper physical properties. 

\subsubsection{\Cp{13} Column density} \label{13cd}

If we assume that \Cp{13} is optically thin, we can estimate its column density as a function of the integrated intensity of the emission. The optical depth as a function of the velocity is:

\begin{align} \label{eq:tau}
  \begin{split}
\tau(v) = \phi(v) \frac{g_{\mathrm u}}{g_{\mathrm l}} \frac{c^{3}}{8 \pi \nu_{ul}^{3} } \, A_{\mathrm{ul}} \, N(\ion{C}{II}) \, \frac{\left(1 - e^{-T_0/T_{\mathrm{ex}}} \right)}{1 + \frac{g_{\mathrm u}}{g_{\mathrm l}} e^{-T_0/T_{\mathrm{ex}}}}, \\
 \end{split}
\end{align}

using a normalized profile function $\int \phi(v) \, dv = 1$. A$_{\mathrm{ul}}$ is the \Cp{} fine structure transition's Einstein A coefficient for spontaneous emission \citep[2.3x10$^{-6}$~s$^{-1}$,][]{2007JPCRD..36.1287W}, $g_{\mathrm u,\mathrm l}$ are the statistical weights of the \Cp{12} levels, with $g_{\mathrm u}$ = 4 and $g_{\mathrm l}$ = 2, $N$(\ion{C}{II}) is the \Cp{} column density, $\nu_{ul}$ is the frequency of the \Cp{} (1900.5369~GHz) and $T_0=h\nu_{\mathrm{ul}} /k$ is the equivalent temperature of the excited level, 91.25~K. \par

Now from Eq.~\ref{eq:detequ}, if the emission is optically thin, $\tau\ll$1, we can approximate the optical depth as $(1-e^{-\tau_{\nu}})\approx \tau_{\nu}$. The integral of the brightness temperature, using Eq.~\ref{eq:tau} for the optical depth, is:

\begin{equation}
\label{eq:inttmb2}
\int \mathrm T_{\mathrm{mb}}(v) dv = \eta_{\phi} T_0 \frac{c^3}{8\pi\nu_{ul}^3} A_{ul} N(\ion{C}{II}) \left( \frac{1}{1+ \frac{g_l}{g_u}e^{T_0/T_{\mathrm{ex}}}} \right)
,\end{equation}

Rearranging Eq.~\ref{eq:inttmb2} for the column density gives:

\begin{equation}
\label{eq:ncii}
N(\ion{C}{II}) = \frac{8\pi\nu_{ul}^3}{T_0 c^3 \eta_{\phi} A_{ul}} \mathrm f(T_{\mathrm{ex}}) \int \mathrm T_{\mathrm{mb}}(v) dv,  
\end{equation}

with $ \mathrm f(T_{\mathrm{ex}}) = 1+ \frac{g_l}{g_u}e^{T_0/T_{\mathrm{ex}}}$. In the high limit $T_{\mathrm{ex}} \rightarrow \infty $ and f($T_{\mathrm{ex}}$) $\rightarrow 3/2$. Higher values occur at lower excitation temperatures. Therefore, we can define a minimum column density as:

\begin{equation}
\label{eq:nciimin}
 N_{\mathrm{min}}(\ion{C}{II}) = \frac{3}{2} \frac{8\pi\nu_{ul}^3}{\eta_{\mathrm{mb}} T_0 c^3 A_{ul}} \int \mathrm T_{\mathrm{mb}}(v) dv
\end{equation}

and the column density as:

\begin{equation} \label{eq:ncii2}
  N(\ion{C}{II}) =  \frac{2}{3} \, \mathrm f(T_{\mathrm{ex}}) \, N_{\mathrm{min}}(\ion{C}{II}) ;\, \frac{2}{3} \, \mathrm f(T_{\mathrm{ex}}) >1
.\end{equation}

\begin{figure}
   \centering
   \includegraphics[width=0.87\hsize]{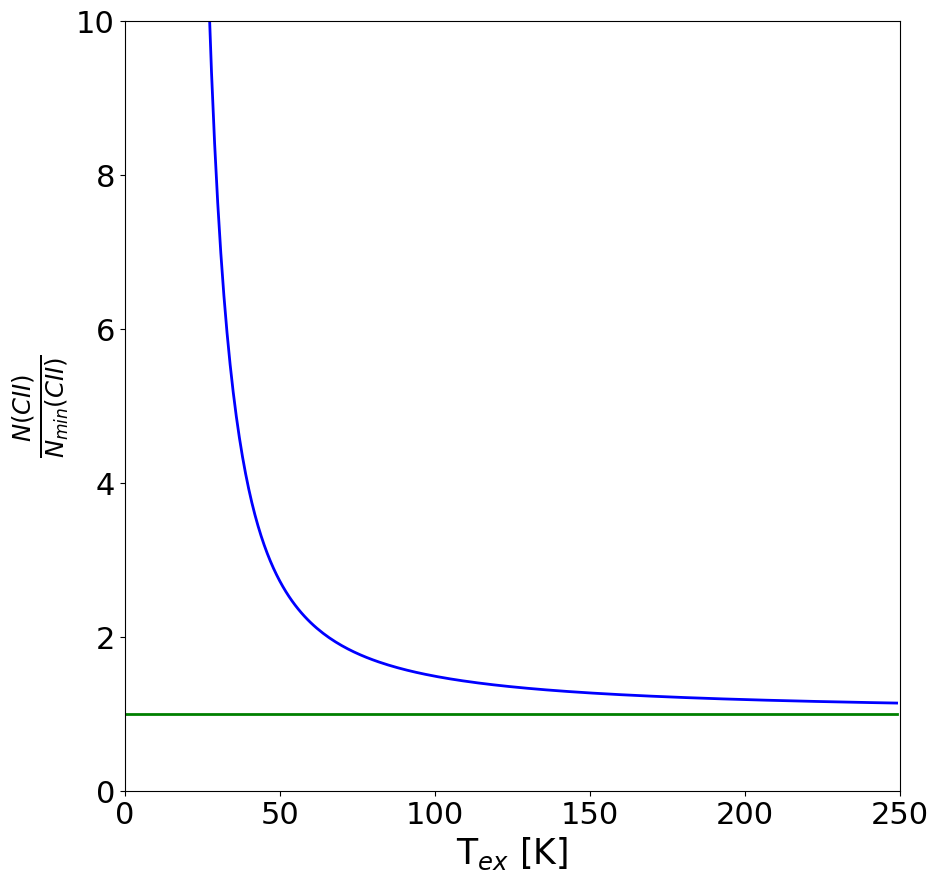}
      \caption{Ratio between $N$(CII) and $N_{\mathrm{min}}$(CII) assuming a beam filling factor of 1 as a function of $T_{\mathrm{ex}}$. Above $T_{\mathrm{ex}}=T_0=91.2~K$, the increase relative to the minimum value, is well below a factor of 2.
            }
         \label{ftex}
   \end{figure}

In Figure~\ref{ftex}, we show how the value of $T_{\mathrm{ex}}$ affects the estimated column density.

In Table \ref{13CIIc}, we list the values derived for the \Cp{13} minimum column density by using Eq.~\ref{eq:nciimin} with the \Cp{13} integrated line intensity (the weighted sum over all observationally resolved hyperfine satellites, see Eq.~\ref{eq:tmbavg}) for all positions in each source. We also list the \Cp{12} minimum column densities obtained by scaling the $N$(\Cp{13}) values with $\alpha^+$. For comparison, we also list the minimum $N$(\Cp{12}) that would be obtained directly from the \Cp{12} integrated intensity in Eq.~\ref{eq:nciimin} under the obviously wrong assumption of optically thin emission. We also give the ratio between the $N$(\Cp{12}) derived from the scaled-up \Cp{13} and the $N$(\Cp{12}) that is derived assuming optically thin emission. This emphasizes how the derived \Cp{12} column density changes by taking into account the optical depth and the self-absorption effects described above in Section~\ref{C12od}. We note that observations without sufficient velocity resolution to reveal the \Cp{13} hyperfine lines and the detailed \Cp{12} line profiles (and, hence, the individual different source components along the line-of-sight) would be limited to the values derived from the line-integrated \Cp{12} profiles.\par

In the following, we always convert the derived \Cp{12} column densities to an equivalent visual extinction, where we use the standard value of $1.2\times10^{-4}$ for the relative abundance of hydrogen to carbon \citep{2008ApJ...680..371W} and the canonical conversion factor between hydrogen column density and visual extinction of $1.87\times 10^{21} \,{\rm cm}^{-2}/A_{\mathrm V}$ \citep{1973A&A....26..257R,1978ApJ...224..132B,1994ApJ...427..274D,1995A&A...293..889P}, hence $A_{\mathrm V} = N(\Cp{12}) * 0.83 * 10^{4}/1.87\times 10^{21}\,{\rm cm}^{-2}$. We note that these equivalent extinctions are a lower limit as they assume that all carbon is the form of C$^{+}$.

        \begin{table*}
   \begin{threeparttable}
  \centering
    \caption{\Cp{13} integrated intensity and minimum column density, \Cp{12} minimum column density scaled-up using $\alpha$ and equivalent extinction. \Cp{12} integrated intensity, 
    minimum column density assuming optically thin emission and equivalent visual extinction. For the derivation of the values, we have used more digits than the one presented here.}
    \begin{tabular}{l|cccc|ccc|c}
      \hline
      \hline
            &      \multicolumn{4}{|c|}{\Cp{13}} & \multicolumn{3}{c|}{Optically thin \Cp{12}} & Ratio  \\
            \hline
\multicolumn{1}{c|}{Positions} & \Cp{13} Int. & N$_{\mathrm{min}}$(\Cp{13}) & N$_{\mathrm{min}}$(\Cp{})\tnote{a} & A$_{\mathrm{v,min}}$\tnote{b} & \Cp{12} Int. & N$_{\mathrm{min}}$(\Cp{})\tnote{c} & A$_{\mathrm{v,min}}$\tnote{d} & $\frac{A_{\mathrm{v,min}}(\Cp{13})\tnote{b}}{A_{\mathrm{v,min}}(\Cp{12})\tnote{d}}$\\
                              & Intensity    &                             &  \Cp{13}                 & \Cp{13}              & Intensity    &  \Cp{12}                         & \Cp{12}              &    \\
                              & (K km/s)     & (cm$^{-2}$)                 & (cm$^{-2}$)               &  (mag.)                    & (K km/s)     & (cm$^{-2}$)               &  (mag.)                    &  \\
          \hline
M43 Pos. 0   & 5.5 & 2.5E16 & 1.7E18 & 7.4 & 283.1 & 1.3E18  & 5.6 & 1.3 \\
M43 Pos. 1   & 4.3 & 1.9E16 & 1.3E18 & 5.7 & 249.2 & 1.1E18  & 4.9 & 1.2 \\
M43 Pos. 2   & 2.6 & 1.2E16 & 7.7E17 & 3.4 & 172.2 & 7.7E17  & 3.4 & 1.0  \\ 
M43 Pos. 3   & 2.6 & 1.1E16 & 7.6E17 & 3.4 & 134.0 & 6.0E17  & 2.6 & 1.3 \\ 
M43 Pos. 4   & 5.5 & 2.5E16 & 1.7E18 & 7.4 & 270.1 & 1.2E18  & 5.3 & 1.4 \\
M43 Pos. 5   & 3.7 & 1.6E16 & 1.1E18 & 4.9 & 227.4 & 1.0E18  & 4.5 & 1.1 \\ 
M43 Pos. 6   & 4.1 & 1.8E16 & 1.2E18 & 5.4 & 237.9 & 1.1E18  & 4.7 & 1.1 \\
\hline
HOR Pos. 0   & 1.2 & 5.3E15 & 3.6E17 & 1.6 & 39.6 & 1.8E17  & 0.8 & 2.0 \\
HOR Pos. 1   & 0.7 & 3.1E15 & 2.1E17 & 0.9 & 11.2 & 5.0E16  & 0.2 & 4.2 \\ 
HOR Pos. 2   & 1.4 & 6.1E15 & 4.1E17 & 1.8 & 26.6 & 1.2E17  & 0.5 & 3.4 \\
HOR Pos. 3   & 1.0 & 4.7E15 & 3.1E17 & 1.4 & 25.7 & 1.1E17  & 0.5 & 2.7  \\
HOR Pos. 4   & 0.3 & 1.2E15 & 8.4E17 & 0.4 & 14.8 & 6.6E16  & 0.3 & 1.3 \\
HOR Pos. 5   & 0.9 & 3.9E15 & 2.6E17 & 1.2 & 14.7 & 6.5E16  & 0.3 & 4.0  \\
HOR Pos. 6   & 1.6 & 7.0E15 & 4.7E17 & 2.1 & 41.5 & 1.9E17  & 0.8 & 2.5  \\
\hline
MonR2 Pos. 1 & 12.2 & 5.5E16 & 3.7E18 & 16.3 & 410.8 & 1.8E18  & 8.1 & 2.0  \\
MonR2 Pos. 2 & 11.4 & 5.1E16 & 3.4E18 & 15.2 & 477.0 & 2.1E18  & 9.5 & 1.6  \\
\hline
M17SW Pos. 0 & 41.6 & 1.9E17 & 7.4E18 & 33.0 & 657.2 & 2.9E18  & 13.1 & 2.5  \\
M17SW Pos. 1 & 39.1 & 1.7E17 & 7.0E18 & 31.1 & 460.1 & 2.1E18  & 9.1 & 3.4  \\
M17SW Pos. 2 & 26.9 & 1.2E17 & 4.8E18 & 21.3 & 458.1 & 2.0E18  & 9.1 & 2.3 \\
M17SW Pos. 3 & 16.5 & 7.4E16 & 2.9E18 & 13.1 & 489.9 & 2.2E18  & 9.7 & 1.3  \\
M17SW Pos. 4 & 45.1 & 2.0E17 & 8.1E18 & 35.9 & 722.7 & 3.2E18  & 14.4 & 2.5   \\
M17SW Pos. 5 & 14.1 & 6.3E16 & 2.5E18 & 11.2 & 521.7 & 2.3E18  & 10.4 & 1.1  \\
M17SW Pos. 6 & 34.3 & 1.5E17 & 6.1E18 & 27.3 & 617.7 & 2.8E18  & 12.3 & 2.2  \\
       \label{13CIIc}
\end{tabular}
\begin{tablenotes}\footnotesize
\item[a] \Cp{12} column density derived from the scaled-up \Cp{13} column density.
\item[b] \Cp{12} equivalent visual extinction derived from the scaled-up \Cp{13} column density by $\alpha^+$ corresponding to each source.
\item[c] \Cp{12} column density derived directly from the \Cp{12} integrated intensity assuming optically thin regime. 
\item[d] \Cp{12} equivalent visual extinction derived directly from the \Cp{12} integrated intensity assuming optically thin regime. 
\end{tablenotes}
\end{threeparttable}
 \end{table*}

Table \ref{13CIIc} shows high column densities and equivalent visual extinctions for all four sources, especially for Mon~R2 and M17~SW. Using the \Cp{13} intensities, the lower limit of the equivalent A$_\mathrm{V}$ reaches up to around 36~magnitudes for one position. We emphasize that these high equivalent column densities and equivalent visual extinctions are derived directly from the \Cp{13} integrated line intensities, assuming low optical depth, that is, by counting \Cp{13} atoms emitting from the upper fine structure state. An excitation temperature below 100~K would further increase the column densities (see Fig.~\ref{ftex}). \par

The beam-averaged equivalent extinctions, which are already lower limits, derived from \Cp{13} are in the range of 1 to a few for the Horsehead PDR, and range up to about 7~A$_\mathrm{V}$ for M43. We note that the Horsehead PDR optical depth estimated above in Section~\ref{C12od} is similar to the one derived for M43, whereas the \Cp{13} column density is lower: the optical depth is given by the column density per velocity element and, hence, the smaller line width in the Horsehead PDR compensates for the lower column density to give a similar optical depth. Now the derived \Cp{12} equivalent extinctions and the ratio between \Cp{13} and \Cp{12} extinctions show that the scaled-up \Cp{13} A$_{\mathrm{v}}$ is similar or higher than the one from the assumed optically thin \Cp{12}, especially in the Horsehead PDR. For M43, the ratios around unity likely result from a combination of optically thin emission in the outer positions, low S/N in the \Cp{13} wings plus the additional \Cp{12} emission not traced by \Cp{13}. For the Horsehead PDR, with its similar optical depth, the higher ratios result from the higher noise that tends to increase the estimation of the \Cp{13} integrated intensity. For Mon~R2 and M17~SW, the column densities derived from \Cp{13} correspond to an equivalent (minimum) A$_\mathrm{V}$ around 20 and up to 36~mag. \par 

We emphasize that the more sophisticated analysis below in Section~\ref{Analysis}, taking into account the optical depth of \Cp{12} implied by the large \Cp{} column, along with the assumption of a multi-layered source structure with background emission and foreground absorption does not change the fundamental result of very high \Cp{} column densities. 
In fact, the column densities derived here from the optically thin \Cp{13} emission are the minimum possible, due to the fact that we have derived the column densities in the high excitation temperature limit in the current analysis. 
The column densities and total equivalent visual extinctions derived in the different scenarios below always add up to at least the amount of material derived from the simple analysis of the \Cp{13} integrated intensities. Typically, there is additional \Cp{12} that not detected in the \Cp{13} emission due to its limited S/N. \par

\subsection{Multi-component analysis: multi-component dual layer model} \label{Analysis}

The discussion in the previous section offers clear evidence of high column density and optically thick emission for all sources and, in addition, absorption features for Mon~R2 and M17~SW. It is also clear that the simple approximation of a single component source is not adequate. Therefore, we follow an approach similar to the one by \citet{2012A&A...542L..16G} as in the case of NGC2024 for deriving the physical properties of the \Cp{} emission. The objective is to explain the \Cp{12} and \Cp{13} main beam temperature line profiles by a composition of multiple Gaussian source components through a least-square fit to the observed profiles. The source is assumed to contain two layers, a background emission layer with a variable number of components adapted to the observed structure of the \Cp{13} and \Cp{12} line profile, and, in the case of absorption features in the \Cp{12} profile, a foreground absorption layer with a different number of Gaussian components. For completeness, we mention that also a single layer, multi-component model gives formally correct fits to the observed line profiles, as is discussed in Appendix~\ref{singlelayer}. However, we have ruled out this scenario as being physically implausible, as discussed in Appendix~\ref{singlelayer}.\par

We use the radiative transfer equation for \Cp{} \citep{1985ApJ...291..755C,1985PhDT.........9S,2012ApJS..203...13G}, where each source component $i$ is characterized by four parameters: the excitation temperature $T_{\mathrm{ex},i}$, the \Cp{12} column density $N_{i}$(\ion{\element[][12]{C}}{II}), its center velocity  $v_{\mathrm{LSR},i}$), and its FWHM velocity width $\Delta v_{\mathrm{LSR},i}$. The \Cp{13} column density is scaled down from the \Cp{12} column density by applying the abundance ratio $\alpha^+$, as specified above in Section~\ref{Observed Sources} ($\alpha^+$ $N_{i}$(\ion{\element[][13]{C}}{II}) = $N_{i}$(\ion{\element[][12]{C}}{II}) ); the abundance ratio is assumed to be the same for each source component. \par

We make three assumptions for the modeling process: \textit{i}) The excitation temperature is the same for the \Cp{12} and all three \Cp{13} hyperfine structure lines. \textit{ii}) \Cp{13} is always optically thin. \textit{iii}) If a \Cp{12} component does not have a visible \Cp{13} counterpart above the noise level, the \Cp{12} component is not affected by self-absorption effects. We use a superposition of Gaussian line profiles. Therefore, the line profile for 
each individual source component $i$ is characterized by its LSR-velocity, $v_{\mathrm{LSR},i}$, and full-width-half-maximum line width $\Delta v_{\mathrm{FWHM},i}$ as:

\begin{equation}  \label{eq:phii} 
\phi_i(v) = 
\dfrac{2\sqrt{\ln(2)}}{ \Delta v_{\mathrm{FWHM},i} \sqrt{\pi}} \exp\left(-\dfrac{(v-v_{\mathrm{LSR},i})^{2}4\ln(2)}{(\Delta v_{\mathrm{FWHM},i})^{2}}\right) 
.\end{equation} 

The combined line profile of each component $i$ for both the \Cp{12} main line  and all three \Cp{13} hyperfine satellites can be written as:

\begin{equation} \label{eq:phic}
\Phi_{i}(v) = \phi_{i}(v) + \sum_{F,F'} \beta_{F\rightarrow F'} \phi_{i}(v-\delta v_{F \rightarrow F'})
,\end{equation}\

where $\delta v_{\mathrm F \rightarrow \mathrm F'}$  is the \Cp{13} hyperfine satellite's velocity offset with respect to \Cp{12} (Table~\ref{table:13CII}), and $\beta_{\mathrm F\rightarrow \mathrm F'} =\frac{s_{\mathrm F \rightarrow \mathrm F'}}{\alpha^{+}}$, as defined in Eq.~\ref{eq:beta}. In Eq.~\ref{eq:phic}, the first term corresponds to the line profile of the \Cp{12} emission and the second one is the combined line profile of the three \Cp{13} satellites. With these definitions, we can define, using Eq.~\ref{eq:tau}, the optical depth profile $\tau_{i}(v)$  for each individual
source component $i$ as:

\begin{equation} \label{eq:tau2}
\tau_{i}(v) = \Phi_{i}(v)  \frac{g_{\mathrm u}}{g_{\mathrm l}} \frac{c^{3}}{8 \pi \nu^{3} } \, A_{\mathrm{ul}} \, N_{i}(\ion{\element[][12]{C}}{II}) \, \frac{\left(1 - e^{-T_0/T_{\mathrm{ex},i}} 
\right)}{1 + \frac{g_{\mathrm u}}{g_{\mathrm l}} e^{-T_0/T_{\mathrm{ex},i}}} 
.\end{equation}

Finally, from Eq.~\ref{eq:detequ}, the observed main beam temperature T$_{\mathrm{mb}}$ is the combination of the emission from the components $i_{b}$ of the background layer, absorbed through the combined absorption by the components $i_{f}$ of the foreground layer, plus the emission from the foreground layer, and is given by:
 
\begin{equation}  \label{eq:tmb} 
\begin{split}
  T_{\mathrm{mb}}(v) = & \left[ \sum_{i_{b}} \mathcal{J}_{\nu}(T_{\mathrm{ex},i_b}) \, \left( 1-e^{-\tau_{i_b}(v)}\right) \right] e^{-\sum_{i_f} \tau_{i_f}(v)} + \\
                       & \sum_{i_f} \mathcal{J}_{\nu}(T_{\mathrm{ex},i_f}) \, \left( 1-e^{-\tau_{i_f}(v)}\right).  \\
\end{split}
\end{equation}

The multi-component fitting is applied to Eq.~\ref{eq:tmb} in a physically motivated iterative process. We first note that the high S/N of the observed \Cp{12} profile does not leave much ambiguity with regard to the center velocity and width of additional Gaussian components. However, the fitting process is degenerate without any further constraints, as multiple 
combinations of T$_{ \mathrm e \mathrm x}$ and  $N_{i}$(\ion{\element[][12]{C}}{II}) exist for the same optical depth. $T_{ \mathrm e \mathrm x}$ and $N_{i}$(\ion{\element[][12]{C}}{II}) are, roughly, inversely proportional to each other, as we can see in Eq.~\ref{eq:tau2} in relation to the optical depth. For the line profiles without absorption notches requiring additional foreground components, $T_{\mathrm{ex}}$ is constrained by the (Rayleigh-Jeans corrected) peak brightness of each Gaussian component, so that it, like the column density, it can be handled as a free fit parameters (this applies to M43 and the Horsehead PDR spectra). For the cases of both background emission and foreground absorption, brighter background emission, namely higher excitation temperature in the background, can be compensated by deeper foreground absorption, hence higher column density and optical depth, in the foreground. We thus have to fix $T_{ \mathrm e \mathrm x}$ to a reasonable value. For the background layer, we have, as a lower boundary to $T_{\mathrm ex}$, the Rayleigh-Jeans corrected, observed T$_{ \mathrm m \mathrm b}$, which, to first order, adds $T_0/2\,=\,45.6\,{\rm K}$ to the brightness temperature. Also, according to PDR models, we expect an excitation temperature of 50~K to a few 100~K maximum. We have therefore decided, for these sources, to fix the $T_{ \mathrm e \mathrm x}$ for the background to values of 30 to 200~K, depending on the source (see Tables~\ref{M43gauss} to \ref{M17gaus} in the Appendix). We do this to keep the background optical depth at reasonable values, namely, close to unity or a bit higher. It is important to note that the fitted excitation temperatures are expected to be higher than the dust temperatures in the PDR due to the lose coupling of the gas and dust thermal balance. Thus, the gas is heated by the photoelectric effect by UV radiation, increasing its   temperature well above the dust temperature.  This is an inherent property of PDRs \citep{1985ApJ...291..722T}. Deep in the cloud, the situation may reverse (see also Sect.~\ref{COdis}). This is theoretically well understood \citep[see eg.,][]{2013A&A...549A..85R} and was observed, for example, for the S140 region by \citet{2015A&A...580A..68K}. \par 

For the foreground layer, $T_{ \mathrm e \mathrm x}$ must be low to act as an absorption layer without significant \Cp{13} emission: an upper boundary is given by the Rayleigh-Jeans corrected brightness temperature in the center of the absorption dips. Lower values result in lower column densities of the absorbing layer, as less material is needed to build up sufficient optical depth. For this reason, we have varied $T_{ \mathrm e \mathrm x}$ for the absorbing layer between 20~K and 45~K providing a kind of minimum column density. The optical depth is insensitive to changes below 20~K. Any change to the excitation temperatures for the same column density affects the optical depth by less than 5$\%$. Above 20~K we can see effects in the optical depth. At temperatures up to 45~K, we also fulfill the assumption that the contribution from the foreground layer to the emission is insignificant. \par

\begin{figure*}
\centering
\begin{subfigure}{\hsize}
  \centering
  \includegraphics[width=0.63\hsize]{{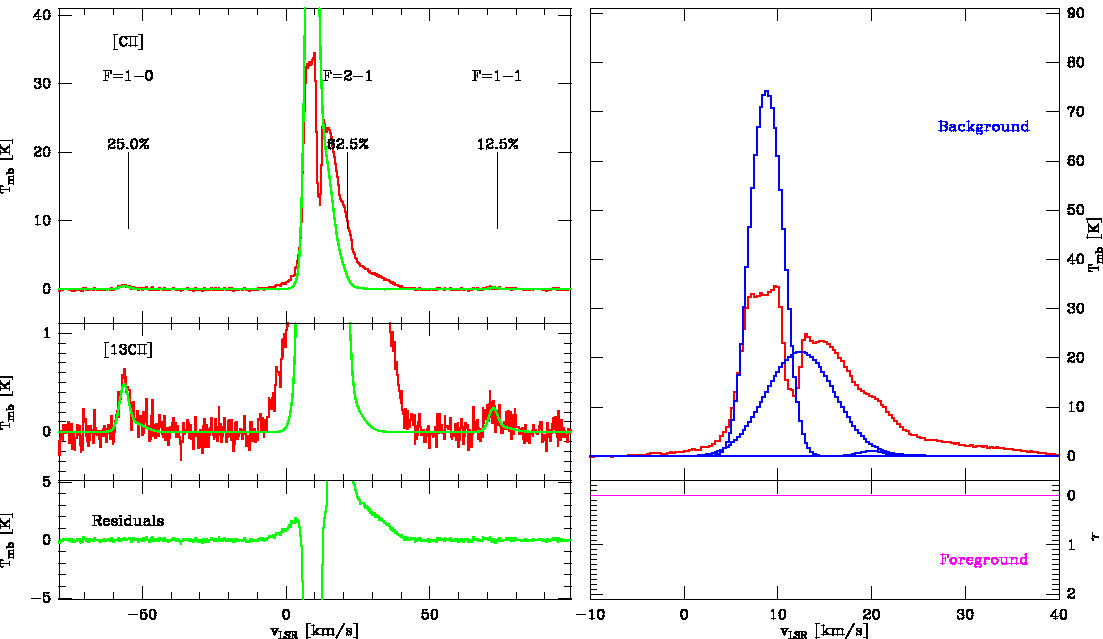}}
  \caption{}
  \label{fig:step1}
\end{subfigure}%
\par
\begin{subfigure}{\hsize}
  \centering
   \includegraphics[width=0.63\hsize]{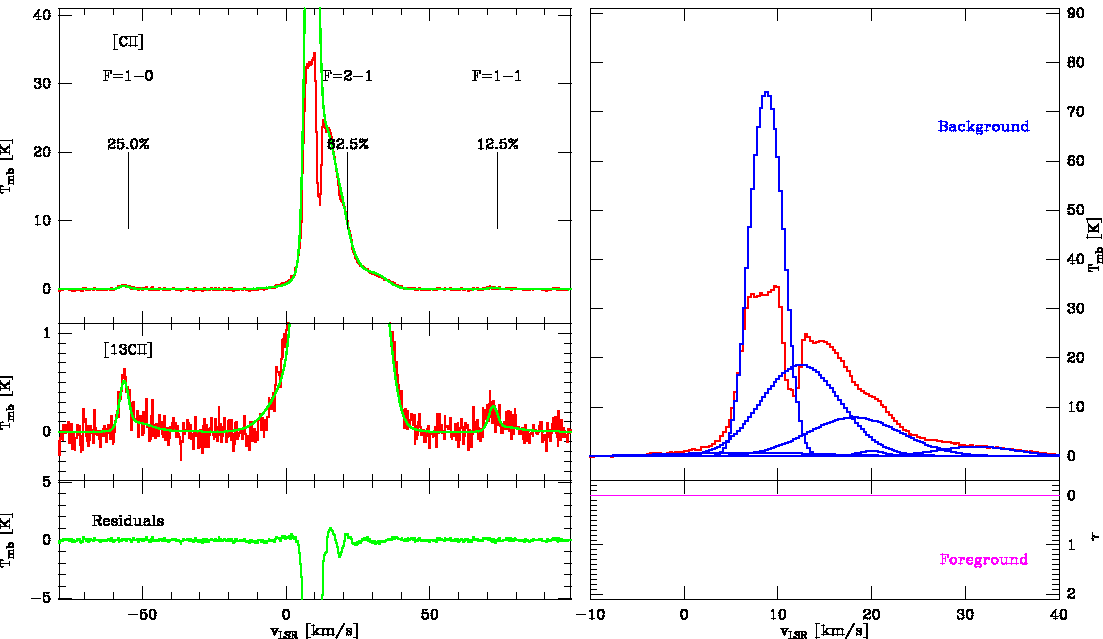}
  \caption{}
  \label{fig:step2}
\end{subfigure}
\par
\begin{subfigure}{\hsize}
  \centering
   \includegraphics[width=0.63\hsize]{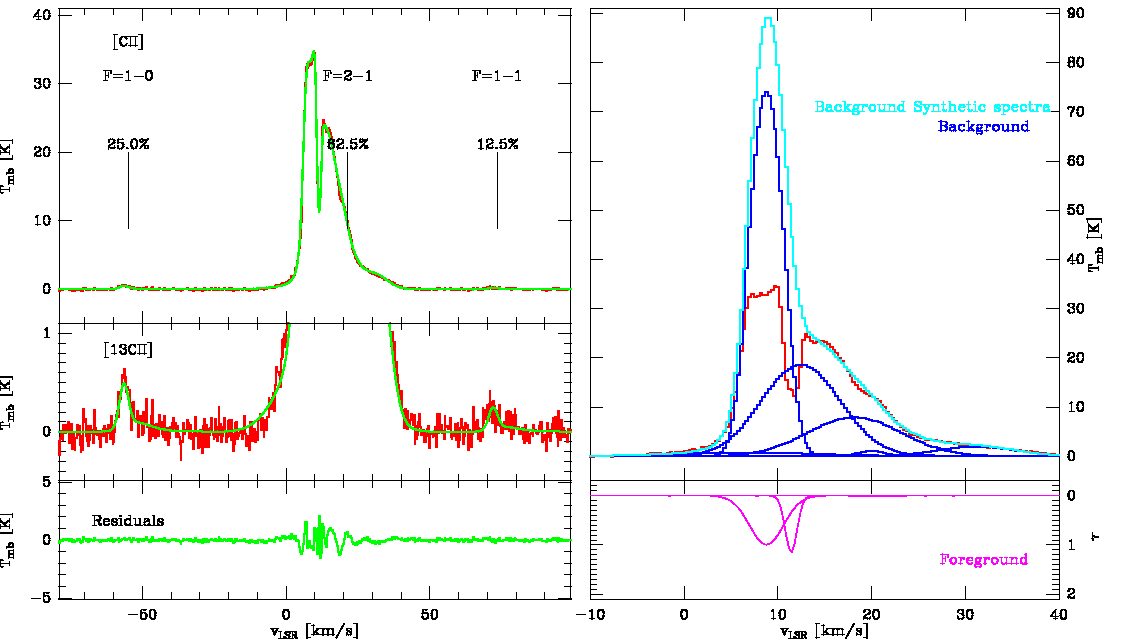}
  \caption{}
  \label{fig:step3}
\end{subfigure}
\caption{
Demonstration of the multi-component fitting procedure, taking the Mon~R2 Pos~1 spectrum as an example. Each plot a) to c) is  structured in the same way. \textit{Left top:} it shows the fitted model in green and the observed spectrum in red. \textit{Middle left:} zoom-in vertically of the fitted model and the observed spectra to better show the \Cp{13} satellites. \textit{Left bottom:} residual between the observed spectra and the model. \textit{Right top:} fitted background emission component in blue, the resulting background emission model from the addition of all the components in cyan and the observed spectrum in red. \textit{Right bottom:} optical depth of each absorbing foreground component (inverted scale)  in pink. \textit{\subref{fig:step1})} fitting of the \Cp{13} emission, masking the velocity range of the \Cp{12} emission. \textit{\subref{fig:step2})} fitting of the remaining \Cp{12} background emission. \textit{\subref{fig:step3})} fitting of the foreground absorbing components. 
}
\label{13CIIsteps}
\end{figure*}

In order to illustrate the iterative fitting procedure, we select position 1 of Monoceros~R2 as an illustration case and describe this complex procedure step by step: \par

 i) Fig.~\ref{fig:step1}: we fit the \Cp{13} emission as originating in part of the background layer, masking the \Cp{12} line, fixing the excitation temperature and leaving the other parameters free. As the fitting function contains simultaneously the matching \Cp{12} line, the \Cp{13} fitting produces a \Cp{12} profile (scaled by the abundance ratio 
$ \mathrm \alpha^{+}$) that overshoots the observed \Cp{12} profile. In this example, we have used an excitation temperature of 160~K for the background component with an abundance ratio of $\alpha^{+}=67$. This way, we keep the background optical depth close to unity. We originally fixed the temperature to 150~K, but an increment to 160~K improved the fit. \par 

ii) Fig.~\ref{fig:step2}: next, we fit the \Cp{12} remaining emission with additional background layer components which, due to their low column density, have a negligible contribution to the \Cp{13} emission, using the smallest possible number of Gaussian components for the fitting. In this example, we  used a $T_{ \mathrm e \mathrm x}$ 
of 150~K for the remaining background components.\par

iii) Fig.~\ref{fig:step3}: as the fitted line profile of these combined background emission components now overshoots the observed one in several narrow velocity ranges, we then fit the foreground absorption features using a fixed and low $T_{\mathrm{ex}}$. For position 1, in the Mon~R2 example, we used a $T_{\mathrm e \mathrm x}$ of 20~K for these foreground components. This step is necessary only if the source is affected by self-absorption.\par

We applied this two-layer, multi-component fitting procedure to the \Cp{12} and \Cp{13} emission in all positions observed in all four sources. We calculate the line center optical depth of each component from the fit-parameters, following Eq.~\ref{eq:tau2}. For Mon~R2 and M17~SW, the fitting requires the inclusion of foreground absorption as 
discussed above. But for M43 and the Horsehead PDR, we have seen that the \Cp{13} line profile follows the \Cp{12} one and their emission peaks at the same velocity. Thus, no absorbing foreground layer is needed, and we can model the \Cp{12} and\Cp{13} emission for these two sources by using only a single background layer with multiple emission components. In this case, we can leave $T_{\mathrm{ex}}$ as a free fit parameter instead of fixing it. \par

We summarize the fitting results in Tables~\ref{M43table} and \ref{M17table} (the full set of fit parameters of each component for all positions of the sources is shown in Appendix~\ref{Doc}), including the $\chi^2$ of the fit result, the excitation temperature for the background layer (T$_{\mathrm{ex,bg}}$) (taken as the excitation temperature of the background component that traces the \Cp{13} emission), and the temperature of the foreground layer component that has the highest optical depth (T$_{\mathrm{ex,fg}}$). We also show the total column density N$_{12}$(\ion{C}{II}) for each layer and the peak optical depth of the component closest in LSR-velocity to the \Cp{13} peak temperature for the 
background ($\tau^{*}_{\mathrm{bg}}$) and foreground layers ($\tau^{*}_{\mathrm{fg}}$) as representative optical depths for each position. We have selected this optical depth as representative for two reasons: The bulk of the \Cp{12} emission comes from the material traced by the \Cp{13} emission (as we can see in Fig.~\ref{fig:step1}), and it is the 
component that experiences the largest self-absorption effects. We also quote the equivalent visual extinction corresponding to the \Cp{} column density as A$_{\mathrm V}$.\par

\subsubsection{M43 analysis}

\begin{figure}
   \centering
   \includegraphics[width=\hsize]{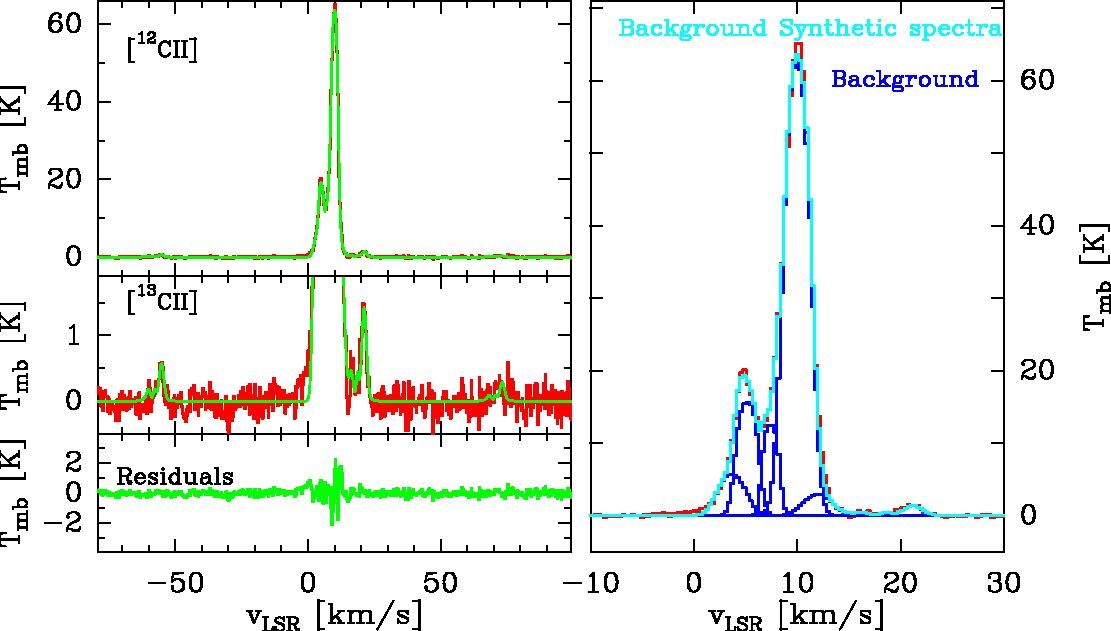}
      \caption{ 
      Same as Fig.~\ref{13CIIsteps}, but for M43 \Cp{12} spectra of position 0 with no foreground absorption.
              }
         \label{M43CIIpixel0}
   \end{figure}

The best-fit $T_{\mathrm{ex}}$ for the \Cp{13} emission is close to 100~K for the different positions, while the $T_{\mathrm{ex}}$ fitted for the \Cp{12} background emission not covered within the \Cp{13} profile is much lower, in the range from 30~K to 70~K (see Fig.~\ref{M43CIIpixel0} for an example of the fitting corresponding to the central observed position). The total \Cp{12} column density for the different positions varies between 1$\times$10$^{18}$ to 4$\times$10$^{18}$~cm$^{-2}$, with an equivalent visual extinction between 4.9 and 18.3~mag (Table~\ref{M43table}). \par

\subsubsection{Horsehead PDR analysis}

\begin{figure}
   \centering
   \includegraphics[width=\hsize]{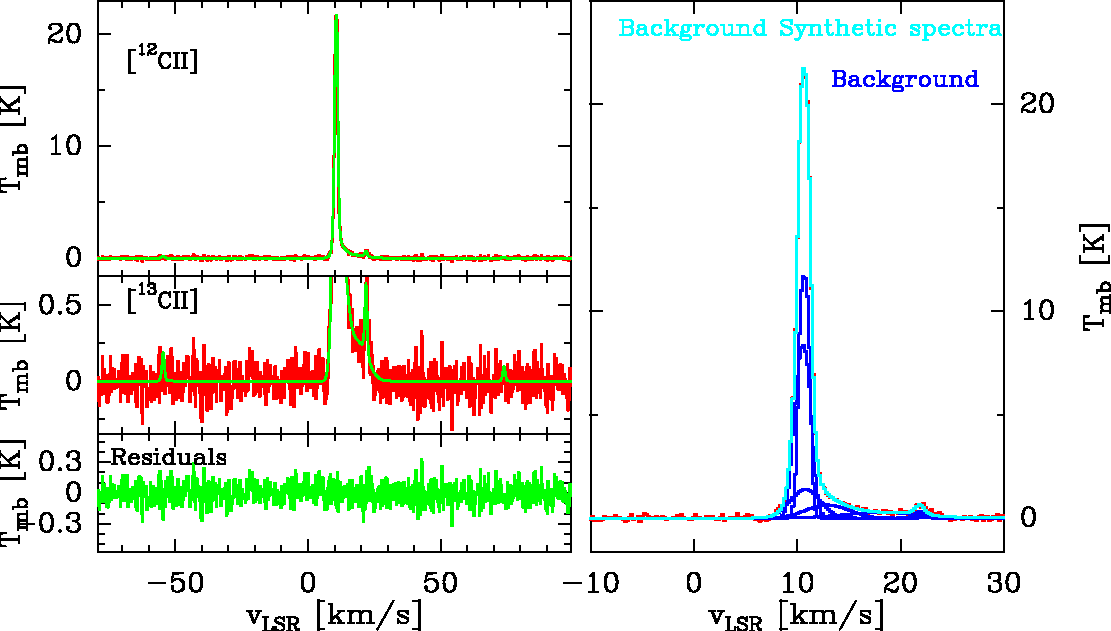}
      \caption{
      Same as Fig.~\ref{13CIIsteps}, but for the Horsehead \Cp{12} spectra of position 6 with no foreground absorption.
              }
         \label{HORCIIpixel0}
   \end{figure}

For the multi-component analysis, we have used the spectra discussed previously in Section~\ref{sub:lineprofile}, without applying the wing subtraction through a polynomial. In the Horsehead PDR, due to the blending 
of the \Cp{13}$_{\mathrm{F=2-1}}$ satellite with the \Cp{12} wing at the higher LSR-velocities (Fig.~\ref{hor_wing}), we first fitted the wing emission with several Gaussian components while masking the \Cp{13}$_{\mathrm{F=2-1}}$ velocity range. This is necessary because the wing overlaps with the \Cp{13}$_{\mathrm{F=2-1}}$ emission and thus affects the \Cp{13} fitting process (we note that the approach here is different from the one done in the zeroth-order analysis, where we subtracted the wing emission by a polynomial fit to obtain the \Cp{13} profile). Then we continue fitting the \Cp{12} and \Cp{13}$_{\mathrm{F=2-1}}$ emission. The excitation temperature which can be left as a free fit parameter in this case, as discussed before, gives a value for all the components at the different positions of around 30~K. For the positions that are located outside the main interface ridge, positions 1, 4, and 5 (Fig.~\ref{fig:sub3}), the \Cp{13}$_{\mathrm{F=2-1}}$ satellite emission is heavily blended with the wing emission, so those column densities fitted for these positions should be considered as a rough estimate because they are not constrained by the  optically thin \Cp{13} emission. We derive a total \Cp{12} column density for the different positions ranging from 3.6$\times$10$^{17}$~cm$^{-2}$ to 1.3$\times$10$^{18}$~cm$^{-2}$ (see Table~\ref{M43table}), which is much lower than for the case of M43 due the smaller line width. The equivalent visual extinctions range from 1.6 to 5.8~mag. Fig.~\ref{HORCIIpixel0} shows as an example the fit results in position 6.\par 

\begin{table*}
    \centering
   \caption{M43 and the Horsehead PDR column density for the background components }
   \label{M43table}
   \begin{threeparttable}
           \begin{tabular}{l c c c c c c }
      \hline
      \hline
\multicolumn{1}{c}{Positions} & \multicolumn{1}{c}{No.}   & \multicolumn{1}{c}{$\chi^2$}  & $T_{\mathrm{ex}}$ & \multicolumn{1}{c}{Background}  & \multicolumn{1}{c}{$\tau^{*}$\tnote{a}} & \multicolumn{1}{c}{Back.}   \\
                              & \multicolumn{1}{c}{Back.} &                               &                   & \multicolumn{1}{c}{N$_{12}$(\ion{C}{II})} &   & \multicolumn{1}{c}{ A$_{\mathrm V}$ \ion{C}{II}}         \\
                              & \multicolumn{1}{c}{Comp.} &                               &              (K)  & \multicolumn{1}{c}{(cm$^{-2}$)}    &      & \multicolumn{1}{c}{(mag.)}             \\
          \hline
    M43 0 & 5 & 1.7 & 110.7 & 4.1E18 & 2.07 & 18.3  \\
    M43 1 & 5 & 1.6 & 97.3 & 2.9E18 & 1.75 & 12.9\\
    M43 2 & 2 & 1.6 & 60.0 & 2.5E18 & 1.43 & 11.2\\
    M43 3 & 2 & 1.2 & 70.0 & 1.1E18 & 0.43 & 4.9 \\
    M43 4 & 6 & 1.1 & 108.4 & 2.7E18 & 1.68 & 12.0 \\
    M43 5 & 4 & 1.5 & 108.0 & 2.0E18 & 0.96 & 9.0 \\
    M43 6 & 5 & 1.1 & 101.04 & 3.0E18 & 1.66 & 13.4 \\
       \hline
   HOR 0 & 4 & 1.2 & 38.0 & 7.9E17 & 2.16 & 3.6 \\
   HOR 1 & 4 & 1.1 & 26.7 & 5.7E17 & 0.78 & 2.5 \\
   HOR 2 & 4 & 1.2 & 37.2 & 8.5E17 & 1.80 & 3.8 \\
   HOR 3 & 4 & 1.4 & 38.0 & 6.2E17 & 1.34 & 2.8 \\
   HOR 4 & 4 & 1.3 & 35.5 & 4.3E17 & 0.75 & 1.9 \\
   HOR 5 & 4 & 1.4 & 48.0 & 3.6E17 & 0.52 & 1.6 \\
   HOR 6 & 5 & 1.0 & 43.0 & 1.3E18 & 2.84 & 5.8   \\
       \hline
\end{tabular}
\begin{tablenotes}\footnotesize
\item[a] $\tau^*$ corresponds to the optical depth of the component closer to the \Cp{13} peak.
\end{tablenotes}
\end{threeparttable}
 \end{table*}

\subsubsection{Monoceros R2 analysis}

For Mon~R2 (see Fig.~\ref{13CIIsteps}), we fixed the background $T_{\mathrm{ex}}$ to 150~K and the foreground $T_{\mathrm{ex}}$ to 20~K. We determined the total \Cp{} column density for both positions and each layer and obtained a total background column density of 4.2$\times$10$^{18}$~cm$^{-2}$ and 4.7$\times$10$^{18}$~cm$^{-2}$ , respectively, and a total foreground column density of 8.3$\times$10$^{17}$~cm$^{-2}$ and 6.4$\times$10$^{17}$~cm$^{-2}$. The equivalent visual extinction in the two positions observed corresponds to 18.7~mag and 21.0~mag for the background, and 3.7~mag and 2.9~mag for the foreground. \par

\subsubsection{M17~SW analysis}

\begin{figure}
   \centering
   \includegraphics[width=\hsize]{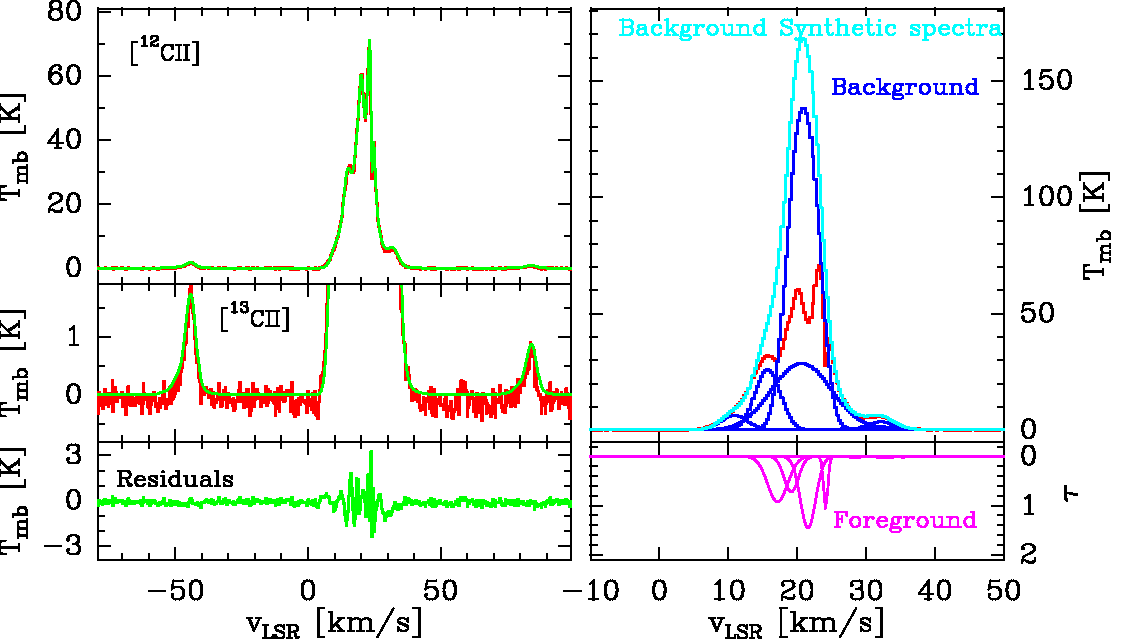}
      \caption{
      Same as Fig.~\ref{13CIIsteps}, but for M17~SW \Cp{12} spectra of position 6.
         }
         \label{M17CIIpixel6}
   \end{figure}

For M17~SW, we fit the \Cp{13} emission using a fixed excitation temperature between 180~K and 250~K for the background.
We selected a higher range, compared to Mon~R2, because the brighter \Cp{13} emission, and correspondingly brighter \Cp{12} background emission, requires a larger temperature for a reasonable optical depth. Otherwise, it would require a larger foreground absorption, even in the line wings. For the foreground components, we use temperatures between 25 and 45~K. Fig.~\ref{M17CIIpixel6} shows the fit results in position 6 as an example. As we can see in Table~\ref{M17table}, the total column density for each of the seven array positions range from 3.0$\times$10$^{18}$~cm$^{-2}$ to 9.2$\times$10$^{18}$~cm$^{-2}$ for the background layer, and from 3.9$\times$10$^{17}$~cm$^{-2}$ to 3$\times$10$^{18}$~cm$^{-2}$ for the foreground layer. The equivalent visual extinction between 13.4~mag and 41.0~mag for the background, and 1.7~mag to 13.4~mag for the foreground. \par

\begin{table*}
  \centering
  \caption{MonR2 and M17~SW column density for the background and foreground components for the double layer model}
  \label{M17table}
  \begin{threeparttable}
  \begin{tabular}{l c c c c c c c c c c c}
      \hline
\hline
\multicolumn{1}{c}{Positions} & \multicolumn{1}{c}{No.}    & \multicolumn{1}{c}{No.}   & \multicolumn{1}{c}{$\chi^2$} & T$_{\mathrm{ex,bg}}$   & \multicolumn{1}{c}{Background}             & \multicolumn{1}{c}{$\tau_{\mathrm{bg}}^{*}$\tnote{a}} & \multicolumn{1}{c}{Back.}                       & T$_{\mathrm{ex,fg}}$ & \multicolumn{1}{c}{Foreground} & \multicolumn{1}{c}{$\tau_{\mathrm{fg}}^{*}$\tnote{a}} & \multicolumn{1}{c}{Fore.}      \\
                              & \multicolumn{1}{c}{Back.}  & \multicolumn{1}{c}{Fore.} &                              &                        & \multicolumn{1}{c}{N$_{12}$(\ion{C}{II})}  &                                              & \multicolumn{1}{c}{A$_{\mathrm V}$ \ion{C}{II}} &                      & \multicolumn{1}{c}{N$_{12}$(\ion{C}{II})}  &                 & \multicolumn{1}{c}{A$_{\mathrm V}$ \ion{C}{II}}  \\
                              & \multicolumn{1}{c}{Comp.}  & \multicolumn{1}{c}{Comp.} &                              & (K)                    & \multicolumn{1}{c}{(cm$^{-2}$)}            &                                              & \multicolumn{1}{c}{mag.}                        &        (K)           & \multicolumn{1}{c}{(cm$^{-2}$)}   &                 &  \multicolumn{1}{c}{mag.}          \\
          \hline
    MonR2 1 & 5 & 2 & 3.1 & 160 & 4.2E18 & 0.98 & 18.7 & 20 & 8.3E17 & 0.99 & 3.7  \\      
    MonR2 2 & 6 & 4 & 4.1 & 150 & 4.7E18 & 1.03 & 21.0 & 20 & 6.4E17 & 1.52 & 2.9  \\
       \hline
    M17SW 0 & 4 & 6 & 1.8 & 250  & 9.2E18 & 1.43 & 41.0 & 40 & 2.0E18 & 1.63 & 9.2 \\
    M17SW 1 & 5 & 4 & 1.2 & 200 & 8.0E18 & 1.61 & 35.6  & 30 & 1.7E18 & 1.52 & 7.6 \\
    M17SW 2 & 4 & 5 & 1.3 & 200 & 5.6E18 & 0.84 & 24.9  & 30 & 3.0E18 & 1.58 & 13.4  \\
    M17SW 3 & 4 & 2 & 3.5 & 180 & 4.4E18 & 0.66 & 19.6  & 25 & 7.7E17 & 1.61 & 3.5    \\
    M17SW 4 & 5 & 5 & 1.9 & 200 & 7.6E18 & 1.14 & 33.9  & 30 & 1.3E18 & 0.90 & 5.8 \\
    M17SW 5 & 4 & 3 & 4.5 & 200 & 3.0E18 & 0.24 & 13.4  & 30 & 3.9E17 & 0.63 & 1.7 \\
    M17SW 6 & 5 & 4 & 1.3 & 250 & 7.7E18 & 1.11 & 34.3  & 45 & 1.8E18 & 1.46 & 8.0 \\
        \hline
\end{tabular}
\begin{tablenotes}\footnotesize
\item[a] $\tau^*$ corresponds to the optical depth of the component closer to the \Cp{13} peak.
\end{tablenotes}
\end{threeparttable}
 \end{table*}

\section{Discussion} \label{Discussion}
 
Considering that the standard PDR models predict an equivalent A$_\mathrm{V}$ of slightly above unity for a single PDR-layer, the \Cp{13} integrated intensity derived in Sect.~\ref{13cd} is consistent with a simple, single layer PDR model only for the case of the Horsehead PDR and M43, and if the emission is fully filling the beam. In the other two sources, as we have seen from the multi-component analysis in Section~\ref{Analysis}, the large equivalent visual extinctions imply at least several, and up to many, PDR layers along the line of sight (and filling the beam) in the framework of standard PDR models. Obviously, this requires the assumption of clumpiness and piling up of many PDR surfaces on clumps along the line of sight. Also, the ratio between the scaled-up \Cp{13} equivalent extinction and the assumed optically thin \Cp{12} show again how \Cp{12} underestimate the column density and the equivalent visual extinction by a factor as high as 3. A key point in this discussion is the \Cp{12}/\Cp{13} abundance ratio assumed for the scaling of the \Cp{13} intensity. \par

\subsection{\Cp{12}/\Cp{13} abundance ratio} \label{abundanceratio}

The observed intensity ratio \Cp{12}/\Cp{13} is lower in the line centers than the assumed abundance ratio for each source. We can interpret this as being due to self-absorption in the line centers as it was done in the multi-component analysis above in Section~\ref{Analysis}. Towards the line wings, the intensity ratio increases. But only in the 
case of M43 it reaches a value close to the assumed abundance ratio for this source (see gray histograms in Figures \ref{M43all} to \ref{M17CIIall}). For Mon~R2 and M17~SW the ratio towards the line wings only reaches up to about half of the assumed abundance. This is, of course, linked to the S/N threshold that we apply to define the useful velocity range: better signal-to-noise would allow us to derive also a ratio further out in the line wings. In fact, for the case of the Horsehead PDR with its relatively weak lines, the signal-to-noise is sufficient only near the line center and we do not observe an increase toward the line wings because we have no valid data there. \par

If the abundance ratio in the source would in fact be lower than the assumed literature value, the derived optical depths would be correspondingly lower. Thus, the important question is whether the derived high optical depths are an artifact, based on an assumed high abundance ratio. Higher S/N would increase the useful velocity range and would thus allow to trace the line intensity ratio further out in the line wings. Thus, it is essential  whether the intensity ratio in this regime keeps increasing until it reaches a plateau at the (assumed) value for the $^{12}$C$^{+}$/$^{13}$C$^{+}$ abundance ratio, namely $\alpha^{+}$. In Appendix~\ref{app:abundanceline} we present a discussion about how the binning, the \Cp{12} optical depth and the 1.5 $\sigma$ threshold affects the \Cp{12}/\Cp{13} abundance ratio  estimation. \par

\begin{figure}
   \centering
   \includegraphics[width=0.9\hsize]{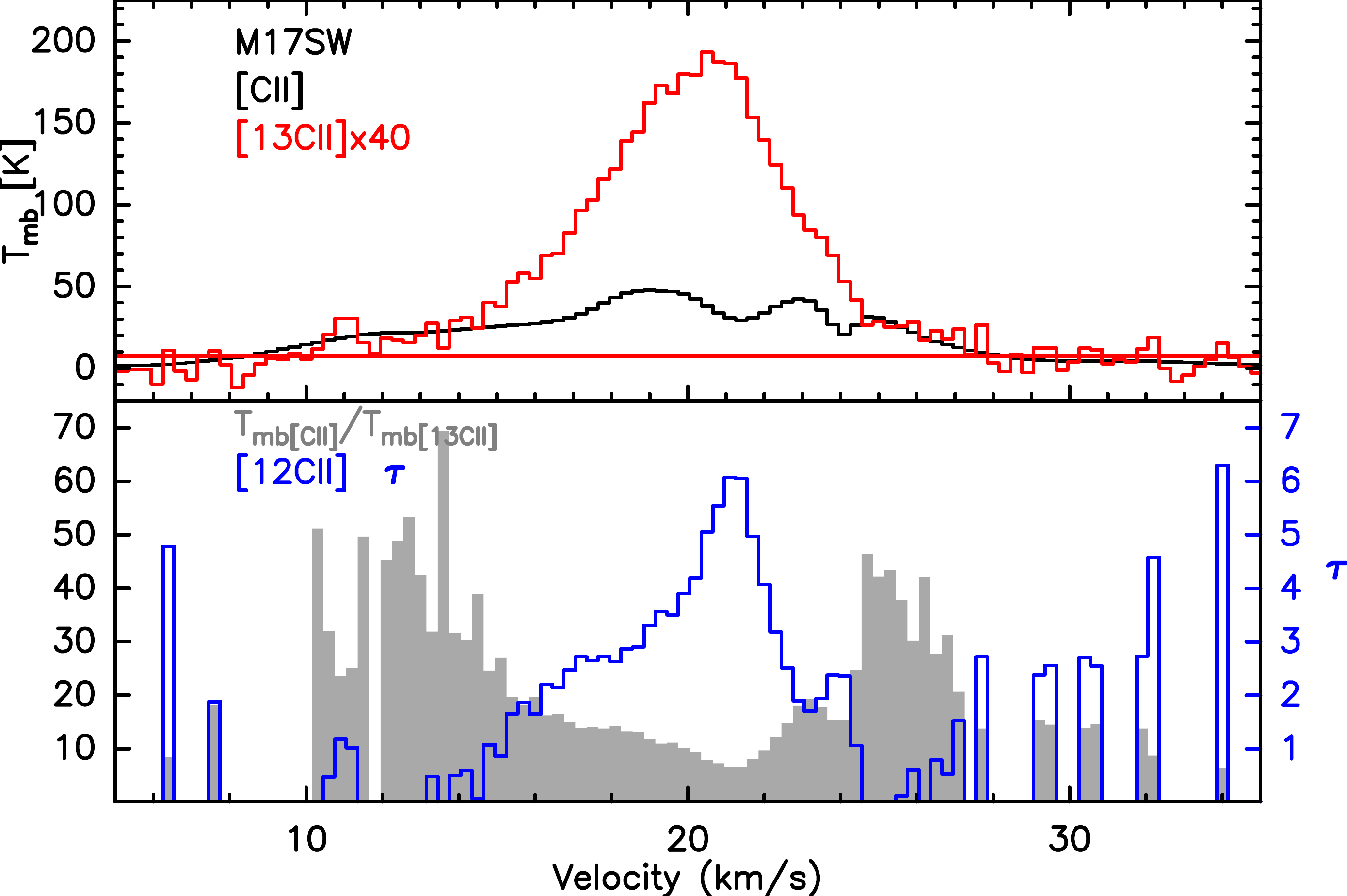}
      \caption{Same as Fig.~\ref{M43all}, but for the M17~SW average spectra of six positions observed by upGREAT.
              }
         \label{M17averageCII}
   \end{figure}

We can check on this for the case of M17~SW by averaging six of the seven positions observed (the ones with high intensity, i.e.,\ excluding position no.~5) and analyzing the average spectrum. Fig.~\ref{M17averageCII} shows this average spectrum, for which the useful velocity range with a \Cp{13} intensity above 1.5 $\sigma$ now extends from 10 to 28~km/s. The intensity ratio in the outer wings clearly rises up to values $\approx$45, close to the assumed abundance ratio of around 40. The high S/N of 18 for the \Cp{13} emission is not necessarily enough to determine $\alpha^+$, but, rather, it may underestimate it by up to 30\% (see Appendix~\ref{app:abundanceline}). This would lead to a corrected intrinsic abundance ratio of 60 instead of 40 for M17~SW. The increased $\alpha^{+}$ value may be a sign of fractionation of the ionic species; in fact, we point out that we have used a conservative lower value for $\alpha$ (see Section~\ref{Observations}), whereas the values from the literature and the Galactocentric distribution for carbon isotopes (detailed above in Section~\ref{Introduction}) are closer to 60. This demonstrates that with high enough S/N in addition to a correction factor, we can directly determine the abundance ratio $\alpha^{+}$ from line intensity ratio observed in the optically thin line wings. \par

With the derived new abundance ratio, we repeat the multi-component analysis for M17~SW. We summarize the fitting results in Table~\ref{M17table60}. The increment in the abundance ratio leads to an increase in the number of components needed for a good fit due to the larger intensity of the scaled-up \Cp{13} line emission requiring more components for a similar excitation (or a higher excitation temperature). This results in an increase between 20\% and 40\% in the derived column density and, proportionally, in the equivalent visual extinction.  \par
   
 \begin{table*}
  \centering
  \caption{M17~SW column density for the background and foreground components for the double layer model assuming an $\alpha^{+}$=60 (compare to Table \ref{M17table}) }
           \label{M17table60}
   \begin{threeparttable}        
  \begin{tabular}{l c c c c c c c c c c c}
      \hline
\hline
\multicolumn{1}{c}{Positions} & \multicolumn{1}{c}{No.}    & \multicolumn{1}{c}{No.}   & \multicolumn{1}{c}{$\chi^2$} & T$_{\mathrm{ex,bg}}$  & \multicolumn{1}{c}{Background}             & \multicolumn{1}{c}{$\tau_{\mathrm{bg}}^{*}$\tnote{a}} & \multicolumn{1}{c}{Back.}                       & T$_{\mathrm{ex,fg}}$ & \multicolumn{1}{c}{Foreground} & \multicolumn{1}{c}{$\tau_{\mathrm{fg}}^{*}$\tnote{a}} & \multicolumn{1}{c}{Fore.}      \\
                              & \multicolumn{1}{c}{Back.}  & \multicolumn{1}{c}{Fore.} &                              &                       & \multicolumn{1}{c}{N$_{12}$(\ion{C}{II})}  &                                              & \multicolumn{1}{c}{A$_{\mathrm V}$ \ion{C}{II}} &                      & \multicolumn{1}{c}{N$_{12}$(\ion{C}{II})}  &                 & \multicolumn{1}{c}{A$_{\mathrm V}$ \ion{C}{II}}  \\
                              & \multicolumn{1}{c}{Comp.}  & \multicolumn{1}{c}{Comp.} &                              &       (K)             & \multicolumn{1}{c}{(cm$^{-2}$)}            &                                              & \multicolumn{1}{c}{mag.}                        &      (K)             & \multicolumn{1}{c}{(cm$^{-2}$)}   &                 &  \multicolumn{1}{c}{mag.}          \\
          \hline
    M17SW 0 & 6 & 5 & 0.9 & 250 & 1.2E19 & 1.81 & 55.3 & 40 & 3.1E18 & 1.41 & 13.7 \\
    M17SW 1 & 7 & 4 & 1.6 & 200 & 1.2E19 & 2.41 & 52.5 & 30 & 2.2E18 & 1.41 & 10.0 \\ 
    M17SW 2 & 5 & 4 & 0.8 & 200 & 7.5E18 & 1.25 & 33.3 & 30 & 2.4E18 & 1.61 & 10.7 \\ 
    M17SW 3 & 9 & 5 & 4.3 & 200 & 5.9E18 & 1.21 & 26.2 & 20 & 1.0E18 & 1.39 & 4.4    \\ 
    M17SW 4 & 9 & 7 & 2.5 & 180 & 1.2E19 & 1.97 & 51.7 & 20 & 1.4E18 & 1.12 & 6.3 \\ 
    M17SW 5 & 5 & 6 & 3.3 & 200 & 3.7E18 & 0.35 & 16.7 & 20 & 7.0E17 & 0.58 & 3.1 \\ 
    M17SW 6 & 7 & 4 & 1.0 & 200 & 1.1E19 & 1.42 & 46.9 & 35 & 1.7E18 & 1.77 & 7.7 \\ 
        \hline
  \end{tabular}
\begin{tablenotes}\footnotesize
\item[a] $\tau^*$ corresponds to the optical depth of the component closer to the \Cp{13} peak.
\end{tablenotes}
\end{threeparttable}
\end{table*}

Unfortunately, we can only do this test for the case of M17~SW. For the second source with significant optical depth even in the line wings and, therefore, an \Cp{12}/\Cp{13} intensity ratio in the wings still significantly below the assumed abundance ratio in the source, namely Mon~R2, we only have the two spectra taken with the GREAT~L2 single pixel instrument and averaging the two does not give a sufficient increase in S/N to allow to trace the line intensity ratio further out in the wings. Future observations are needed to check on the assumed abundance ratio also in this and the other sources, in addition to further analysis for a study of this possible indication of fractionation in M17~SW. \par 

\subsection{Comparison between \Cp{} and other tracers}

\subsubsection{CO Molecular emission} \label{COdis}

To compare \Cp{12}-emitting gas with the molecular gas traced by CO and its isotopologues, we used the low-J CO rotational line data, including the rare isotopologues of C\element[][18]{O} 1-0 and  C\element[][17]{O} 1-0, in this part of the study. We compared the molecular and ionized material for M43, Mon~R2, and M17~SW respectively. \par

\begin{figure}
    \centering
    \includegraphics[width=0.95\hsize]{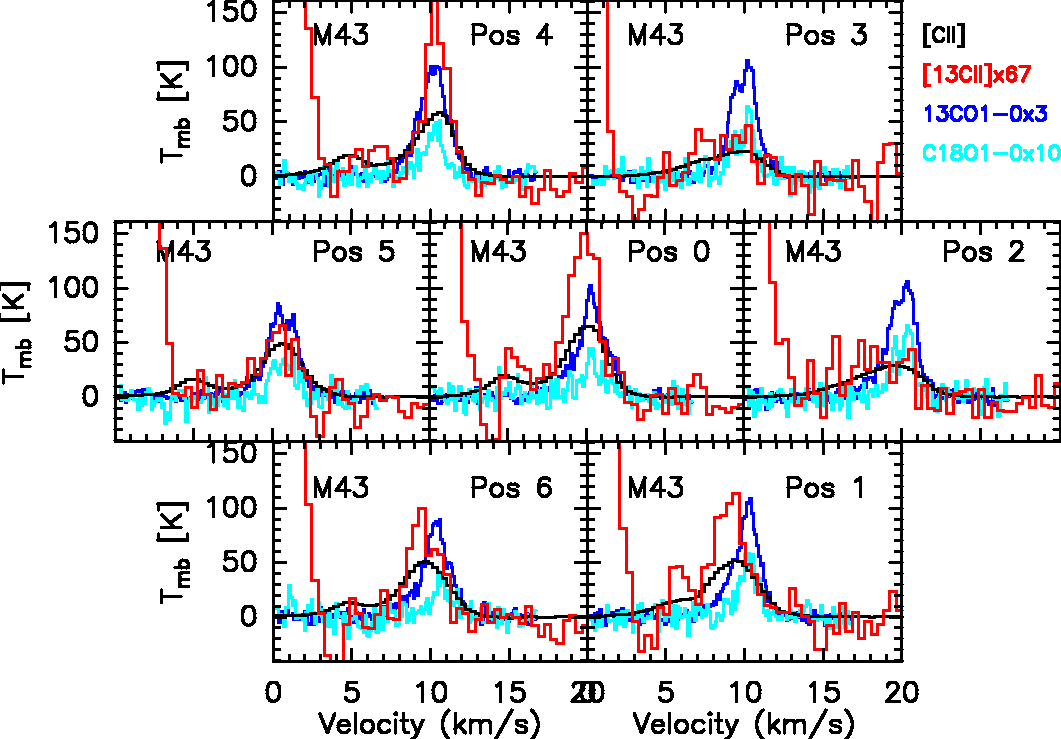}
        \caption{
        M43 line profile comparison between \Cp{12} (in black), \Cp{13} (in red), scaled-up by $\alpha^{+}$, CO (1-0) (in green), \element[][13]{CO} (1-0) (in blue), and 
        C\element[][18]{O} (1-0) (in cyan). CO and its isotopologues has been scaled-up by the factors indicated only to be compared with \Cp{13}.
                }
          \label{M43CO}
    \end{figure}

\begin{figure}
    \centering
    \includegraphics[width=0.95\hsize]{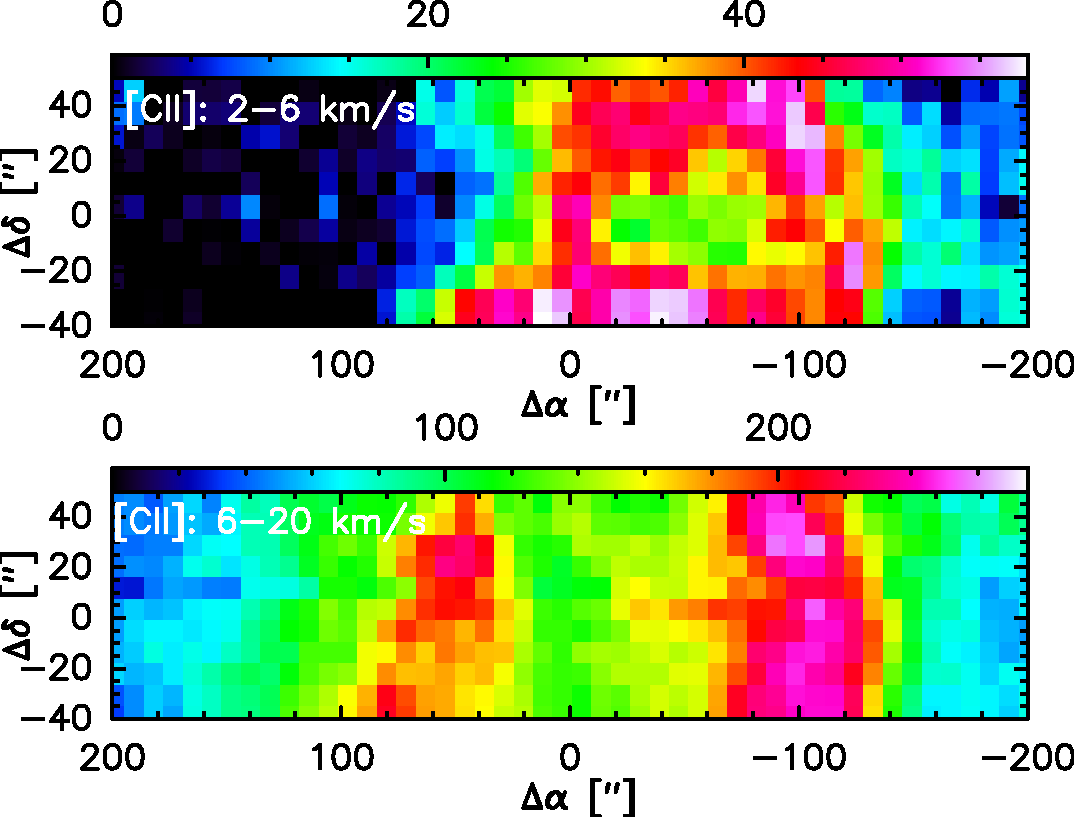}
        \caption{
         \textit{Top:} M43 \Cp{} integrated intensity map between 2 and 6~km/s. \textit{Bottom:} same as before but between 6 and 20~km/s.
                }
          \label{M43sepatared}
    \end{figure}

For M43, we use CO data observed with the Combined Array for Millimeter-Wave Astronomy (CARMA), within the CARMA-NRO Orion Survey project \citep[][]{2018ApJS..236...25K,2019A&A...623A.142S}. The molecular lines observed are \element[][13]{CO} J = 1-0  and C\element[][18]{O} J = 1-0 for the seven \Cp{} positions. In Figure \ref{M43CO}, we compare the different CO isotopic line profiles against the \Cp{12} and \Cp{13} emission. The line profile between the CO isotopologues and \Cp{} tend to be similar for the main emission located at 10~km/s, with the \Cp{} peak shifted 
to the blue part of the spectra. On the other hand, there is no molecular counterpart to the secondary peak at 4~km/s. From the \Cp{} integrated intensity map (Fig.~\ref{M43sepatared}), we can see that this secondary peak forms a ring-like structure, and from the comparison against [\ion{N}{II}] done below in Section~\ref{NIIsec} (Fig.~\ref{NII_all}), the [\ion{N}{II}] also peaks at 4~km/s. As a result, the material at the secondary peak would correspond to ionized material; the material at 10~km/s would be the molecular region traced by CO. \par

\begin{figure}
   \centering
   \includegraphics[width=0.95\hsize]{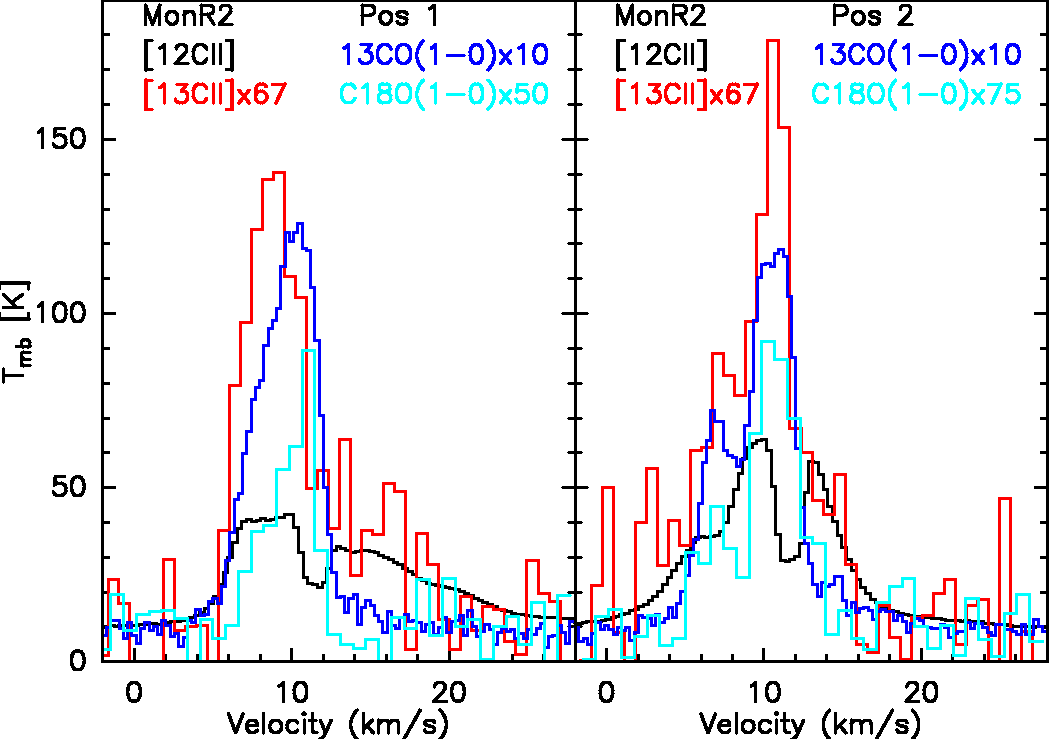}
      \caption{Mon~R2 comparison between \Cp{12} in black, \Cp{13} in red scaled-up using $\alpha$, \element[][13]{CO} (1-0) 
      in blue and C\element[][18]{O} (1-0) in cyan. CO and its isotopologues has been scaled-up only to be compared with \Cp{13}. \Cp{13} has been smoothed for display purposes. 
              }
         \label{MonR2CO}
   \end{figure}

For Mon~R2, we use the published data by \citet{2012A&A...543A..27G} at the two positions observed in \Cp{}. The molecular lines are \element[][13]{CO} J = 1-0 and C\element[][18]{O} J = 1-0 observed with the EMIR receiver \citep{2012A&A...538A..89C} at the IRAM 30 m telescope. The CO isotopologues (Fig.~\ref{MonR2CO}) show a similar profile to the \Cp{13} emission at the 10~km/s peak, with \Cp{} more extended at higher velocities. There is no molecular emission at 15~km/s, so it is safe to assume that this material could be correlated with ionized gas. \par

\begin{figure}
   \centering
   \includegraphics[width=0.90\hsize]{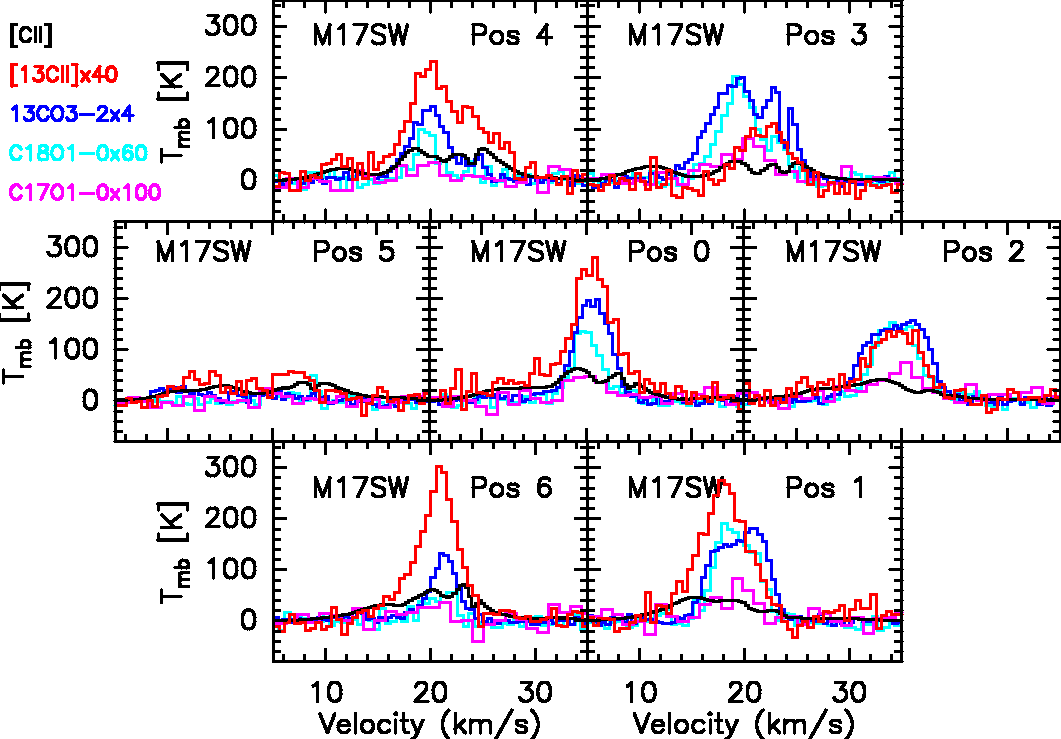}
      \caption{M17~SW comparison between \Cp{12} in black, \Cp{13} in red scaled-up using $\alpha$, \element[][13]{CO} (3-2) 
      in blue, C\element[][18]{O} (1-0) in cyan and C\element[][17]{O} (1-0) in purple. CO and its isotopologues has been scaled-up only to be compared with \Cp{13}. 
              }
         \label{M17CO}
   \end{figure}

For M17~SW, we use the published data from \citet{2015A&A...575A...9P,2015A&A...583A.107P} at the seven positions observed in \Cp{}. The CO isotopologue lines are \element[][13]{CO} J = 3-2, observed with FLASH \citep{2006A&A...454L..21H} at the APEX telescope \citep{2006A&A...454L..13G}, and C\element[][18]{O} J = 1-0 and C\element[][17]{O} J = 1-0 observed with the EMIR receiver at the IRAM 30 m telescope. The spectra have a similar angular resolution, therefore they are directly comparable. Also, the comparison with low-J CO is appropriate in this case, the foreground absorption layer is composed by cold gas, hence gas traced by low-J CO emission. Figure.~\ref{M17CO} shows that the molecular emission is associated with the \Cp{} gas in the central \Cp{13} emission that peaks at 20~km/s. But at lower velocities (lower than 15~km/s), there is no molecular emission. From the comparison against [\ion{N}{II}], as can be seen below in Section~\ref{NIIsec}, [\ion{N}{II}] peaks at 0~km/s, with a long tail from 0 to 20~km/s. This shows that the ionized gas is located at 0~km/s and has only weak \Cp{} emission associated. For larger velocities, the emission gets associated with the molecular region, showing the transition from the ionized to the molecular regime. \par

This scenario was already studied by \citet{2015A&A...575A...9P}. They found correlations between the \Cp{} and \ion{H}{I} emission at 10~km/s, with molecular material at 20~km/s and ionized at 30 km/s from the residuals. From our observations, we have \ion{H}{II} at 0~km/s (traced by [\ion{N}{II}]), \ion{H}{I} at 10~km/s and molecular H$_2$ at 20~km/s. For the velocities higher than 25~km/s, we can only affirm that \Cp{} does not correlate with any other tracer. \par
   
From the C\element[][18]{O} J = 1-0 observations, we estimate the C\element[][18]{O} column density for all available sources and each position, following \citet{2015PASP..127..266M} and \citet{2016A&A...587A..74S}. For the rotational excitation temperature, we used values based on the dust temperatures from the Herschel Gould Belt \citep{2010A&A...518L.102A} and HOBYS \citep{2010A&A...518L..77M} imaging key programs and published in \citet{2015A&A...577L...6S} for M43, \citet{2017A&A...607A..22R} for Mon~R2, and Schneider, N., priv. comm., for M17~SW. The typical dust temperatures in regions of peak \Cp{} emission are: 16~K for M43, 26~K for Mon~R2, and 35 K for M17~SW, respectively. For the rotational excitation temperature, we used  the dust temperatures for M43 and Mon~R2, but for M17~SW, we  used a value of 25~K, which is lower than the dust. We lowered this value because we expected the molecular material to be located in the UV-shielded core, whereas the dust is located in the UV-heated out layers. \par

Using these as the C\element[][18]{O} excitation temperature (assuming the temperatures as an upper limit for the molecular gas), we derive the C\element[][18]{O} column densities listed in Table~\ref{M17COtable}, also converted
to an equivalent H$_{2}$ column density assuming an \element[][16]{O}/\element[][18]{O} ratio of 490 for M43, 500 for Mon~R2 and 425 for M17~SW, according to \citet{1994ARA&A..32..191W}, and a CO/H$_{2}$ ratio of 1.2$\times$10$^{-4}$ \citep{2008ApJ...680..371W}. Here we assume that all carbon in the molecular region is in molecular form as CO. Finally, we also list the equivalent visual extinction knowing that 2N(H$_{2}$) = 1.87$\times$10$^{21}$ cm$^{-2}$ A$_{\mathrm V}$. As an independent verification, we also checked the dust column densities derived from Herschel. The dust column densities agree within 30\% with the values determined from C\element[][18]{O} using these excitation temperatures. \par     

 \begin{table*}
  \centering
    \caption{M43, Mon~R2, and M17~SW C\element[][18]{O} 1-0 column density and equivalent visual extinction and the \Cp{12} equivalent visual extinction for comparison}
  \begin{tabular}{l c c c c c }
      \hline
      \hline
\multicolumn{1}{c}{Positions} & \multicolumn{1}{c}{N(C\element[][18]{O})} & \multicolumn{1}{c}{N(H$_{2}$)} & \multicolumn{1}{c}{A$_{\mathrm V}$} & A$_{\mathrm V, \mathrm{bg}}$ & A$_{\mathrm V,\mathrm{fg}}$         \\
                              &                                           & \multicolumn{1}{c}{C\element[][18]{O}} & \multicolumn{1}{c}{C\element[][18]{O}} & \Cp{12} & \Cp{12}  \\
                              & \multicolumn{1}{c}{(cm$^{-2}$)}           & \multicolumn{1}{c}{(cm$^{-2}$)}        & (mag.)   & (mag.)  &  (mag.)         \\
          \hline
M43 0 & 1.16E16 & 4.75E22 & 50.8 & 18.3 & - \\
M43 1 & 1.44E16 & 5.88E22 & 62.9 & 12.9 & - \\
M43 2 & 1.50E16 & 6.13E22 & 65.6 & 11.2 & - \\
M43 3 & 1.42E16 & 5.80E22 & 62.1 & 4.9 & - \\
M43 4 & 9.89E15 & 4.03E22 & 43.2 & 12.0 & - \\
M43 5 & 7.46E15 & 3.05E22 & 32.6 & 9.0 & -  \\
M43 6 & 8.00E15 & 3.27E22 & 34.9 & 13.4 & -  \\
          \hline
MonR2 1 & 8.98E15 & 3.74E22 & 40.0 & 18.7 & 3.7 \\
MonR2 2 & 9.99E15 & 4.16E22 & 44.5 & 21.0 & 2.9 \\
\hline          
M17SW 0 & 1.65E16 & 5.83E22 & 62.3 & 41.0 & 9.2 \\
M17SW 1 & 2.79E16 & 9.88E22 & 105.6 & 35.6 & 7.6\\
M17SW 2 & 2.74E16 & 9.72E22 & 103.9 & 24.9 & 13.4 \\
M17SW 3 & 3.15E16 & 1.11E23 & 119.1 & 19.6 & 3.5 \\
M17SW 4 & 1.38E16 & 4.88E22 & 52.2 & 33.9 & 5.8 \\
M17SW 5 & 2.05E15 & 7.28E21 & 7.8 & 13.4 &  1.7 \\
M17SW 6 & 5.10E15 & 1.81E22 & 19.3 & 34.3 & 8.0 \\
       \hline
  \end{tabular}
         \label{M17COtable}
\end{table*}

 We can now compare the equivalent extinctions (or for that matter, the derived H$_2$ column densities) derived from \Cp{} and from the CO isotopologues lines  with the ones derived from the multi-component analysis for \Cp{}. For M17~SW, in positions 5 and 6, the \Cp{} equivalent A$_\mathrm{V}$ is higher for both model scenarios than the one 
 estimated from C\element[][18]{O}. This comes as no surprise because, based on velocity channel maps between the molecular and ionized line observations \citep{2015A&A...575A...9P}, it is  known that these positions are located off the main molecular ridge and hence dominated by PDR material. \par
 
 For the other positions, the equivalent A$_\mathrm{V}$ of the \Cp{} layer  estimated for the double-layer \Cp{} emission model gives a lower equivalent \Cp{} column density than the ones derived from C\element[][18]{O}, on average 25\% of the molecular column  density. Hence, a worthwhile part of the hydrogen gas is also in atomic form, associated to the \Cp{} emission in the PDR. It is important to notice that for both cases, we have assumed as a simplification that all carbon is in atomic or molecular form, so the equivalent visual extinctions estimated here are lower, counting only the fraction of the material traced by the respective species. \par

For Mon~R2, the situation is similar. The equivalent visual extinction of the molecular gas traced by C\element[][18]{O} is 40.0 for position 1 and 44.5~mag for position 2, whereas the double layer scenario has significantly lower, although still relatively high, equivalent visual extinctions, 21~mag for the warm background and 3~mag for the foreground \Cp{} emission. A comparison with the single layer model is given in Appendix~\ref{singlelayer}.

\subsubsection{[\ion{N}{II}] Results}  \label{NIIsec} 

\begin{figure}
   \centering
   \includegraphics[width=0.90\hsize]{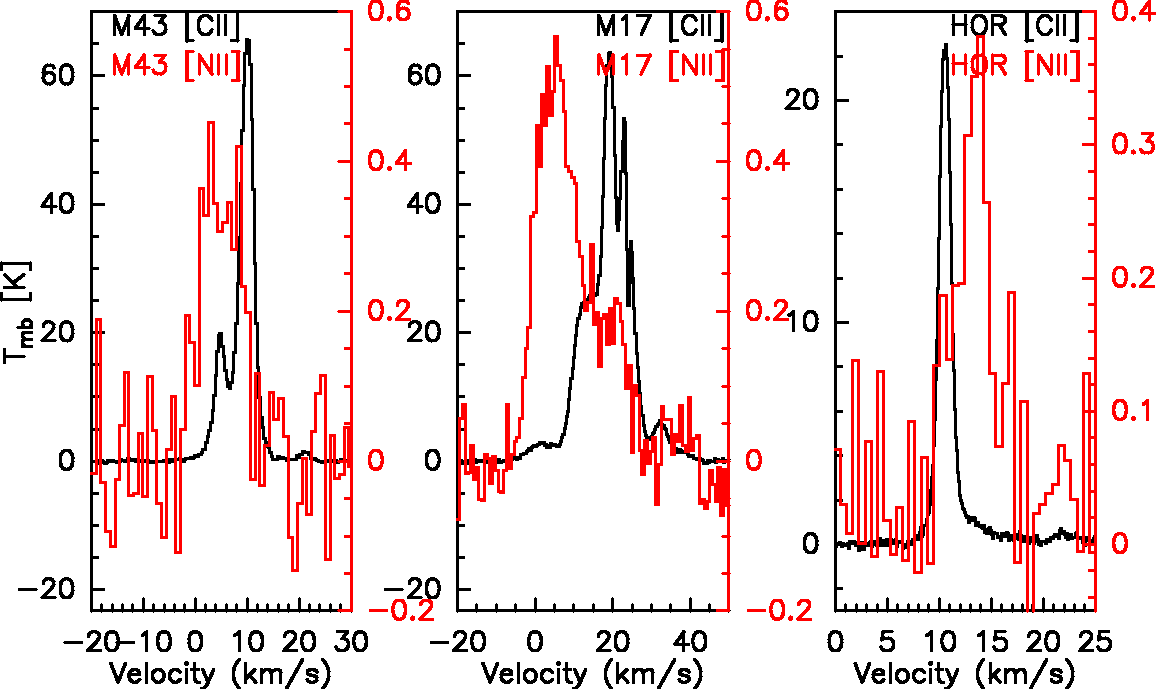}
      \caption{\Cp{} and [\ion{N}{II}] emission for the central pixel for the 3 sources, M43, M17~SW and the Horsehead PDR.}          
         \label{NII_all}
   \end{figure}

We observed the [\ion{N}{II}] 205~$\mu$m line for the central positions of M43, the Horsehead PDR, and M17~SW. For M43, the [\ion{N}{II}] emission is shifted to the blue side of the spectra with respect to the \Cp{12} peak and it has a T$_{\mathrm{peak}}$ of $\sim$ 0.5~K at 4~km/s (Fig.~\ref{NII_all}). For the Horsehead PDR, the [\ion{N}{II}] emission is 
shifted to the red side of the spectra with respect to the \Cp{12} peak. At the M17~SW peak, the [\ion{N}{II}] emission is shifted to the blue side of the spectra with respect to the \Cp{12} emission. The different velocity distribution is an indication that the [\ion{N}{II}] emission originates in a separate component of the cloud, which is likely to be the \ion{H}{II} region (Fig.~\ref{NII_all}).

Based on \citet{2015A&A...576A...1L}, we can estimate the [\ion{N}{II}] column density for the $^{3}$P$_{1}$-$^{3}$P$_{0}$ transition at 205~$\mu$m in the optically thin limit as:

\begin{equation}
 N(\mathrm{N^+}) = 1.9763 \times 10^{15} I([\ion{N}{II}]) / f_1
,\end{equation}

with $N$(N$^+$) the column density of ionized nitrogen in cm$^{-2}$, $I$([\ion{N}{II}]) the integrated intensity of [\ion{N}{II}] in~K~km~s$^{-1}$ and $f_1$ is the fractional population of N$^+$ in the $^{3}$P$_{1}$ state. The fractional population of the different states depends directly on the electron density of the gas as electrons are the main collisional partner of N$^+$ due to its high ionization potential of 14.5~eV. For a kinetic temperature of 8000~K, $f_1$ peaks at 0.40 with an electron density of 100~cm$^{-3}$ \citep{2015ApJ...814..133G}. We assume these values for the estimate of the column density as a lower limit. If we assume that all the nitrogen is ionized, we can estimate the 
ionized hydrogen column density and its equivalent visual extinction. Using a N/H abundance ratio of 5.1 $\times$ 10$^{-5}$ \citep{2007ApJ...654..955J}, we derive the values given in Table~\ref{NIItable}. \par
 
\begin{table}
  \centering
  \caption{M43, the Horsehead PDR, and M17~SW [\ion{N}{II}] column densities and equivalent extinction for the central pixel of the upGREAT array.}
  \begin{tabular}{l r r r }
\hline
\hline
\multicolumn{1}{c}{Sources} & \multicolumn{1}{c}{N(\ion{N}{II})} & \multicolumn{1}{c}{N(H$^+$)}    & \multicolumn{1}{c}{A$_{\mathrm V}$}   \\
                            &                                    & \multicolumn{1}{c}{[\ion{N}{II}]} & \multicolumn{1}{c}{[\ion{N}{II}]}  \\  
                            & \multicolumn{1}{c}{(cm$^{-2}$)}    & \multicolumn{1}{c}{(cm$^{-2}$)}  & (mag.)   \\  
\hline                            
M43                         & 1.43E16 & 2.80E20 & 0.14             \\
HOR PDR                     & 1.21E16 & 2.37E20 & 0.13             \\
M17~SW                      & 3.84E16  & 7.53E20 & 0.40            \\
\hline
  \end{tabular}
         \label{NIItable}
 \end{table}         

The [\ion{N}{II}] emission for the three sources has a much lower column density and hence corresponds to an equivalently lower visual extinction, compared to \Cp{}. Thus, we see that the [\ion{N}{II}] emission is consistent with its origin in the \ion{H}{II} region. When the \ion{H}{II} region is visible in the optical, it is located in front of the 
molecular emission and its emission is expected to be displaced to the blue side of the spectra; this is indeed what we find for M43 and M17~SW. The Horsehead PDR, in contrast, is visible as a dark cloud against the background \ion{H}{II} region, which is located behind the molecular cloud; correspondingly, its [\ion{N}{II}] emission is red-shifted
with respect to \Cp{}. \par

An interesting question is whether the ionized emission originates from inside the \ion{H}{II} region, or whether it is part of an ionized photoevaporation flow. Such flows are made up of ionized material that comes from the molecular region and flow back into the ionized region as a result of the ionization front hitting the molecular region. As a result, one
expects a shift in the velocity of its emission with respect to the molecular gas. \citet{2016MNRAS.457.3593F} estimated a shift in velocity of 1~km/s for the case of a protoplanetary disk for the ionized emission when molecules are dissociated. \citet{2005ApJ...627..813H} estimated a shift of 10~km/s over large volumes of material, and even up to 
17 km/s in dense and magnetized molecular globules \citep{2009MNRAS.398..157H}. \par

We find that the [\ion{N}{II}] emission is shifted 8~km/s for M43 and 15~km/s for M17~SW with respect to the molecular emission, within the range discussed above in Section~\ref{COdis}. For the Horsehead PDR, the [\ion{N}{II}] emission is shifted 5~km/s with respect to \Cp{}. The most probable scenario is an origin of the [\ion{N}{II}] emission that is a combination of ionized gas from the \ion{H}{II} region, and from a photoevaporation flow. For M17~SW in particular, it could be that the emission peaking at 0~km/s originates from the \ion{H}{II} region, whereas the emission at 15~km/s where there is a secondary peak or even a plateau, originates from the photoevaporation flow. However, this is speculative and more evidence is required. \par
      
\subsubsection{M17~SW \ion{H}{I} comparison}
 
\begin{figure}
   \centering
   \includegraphics[width=0.90\hsize]{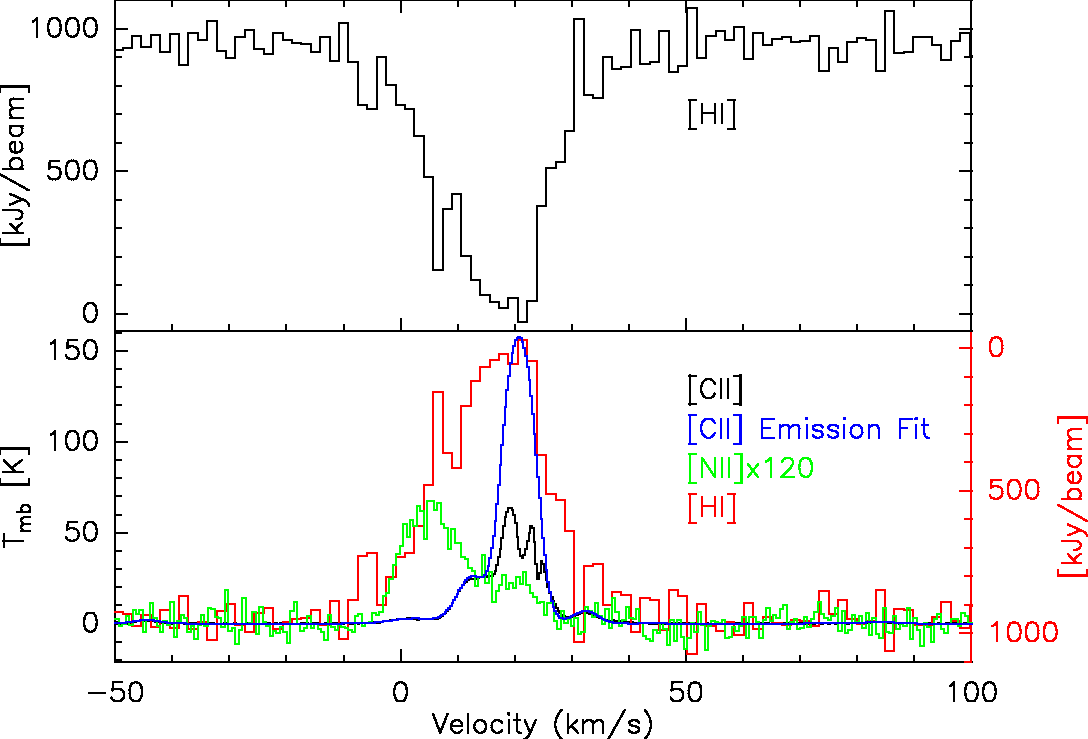}
      \caption{\textit{Top -} M17~SW position 0 \ion{H}{I} absorption spectra. \textit{Bottom -} M17~SW position 6 \Cp{} spectra for the observed and calculated background 
      emission respectively, [\ion{N}{II}] and an inverted \ion{H}{I} spectra for profile comparison.  
              }
         \label{M17HI}
   \end{figure}

In the case of M17~SW, we can extract an \ion{H}{I} absorption spectrum from the \ion{H}{I}/OH/Recombination line survey of the inner Milky Way \citep[THOR,][]{2016A&A...595A..32B}, and compare the velocity profile with its respective counterparts in \Cp{} and [\ion{N}{II}]. The position corresponds to \Cp{} position 0 (Fig.~\ref{fig:sub2}). We can see in Fig.~\ref{M17HI} that the \ion{H}{I} emission is much more extended in velocity, embracing both the \Cp{} and [\ion{N}{II}] emission. The \ion{H}{I} spectrum peaks at 20~kms/s, the same velocity as the \Cp{} peak emission, and it has an asymmetric profile with a long wing at lower velocities, matching at 0~kms/s the low-intensity emission of the weak [\ion{N}{II}]. 

\subsection{Origin of the gas}

The high column densities of the warm background layer in M43, Mon~R2, and M17~SW are difficult to explain in the context of standard PDR-models and ISM phases. In this scenario, the C$^+$ layer in a single PDR layer has typically an A$_{\mathrm v}$ of a few \citep{1997ARA&A..35..179H}. The large values for the C$^+$ column density derived here, with equivalent visual extinctions up to 41~mag, then requires tens of layers of C$^{+}$ stacked on top of each other along the line of sight. This may be possible if the cloud material is very clumpy and fractal with a large fraction of the total cloud material being located in UV-affected clump surfaces. This scenario can be possible, in particular, for M17~SW with its edge-on geometry and known complex structure. Non-standard PDR scenarios might apply; for example, \citet{2007ApJ...658.1119P} proposed that in M17~SW, a high column density of the PDR layer can be obtained by including magnetic fields that would raise the pressure and density of the heated gas. \par

For the foreground material, the situation is much more puzzling: the explanation of the observed line profiles needs ionized gas with high column densities of C$^+$ at a $T_{\mathrm{ex}}$ much lower than possible in PDR scenarios. We can only speculate about its origin: a high X-ray or cosmic ray emission might keep a high fraction of ionized carbon in dense, cold clouds, as long as cooling is efficient to avoid heating through the ionization. For all three sources in which foreground \Cp{12} absorption is observed, namely the two presented in this paper--Mon~R2 and M17~SW, and also NGC~20204 \citep{2012A&A...542L..16G}--there is evidence for a high X-ray or CR flux: M17~SW shows strong emission 
of X-rays due to the large number of young stars \citep{2007ApJS..169..353B}, Mon~R2, has an enhanced X-rays flux coming from T Tauri stars \citep{1998A&A...331..193G, 2003PASJ...55..635N}, and NGC-2024 also exhibits strong X-ray activity \citep{2003ApJ...598..375S}.  \par

\begin{figure}
   \centering
   \includegraphics[width=0.90\hsize]{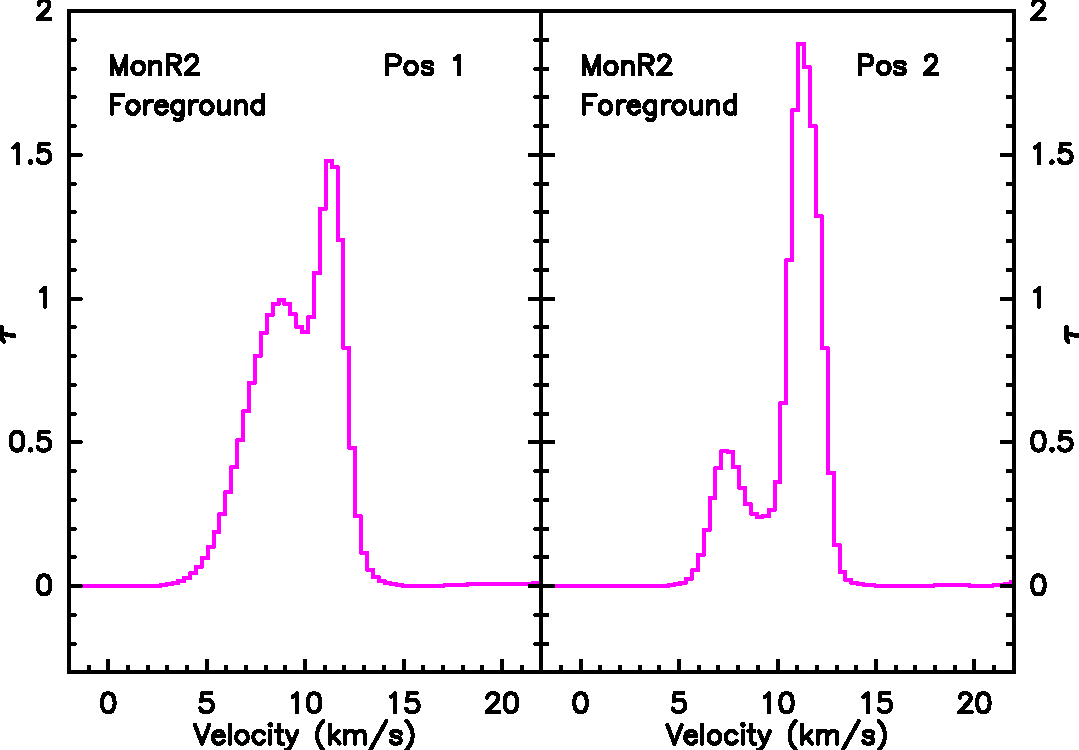}
      \caption{Mon~R2 line profile for the optical depth for the foreground component derived from the multi-component double layer model in pink. The two positions are 20\arcsec ($\sim$ 0.1~pc) apart from each other.
              }
         \label{fig:MonR2fore}
   \end{figure}

\begin{figure}
   \centering
   \includegraphics[width=0.90\hsize]{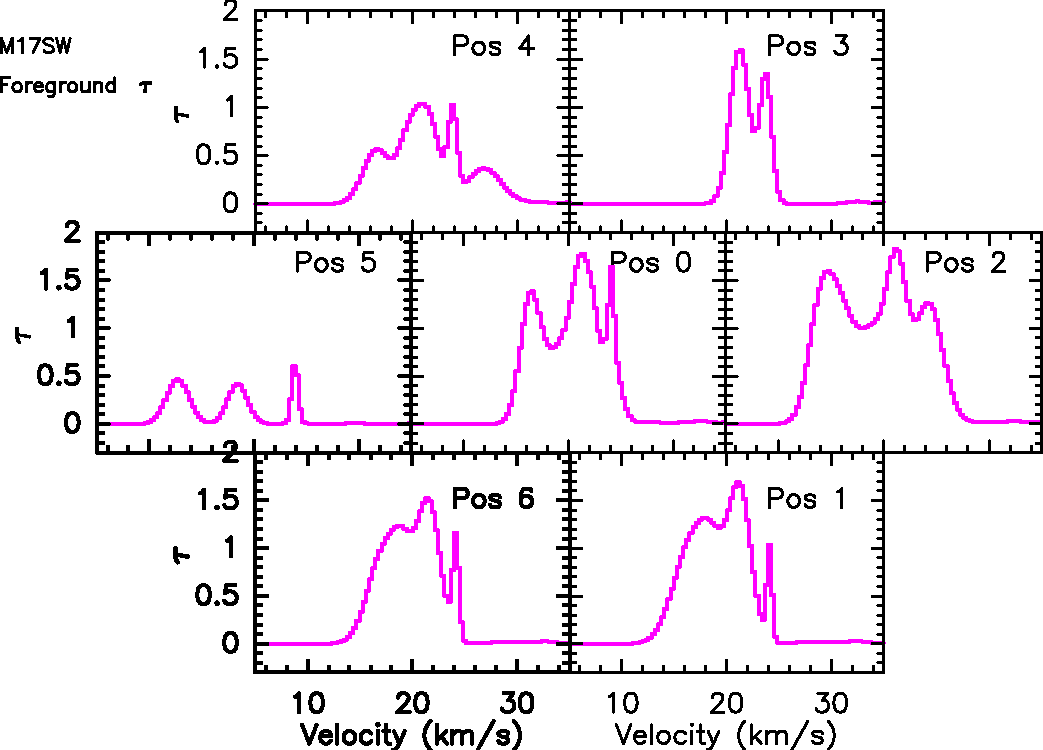}
      \caption{Same as Fig.~\ref{fig:MonR2fore} but for M17~SW. The positions are 30\arcsec ($\sim$ 0.3~pc) apart from each other. 
              }
         \label{fig:M17fore}
   \end{figure}

Even if the nature of the high column density foreground gas is unknown, we have evidence that it is not diffuse, and hence low $T_{\mathrm{ex}}$, ionized gas. We find strong variations in both the line profiles and the absorption patterns between the different observed positions, separated by only 30\arcsec. In Figures~\ref{fig:MonR2fore} and \ref{fig:M17fore}, we show the foreground optical depth line profile derived from the multi-component fits for Mon~R2 and M17~SW, respectively. Even if certain positions share similarities, there are variations in intensity and velocity between the different components. The separation between the positions corresponds to a distance of 0.27~pc for M17~SW and 0.12~pc for Mon~R2. The spatial extent of the absorption feature thus has to be of this order. In combination with the values of the column density of $\sim$ 7 $\times$ 10$^{21}$~cm$^{-2}$ for the foreground gas derived above in Section~\ref{Analysis} (and being a lower limit), the absorbing layer has thus to have, as minimum, a density around 1.3 
$\times$ 10$^{4}$~cm$^{-3}$ for M17~SW and 2.9 $\times$ 10$^{4}$ cm$^{-3}$ for Mon~R2. \par 

\section{Summary}  \label{Conclusions}

Our observations and analysis confirm the long-standing suspicion \citep{1980ApJ...240L..99R,2016A&A...590A..43L}, already proven for the single case of Orion-B \citep{2012A&A...542L..16G}, that \Cp{12} emission is heavily affected by high optical depth or self-absorption effects. The observed \Cp{13} emission, if scaled-up by the abundance ratio of $^{12}$C/$^{13}$C ($\alpha^{+}$) in all cases overshoots the observed \Cp{12} intensities, giving the first indication that the \Cp{12} emission has significant optical depth. A zeroth-order analysis, assuming a homogeneous single layer source, gives an optical depth of the \Cp{12} emission for the two sources M43 and the Horsehead PDR of about 2; for the other two sources, Mon~R2 and M17~SW, the thus derived optical depths are much higher, around 5 to 7. \par 

The integrated \Cp{13} intensities give a strict lower limit to the \Cp{13} upper state column density, valid if the line \Cp{13} line is assumed to be optically thin. In the limit of high excitation temperatures of the C$^+$ fine structure levels, we can also derive the minimum \Cp{13} column density, and from this the \Cp{12} column density 
with the abundance ratio $\alpha^+$. Assuming a $T_{\mathrm{ex}}$ well above 100~K, the thus derived value is a lower limit, rapidly increasing for lower temperatures. Now, ignoring our knowledge of the observed \Cp{13} emission and following the standard approach of assuming that the \Cp{12} emission is optically thin, the thus-derived minimum column density derived from the velocity integrated \Cp{12} line systematically underestimates the C$^+$ column density derived from the observed \Cp{13} line, by a factor as high as 4 (however, the complex line profiles in the latter sources show a clear indication of self-absorption or an otherwise non-homogeneous source structure). \par

When fitting the \Cp{13} and \Cp{12} emission simultaneously in a multi-component source model, the emission of the first two sources, M43 and the Horsehead PDR, can be fitted by emission components only and these fits also allow for the fit of the excitation temperature as a free parameter. The resulting $T_{\mathrm{ex}}$ is around 100~K, leading to\ a higher column density than the one derived from the \Cp{13} integrated intensity using the high $T_{\mathrm{ex}}$ limit.
For the other two sources, the complex line profiles with the apparent self-absorption notches visible in \Cp{12} require the inclusion of a low temperature-absorbing foreground layer. These double-layer, multi-component fits reproduce the combined \Cp{13} and \Cp{12} profiles. Due to the high number of free parameters, it is required that we assume a
fixed value for $T_{\mathrm{ex}}$ both for the background and the foreground layer. For the background components, we used values typical for the C$^+$ layer in PDRs, from 150 to 250~K, as discussed above in Section~\ref{Analysis}; for the foreground, the brightness temperatures in the center of the absorption notches gives an upper limit to the $T_{\mathrm{ex}}$ between 20 and 45~K. In these fits, the \Cp{12} optical depth of the individual components are much lower than the ones derived in the zeroth-order analysis, covering a range up to 2. This is plausible because the unjustified assumption of a constant, uniform $T_{\mathrm{ex}}$ in the zeroth-order analysis, which is clearly not applicable with the complex line profiles, is then released. \par

The total C$^+$ column densities derived for the sources that present absorption dips are slightly larger in the multi-component analysis compared to the ones derived from the \Cp{13} integrated intensity. The bulk of the column density is, of course, constrained by the \Cp{13} emission, but the multi-component source models now adds additional components, which are not visible in \Cp{13}, as their optical depth or excitation temperature are too low for being visible above the noise level. This includes the additional emission components and the foreground absorption components. The latter typically contribute 10\% to 50\% of the total column density. We also took into account a second scenario for the source model, namely a single layer, pure emission, model. But we discard it as physically implausible and also because the individual components traced  in \Cp{} and CO isotopologues do not match. \par

The value of the isotopic abundance ratio $\alpha^+$, where we have used the literature values for the different sources, is an important parameter. A lower ratio would imply a lower \Cp{12}/\Cp{13} intensity ratio in the optically thin limit and correspondingly lower optical depths derived in the zeroth-order analysis and vice versa for a higher value. Although the S/N is not sufficient in the individual spectra of the present observations, we show that for the average spectrum of M17~SW, the S/N is high enough that we can, in fact, derive the value of the abundance ratio from the observed intensity ratio in the optically thin line wings; after applying a correction for the observational bias in the ratio of high and low S/N spectra bins, the line ratio derived for the M17SW average spectrum shows a value for $\alpha^{+}$ slightly higher than the $^{12}$C/$^{13}$C ratio from the literature, which is derived from molecular isotopic species. This may indicate fractionation effects for the ionic and the molecular isotopic species. Higher signal-to-noise observations both for the ionic fine structure lines and the molecular lines may resolve this issue in the future. \par

The \Cp{12} and \Cp{13} study presented here shows that the origin of \Cp{} emission is somewhat more complex than simple model scenarios would suggest. The complex line profiles and high optical depth visible in particular in the bright sources of strong \Cp{} emission in the Milky Way reveal substantially higher column densities of \Cp{} than estimated in the optically thin approximation from integrated line profiles. The self-absorption implies significant column densities of C$^+$ at low temperatures that are, at present, of unknown origin. Therefore, physical parameters derived assuming a standard PDR scenario and considering only line integrated intensities and ratios of velocity-unresolved spectra (which ignores these effects) must be regarded with caution. How empirical correlations between the \Cp{} integrated line intensity and bulk parameters like star formation rate can be explained on the background of this more complex 
origin of the \Cp{} emission will be an interesting issue to resolve, as well a basis for studying how the other cooling line, [\ion{O}{I}], once it is observed at a high spectral resolution, is  affected by self-absorption effects.

\begin{acknowledgements}
This work is based on observations made with the NASA/DLR Stratospheric Observatory for Infrared Astronomy (SOFIA). SOFIA is jointly operated by the Universities Space Research Association, Inc. (USRA), under NASA contract NAS2-97001, and the Deutsches SOFIA Institut (DSI) under DLR contract 50 OK 0901 to the University of Stuttgart. The work is carried out within the Collaborative Research Centre 956, sub-projects A4 and C1 (project ID 184018867), funded by the Deutsche Forschungsgemeinschaft (DFG). R.S. acknowledges support by the French ANR and the German DFG through the project "GENESIS" (ANR-16-CE92-0035-01/DFG1591/2-1). We thank Shuo Kong and the CARMA-NRO Orion team for sharing their CO data. We thank Sandra Treviño-Morales for sharing the Mon~R2 CO data.
\end{acknowledgements}

\bibliographystyle{aa} 
\bibliography{bibpap} 

\begin{appendix}

\section{Observational parameters}\label{app:dpeac}

Here we give the observational parameters in detail for all sources and observed positions of all observed spectra for a detailed reference. The tables contain the absolute and relative coordinates (with respect to the source coordinate) for each observed position, plus the rms noise in K. The mosaics show the \Cp{} spectra observed for the four sources in T$_{\mathrm{mb}}$. The black boxes represent the windows used for the baseline subtraction.

\begin{table}[h]
  \centering
    \caption{M43 positions}
  \begin{tabular}{l r r r r r }
      \hline
      \hline
      & RA      & DEC                     & Rel.       & Rel.       & rms   \\
            & (J2000) & (J2000)                 & Offset     & Offset     &        \\
            & (h:m:s) & (\degr:':'') & l ('') & b ('') & (K)      \\
            \hline
Pos.0  & 5:35:24.16 & $-$5:15:34.14 & $-$107.6 & 28.5  & 0.18         \\
Pos.1  & 5:35:23.18 & $-$5:15:59.82 & $-$122.3 & 2.8 & 0.20         \\
Pos.2  & 5:35:22.19 & $-$5:15:36.07 & $-$137.1 & 26.5  & 0.20         \\
Pos.3  & 5:35:23.35 & $-$5:15:03.49 & $-$119.7 & 59.1  & 0.17         \\
Pos.4  & 5:35:25.39 & $-$5:15:07.15 & $-$89.3 & 55.5  & 0.18         \\
Pos.5  & 5:35:26.42 & $-$5:15:35.11 & $-$73.9 & 27.5   & 0.16         \\
Pos.6  & 5:35:25.23 & $-$5:16:01.35 & $-$91.7 & 1.3 & 0.15         \\
        \hline
  \end{tabular}
\end{table}  
   
\begin{table}[h]
  \centering
    \caption{Horsehead PDR }
  \begin{tabular}{l r r r r r }
      \hline
      \hline
      & RA      & DEC                     & Rel.       & Rel.       & rms   \\
            & (J2000) & (J2000)                 & Offset     & Offset     &        \\
            & (h:m:s) & (\degr:':'') & l ('') & b ('') & (K)      \\
            \hline
Pos.0  & 5:40:53.91 & $-$2:27:14.09 & $-$5.8 & 45.9  & 0.08  \\
Pos.1  & 5:40:52.11 & $-$2:27:29.27 & $-$32.4 & 30.7 & 0.10  \\
Pos.2  & 5:40:52.14 & $-$2:26:59.76 & $-$32.1 & 60.2  & 0.10  \\
Pos.3  & 5:40:53.98 & $-$2:26:43.54 & $-$4.4 & 76.5  & 0.10   \\
Pos.4  & 5:40:55.80 & $-$2:27:00.60 & 22.8 & 59.4  & 0.09  \\
Pos.5  & 5:40:55.75 & $-$2:27:32.54 & 22.1 & 27.5 & 0.08   \\
Pos.6  & 5:40:53.91 & $-$2:27:45.24 & $-$5.6 & 14.8 & 0.09   \\
        \hline
  \end{tabular}
\end{table}

\begin{figure}[ht]
   \centering
   \includegraphics[width=1\hsize]{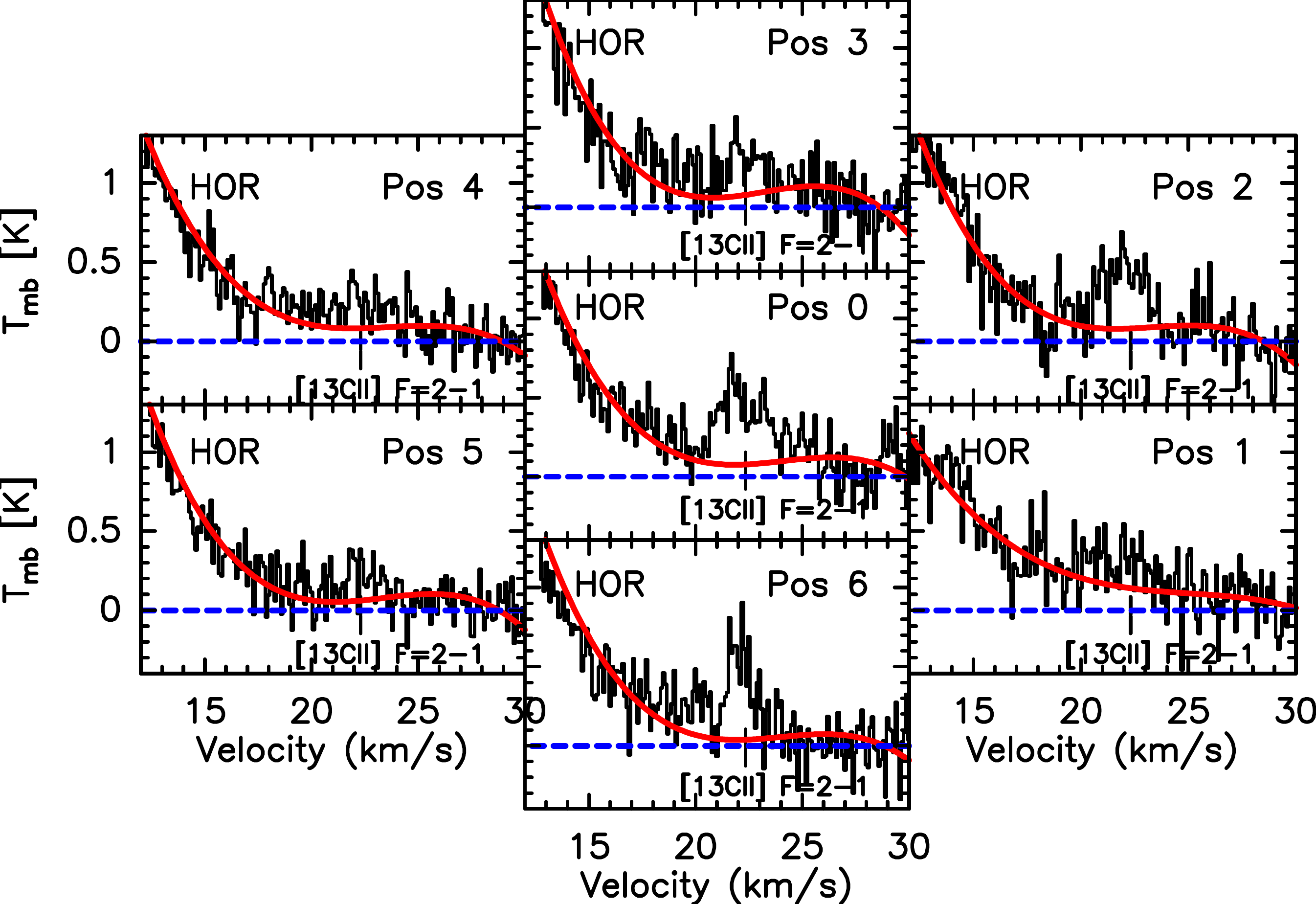}
      \caption{ Horsehead PDR \Cp{13} F=2-1 emission with the \Cp{12} wing subtracted in red with a baseline of order 3. The blue dashed line represents the zero intensity level.
              }
         \label{HORwingall}
   \end{figure}

The weak, but extended, blue-shifted line wing observed in the Horsehead PDR, requires its subtraction from the spectra to derive a proper \Cp{13} $F=2\rightarrow 1$ hyperfine satellite spectrum. In Fig.~\ref{HORwingall}, we show the polynomial fit applied to subtract the wing emission. \par 
   
\begin{table}
  \centering
    \caption{Monoceros~R2}
  \begin{tabular}{l r r r r r }
      \hline
      \hline
      & RA      & DEC                     & Rel.       & Rel.       & rms   \\
            & (J2000) & (J2000)                 & Offset     & Offset     &        \\
            & (h:m:s) & (\degr:':'') & l ('') & b ('') & (K)      \\
            \hline
Pos.1  & 6:07:46.21 & $-$6:23:03.01 & 0 & 5  & 0.17         \\
Pos.2  & 6:07:44.87 & $-$6:23:03.01 & -20 & 5 & 0.3         \\
        \hline
  \end{tabular}
\end{table}

\begin{table}[H]
  \centering
    \caption{M17~SW
}
  \begin{tabular}{l r r r r r }
      \hline
      \hline
     & RA      & DEC                     & Rel.       & Rel.       & rms   \\
            & (J2000) & (J2000)                 & Off.     & Off.     &        \\
            & (h:m:s) & (\degr:':'') & l ('') & b ('') & (K)      \\
            \hline
Pos.0  & 18:20:23.46 & $-$16:12:02.01 & $-$60.6 & $-$1.1   & 0.32   \\
Pos.1  & 18:20:22.45 & $-$16:12:30.20 & $-$75.1 & $-$29.3  & 0.33   \\
Pos.2  & 18:20:21.34 & $-$16:12:05.00 & $-$91.1 & $-$4.1   & 0.23   \\
Pos.3  & 18:20:22.32 & $-$16:11:35.89 & $-$76.9 & 25.0   & 0.22   \\
Pos.4  & 18:20:24.61 & $-$16:11:34.82 & $-$44.0 & 26.1  & 0.31    \\
Pos.5  & 18:20:25.70 & $-$16:12:02.77 & $-$28.2 & $-$1.9  & 0.18    \\
Pos.6  & 18:20:24.66 & $-$16:12:29.23 & $-$43.3 & $-$28.3 & 0.31    \\
        \hline
  \end{tabular}
\end{table}

\section{Off Contamination corrections procedure} \label{Offcp}

The ubiquitous presence of C$^+$ in the ISM and the new conditions of observation established by the use of multi-pixels arrays, together with the large size of the observed sources, have brought the challenge of selecting good off-source (OFF) of blank sky (SKY) positions for the whole array. Using the observation of the OFF against a second far away OFF in some cases, or the SKY-HOT (S-H) spectra in others, we can detect even weak contamination in \Cp{}, and correct the observed spectra by adding this emission into the ON-OFF spectra. \par

\begin{figure}[H]
   \centering
   \includegraphics[width=1\hsize]{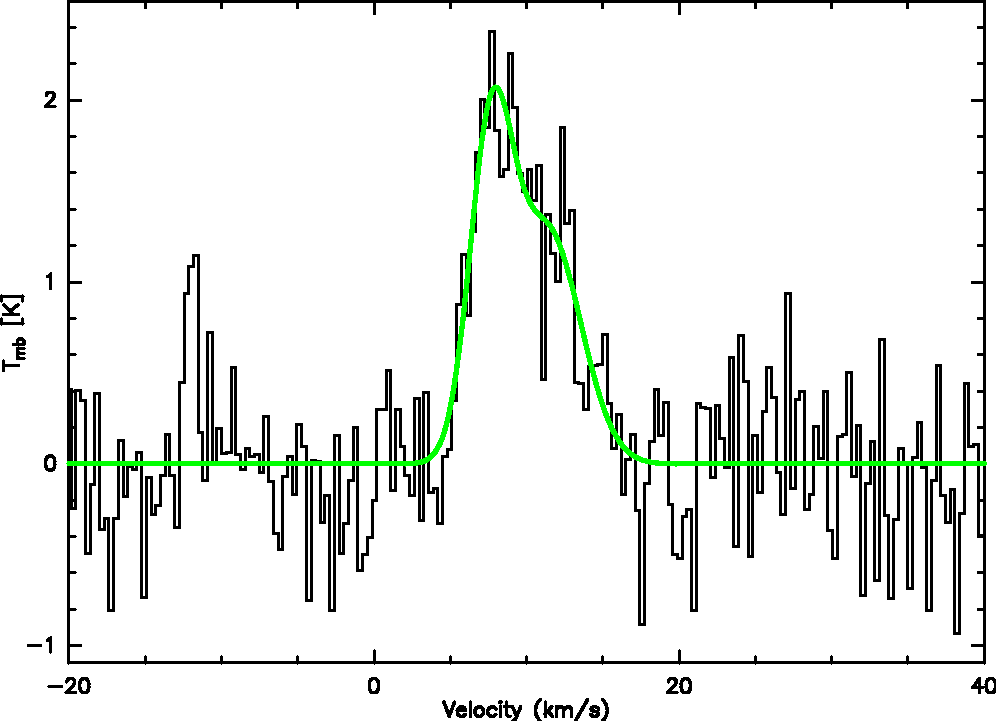}
      \caption{ M43 S-H observation of position 0. The green fit represents the multi-component Gaussian profile fitted to the OFF and added back into the ON-OFF spectra.
              }
         \label{M43SH}
   \end{figure}

\begin{figure}[H]
   \centering
   \includegraphics[width=1\hsize]{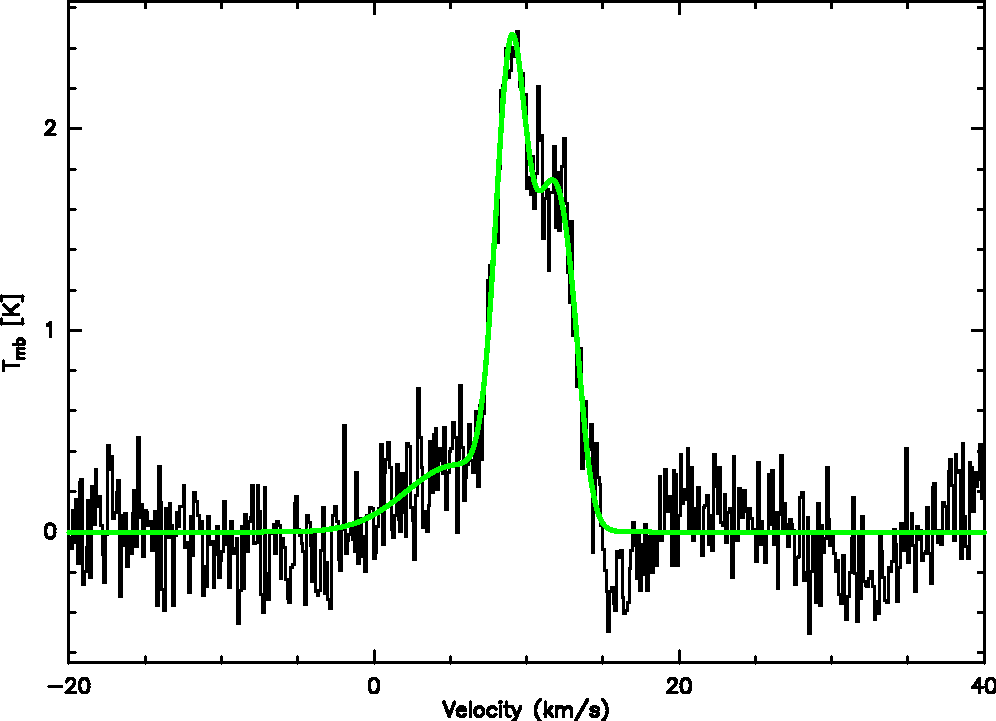}
      \caption{ Mon~R2 S-H observation. As in Figure \ref{M43SH}, the green line shows the OFF-correction model.}
         \label{MonR2SH}
   \end{figure}

\begin{figure}[H]
   \centering
   \includegraphics[width=1\hsize]{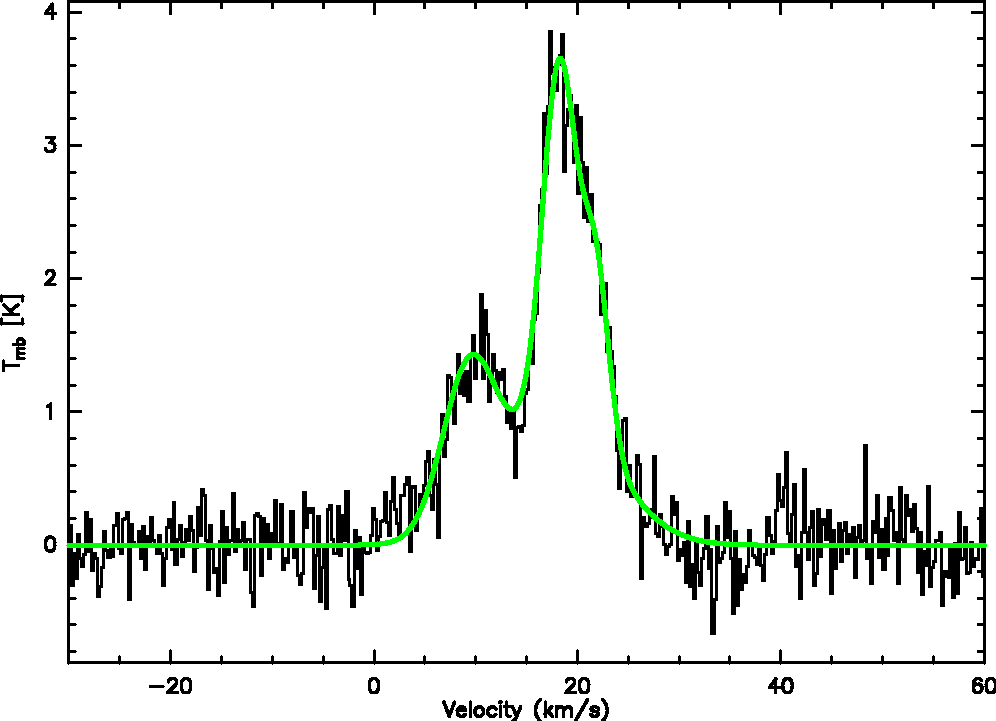}
      \caption{ M17~SW OFF observation of position 6. As in Figure \ref{M43SH}, the green line shows the OFF-correction model.
              }
         \label{M17OFF}
   \end{figure}

The direct addition of the identified OFF emission would add the relatively high noise of each spectral element into the corrected spectra. This can be avoided by modeling the line profile of the identified OFF-emission to correct the ON-OFF spectra. For this reason, we decided to fit a model composed of multiple Gaussians. In Figures~\ref{M43SH}, \ref{MonR2SH}, and \ref{M17OFF}, we show examples of the OFF spectra and their respective Gaussian multi-component models. \par

We added the different OFF Gaussian profiles fitted for each pixel of the array to each contaminated spectrum channel by channel, correcting the contamination and recovering the lost emission. \par

\begin{table}[H]
  \centering
    \caption{Gaussian parameters for the OFF positions}
    \label{offpos}
  \begin{tabular}{l r r r r }
      \hline
      \hline
Source      & Gaussian   & Position & Width  & T$_{\mathrm{peak}}$  \\
            & number     &          & (FWHM) &             \\
          \hline
M43 Pos. 0  & 1          & 11.4    & 5.1    & 1.27         \\
M43 Pos. 0  & 2          & 7.7     & 3.3    & 1.74         \\
\hline
M43 Pos. 1  & 1          & 12.6    & 5.3    & 0.88         \\
M43 Pos. 1  & 2          & 7.3     & 3.9    & 1.71         \\
\hline
M43 Pos. 2  & 1          & 8.2     & 2.8    & 1.71         \\
M43 Pos. 2  & 2          & 10.8    & 5.8    & 1.11         \\
\hline
M43 Pos. 3  & 1          & 11.0    & 6.2    & 0.59         \\
M43 Pos. 3  & 2          & 9.5     & 3.8    & 1.68         \\
\hline
M43 Pos. 4  & 1          & 11.2    & 5.2    & 0.86         \\
M43 Pos. 4  & 2          & 8.8     & 3.4    & 1.81         \\
\hline
M43 Pos. 5  & 1          & 12.8    & 4.3    & 0.75         \\
M43 Pos. 5  & 2          & 8.0     & 3.3    & 1.16         \\
\hline
M43 Pos. 6  & 1          & 9.9     & 3.2    & 1.11         \\
M43 Pos. 6  & 2          & 6.8     & 2.5    & 1.88         \\
       \hline
MonR2  & 1          & 12.1    & 2.8    & 1.71         \\
MonR2  & 2          & 5.2     & 8.0    & 0.46         \\
MonR2  & 3          & 10.2    & 3.4    & 0.40         \\
MonR2  & 4          & 9.0     & 2.4    & 2.13         \\
       \hline
M17SW Pos. 0  & 1          & 20.5   & 4.8    & 2.18         \\
M17SW Pos. 0  & 2          & 15.2   & 12.7   & 0.93         \\
M17SW Pos. 0  & 3          & 11.1   & 3.6    & 0.76         \\
\hline
M17SW Pos. 1  & 1          & 20.1    & 4.5   & 1.58         \\
M17SW Pos. 1  & 2          & 15.2    & 10.8  & 0.96         \\
M17SW Pos. 1  & 3          & 10.7    & 2.6   & 0.99         \\
M17SW Pos. 1  & 4          & 21.6    & 1.1   & 0.29         \\
\hline
M17SW Pos. 2  & 1          & 19.6    & 4.1   & 2.62         \\
M17SW Pos. 2  & 2          & 20.3    & 14.6  & 0.55         \\
M17SW Pos. 2  & 3          & 10.5    & 4.4   & 0.71         \\
M17SW Pos. 2  & 4          & 23.0    & 2.1   & 0.14         \\
\hline
M17SW Pos. 3  & 1          & 20.5    & 4.4   & 3.53         \\
M17SW Pos. 3  & 2          & 10.2    & 7.1   & 0.42         \\
M17SW Pos. 3  & 3          & 9.2     & 4.1   & 0.55         \\
M17SW Pos. 3  & 4          & 24.4    & 2.0   & 0.56         \\
\hline
M17SW Pos. 4  & 1          & 20.6    & 3.9    & 4.70         \\
M17SW Pos. 4  & 2          & 11.3    & 7.5    & 0.83         \\
M17SW Pos. 4  & 3          & 10.6    & 7.1    & 0.47         \\
\hline
M17SW Pos. 5  & 1          & 20.2    & 5.4    & 2.62         \\
M17SW Pos. 5  & 2          & 12.5    & 11.9   & 0.39         \\
M17SW Pos. 5  & 3          & 15.0    & 9.7    & 1.04         \\
M17SW Pos. 5  & 4          & 10.6    & 2.4    & 1.17         \\
M17SW Pos. 5  & 5          & 21.2    & 2.3    & 1.11         \\
\hline
M17SW Pos. 6  & 1          & 19.0     & 5.9    & 2.99         \\
M17SW Pos. 6  & 2          & 23.5     & 7.3    & 0.16         \\
M17SW Pos. 6  & 3          & 17.4     & 14.1   & 0.45         \\
M17SW Pos. 6  & 4          & 9.7      & 5.4    & 1.20         \\
       \hline
  \end{tabular}
         \label{gaussians}
\end{table}

\section{Measuring the abundance ratio from optically thin line wings} \label{app:abundanceline}

As the optical depth approaches a value of zero in the line wings we can measure the underlying abundance ratio $\alpha^+$ from the ratio of the intensities T$_{\mathrm{mb},12}$/T$_{\mathrm{mb},13}$ in the line wings where $\tau_{12} \ll 1$
(see Eq.~\ref{eq:avgtau}). However, the intensity of the \Cp{13} line also becomes low there, so that it is difficult to accurately measure the value of the denominator.\par

\begin{figure}[ht]
\centerline{\includegraphics[angle=90,width=0.9\columnwidth]{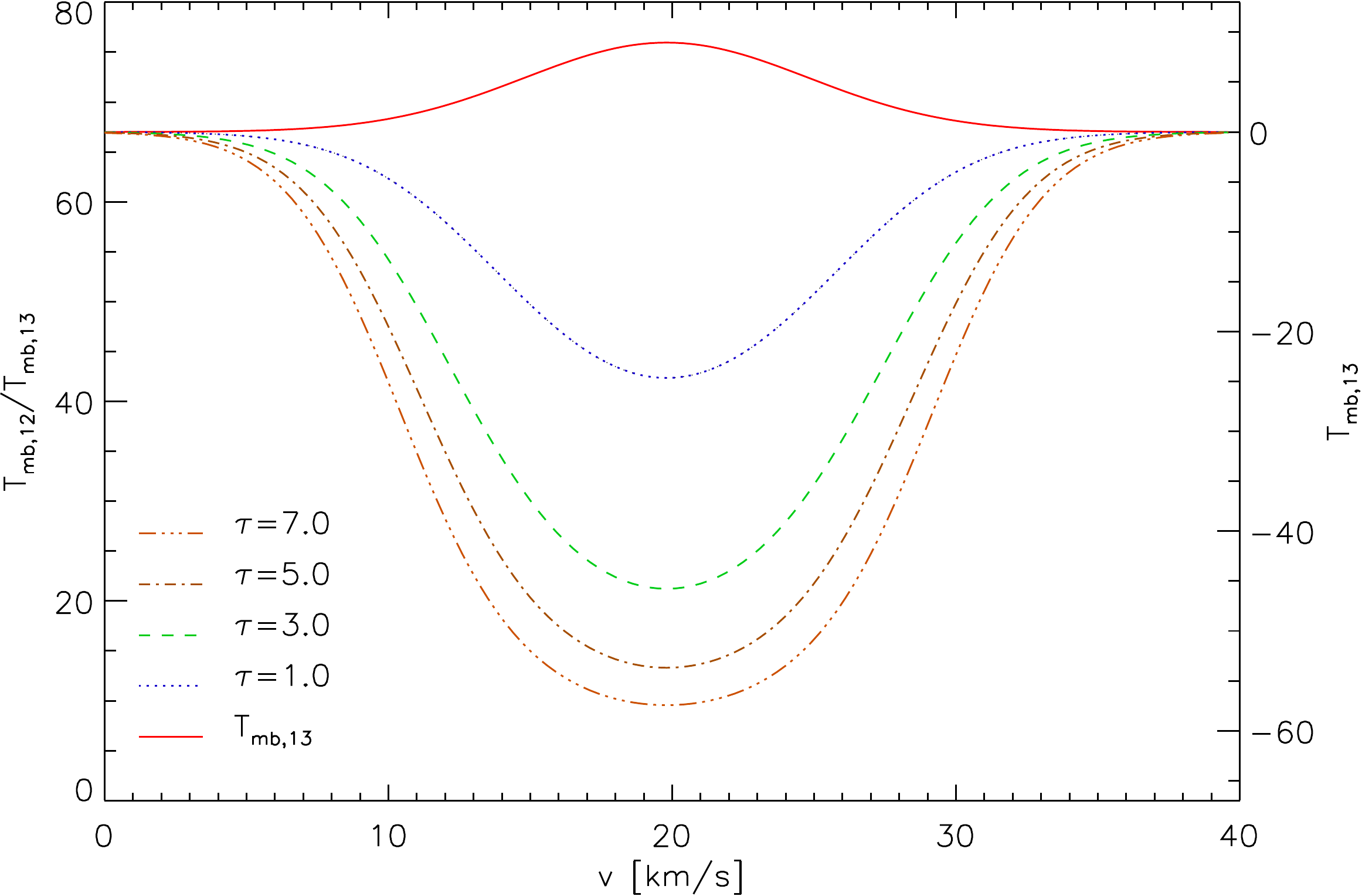}}
\caption{.
\label{fig_nonoise}}
\end{figure}

Figure~\ref{fig_nonoise} visualizes Eq.~\ref{eq:avgtau} for a Gaussian line with $\Delta v=12.5$~\kms{}, and assumed abundance ratio $\alpha^+=67$, and different line center optical depths of the \Cp{12} line. The upper graph shows the intensity of the \Cp{13} line for comparison. We see the prominent decrease in the line ratio T$_{\mathrm{mb},12}$/T$_{\mathrm{mb},13}$ in the line center for the optically thick lines dropping from $\alpha^+$ to values as low as 10 for a line-center optical depth of seven. From the graphs we can estimate in which part of the line wing we can reliably measure the underlying abundance ratio. Using a criterion of values above 65 we see that for $\tau_{12}=1$ the correct abundance ratio is only measured in the part of the wing that drops below 6\,\% of the peak intensity. For $\tau_{12}=3$ one has to follow the wing to less than 3\,\% of the peak intensity and for $\tau_{12}=7$ even down to 1\,\% of
the peak. Using a weaker reliability criterion one can somewhat relax the intensity limits but this simple computation shows that the \Cp{13} sensitivity requirements are very tight to use the line wing channels for a reliable measurement of the elemental abundance ratio. \par

Any real measurement is affected by noise that adds an additional uncertainty to the measurement of the wing intensities. To minimize the noise impact, we only used channels with an intensity above $1.5~\sigma$ of the noise in the determination of the line ratio. However, this approach introduces a bias to the determination of the inherent isotopic ratio $\alpha^+$ as it tends to ignore low \Cp{13} intensities thereby producing too high $\alpha^+$ values from the remaining high \Cp{13} intensities. \par

\begin{figure}
\centerline{\includegraphics[width=0.9\columnwidth]{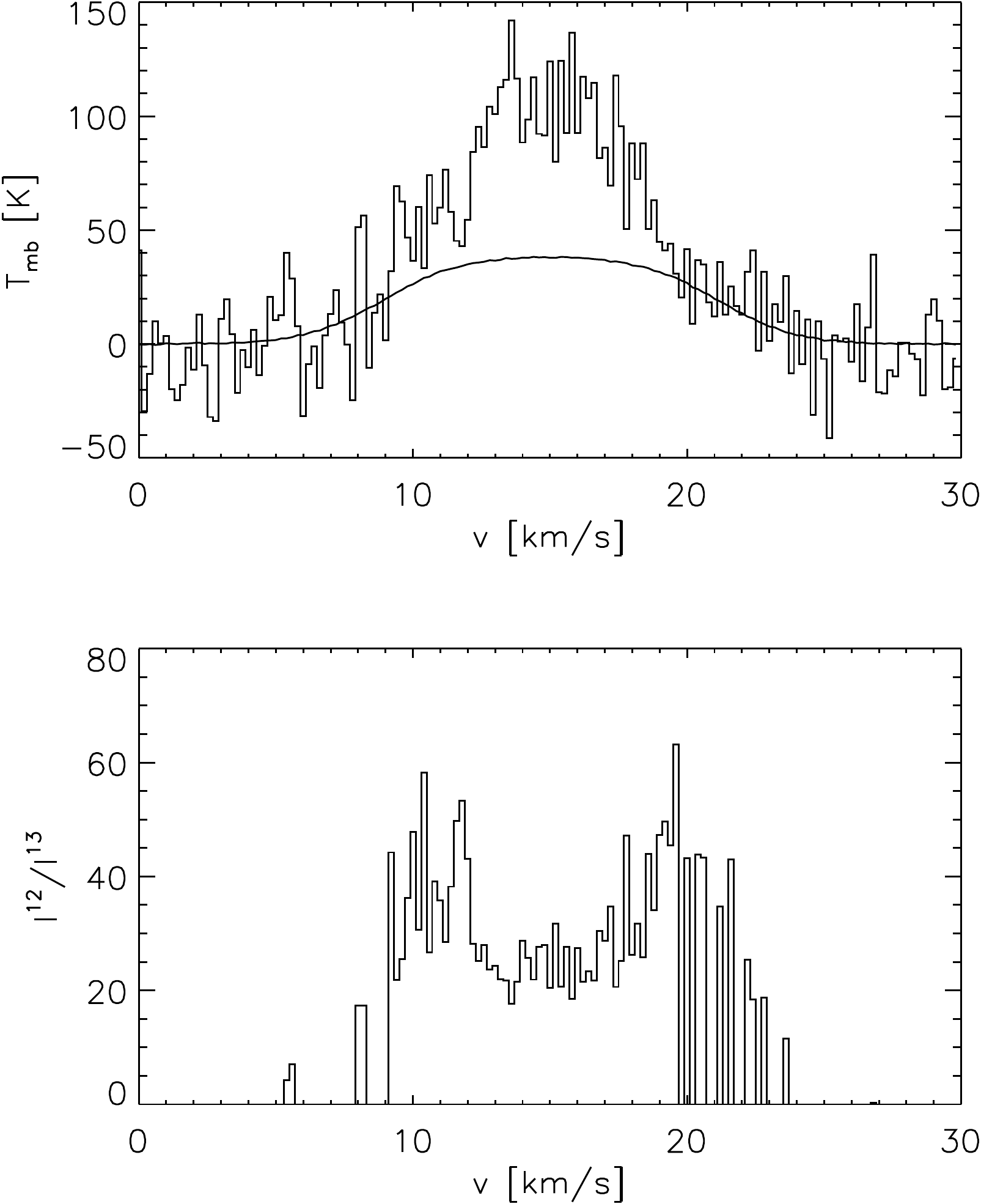}}
\caption{Comparison of the simulated line profiles  of \Cp{12} and \Cp{13} (top) and the line ratio (bottom) equivalent to the plots in Figs.~\ref{M43all}-\ref{M17CIIall}. An intrinsic abundance ratio $\alpha^+=67$ is assumed. The underlying optical depth profile is Gaussian with a width $\Delta v=12.5$~\kms{} and a peak value of $\tau_{12}=3$. Similar to the observations of Mon~R2, noise with $\sigma=0.25$~K was added and the line ratio is only measured above 
$1.5 \sigma$. 
\label{fig_noisyspectra}}
\end{figure}

To study this effect for the typical conditions of our observations, we simulated parameters similar to the Mon~R2 data, excluding the narrow peak in the center, assuming a single layer with a Gaussian profile that covers the whole velocity range of emission, not one of the fitted individual emission components, corresponding to a total width of $\Delta v=12.5$~\kms{}. The peak optical depth in \Cp{} of that component is assumed to be 3.0 and we use an intrinsic abundance ratio $\alpha^+=67$. The velocity resolution is 0.3~\kms{} and similar to the observations of Mon~R2, noise with
$\sigma=0.25$~K was added, corresponding to a signal-to-noise ratio of 7.2 for \Cp{13}. Figure~\ref{fig_noisyspectra} shows the resulting line profiles equivalent to Figs.~\ref{M43all}-\ref{M17CIIall}. The lower plot of the figure shows
the line ratio measured only above $1.5 \sigma =0.37$~K. We see the typical depression of the intensity ratio in the line center that is also clearly visible in Fig.~\ref{M17CIIall} for M17~SW, but in the line wings there is only a single channel left that comes close to the underlying abundance ratio. All other channels underestimate the ratio and those channels that would overestimate the ratio are blanked by the noise cut-off limit. Eye fitting of the wing line ratio would result in estimates of $\alpha^+ \approx 40$ instead of the underlying value of 67 used to compute the line profiles. \par

There is no usable channel range where the optical depth is small enough to reflect the intrinsic abundance ratio but the signal is still above the noise limit. Numerical experiments with lower limits add some channels to the ratio plot that overshoot the intrinsic abundance ratio but also channels that undershoot. The ratio plot becomes much noisier and one does not really obtain a better estimate of the intrinsic $\alpha^+$.  \par
 
A better estimate of the abundance ratio may, however, come from a systematic correction of the measured wing ratio. For this purpose, we ran a five-dimensional parameter study varying the characteristic quantities of the problem. These are the intrinsic abundance ratio $\alpha^+$, the line-center optical depth of the \Cp{12} line, the signal-to-noise ratio of the accumulated \Cp{13} line, the velocity resolution of the observations relative to the line width, and the intensity limit used for the line ratio determination in units of the noise standard deviation. The only fixed assumption of the simulation is a Gaussian velocity distribution of the emitting material. In the resulting line ratio
plot, we located the peak and measured the average ratio in the ten channels around the peak. \par

\begin{figure}
\centerline{\includegraphics[angle=90,width=0.9\columnwidth]{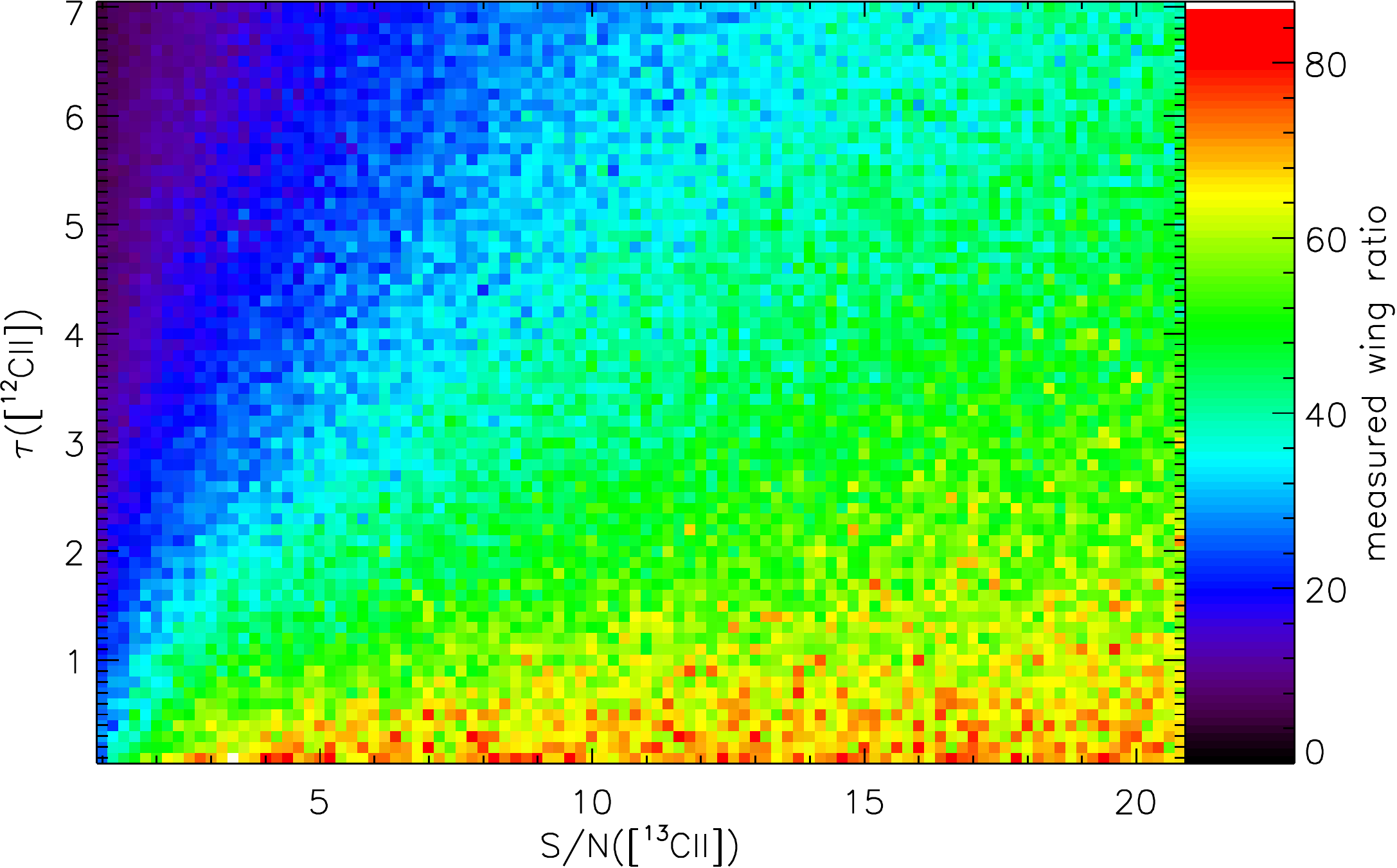}}
\caption{Measured wing intensity ratio as a function of the \Cp{12} line
center optical depth and the signal-to-noise ratio of the \Cp{13} line.
Here, we assumed an abundance ratio $\alpha^+=67$, a channel width of 
$0.024 \Delta v$, and an intensity limit of $1.5 \sigma$ of the noise.
The noisy structure of the plot results from the random noise addition to
the individual spectra.
\label{fig_parameterscan_default}}
\end{figure}

\begin{figure}
\centerline{\includegraphics[angle=90,width=0.9\columnwidth]{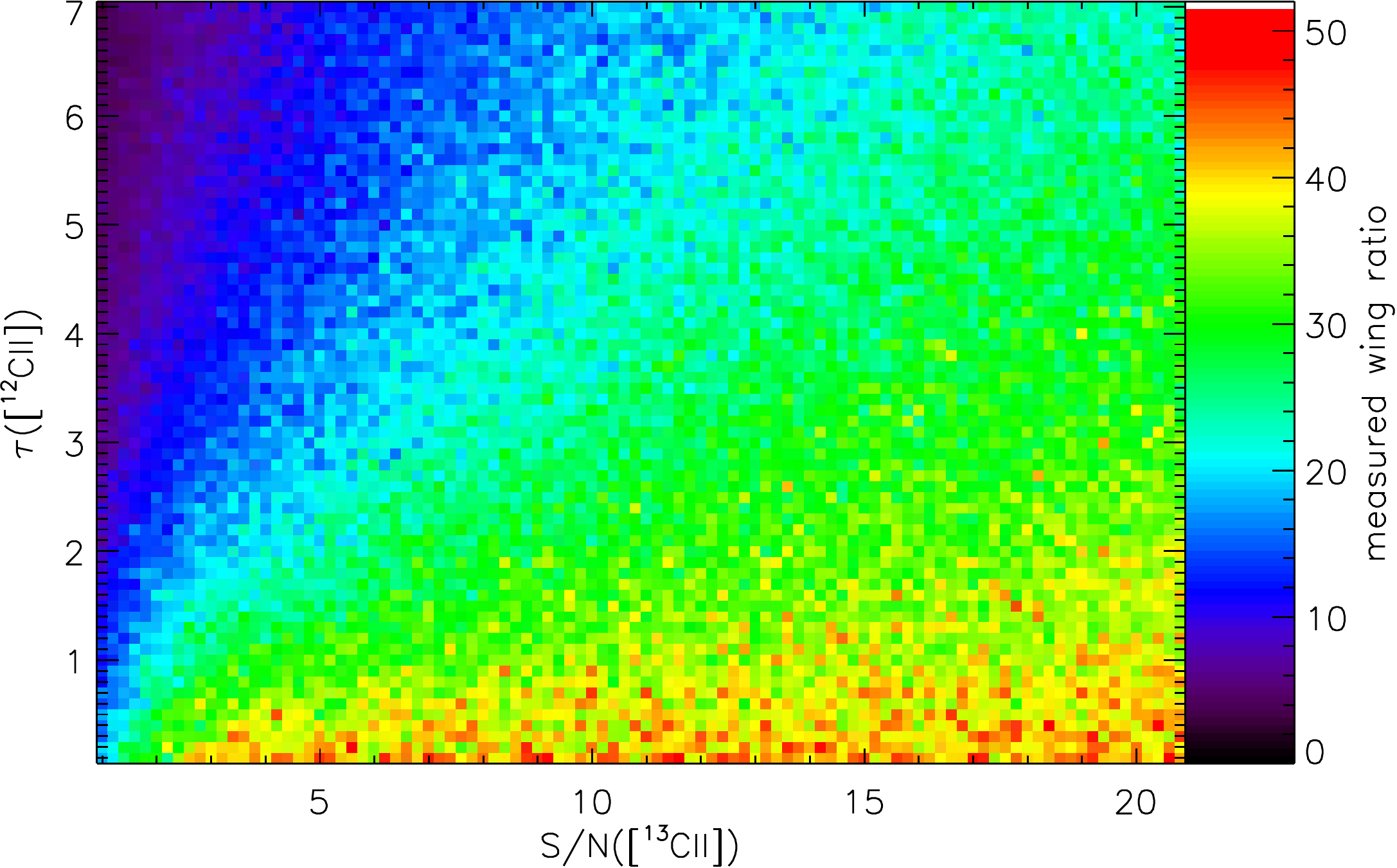}}
\caption{Like Fig.~\ref{fig_parameterscan_default} but for 
an abundance ratio $\alpha^+=50$.
\label{fig_parameterscan_alpha50}}
\end{figure}

\begin{figure}
\centerline{\includegraphics[angle=90,width=0.9\columnwidth]{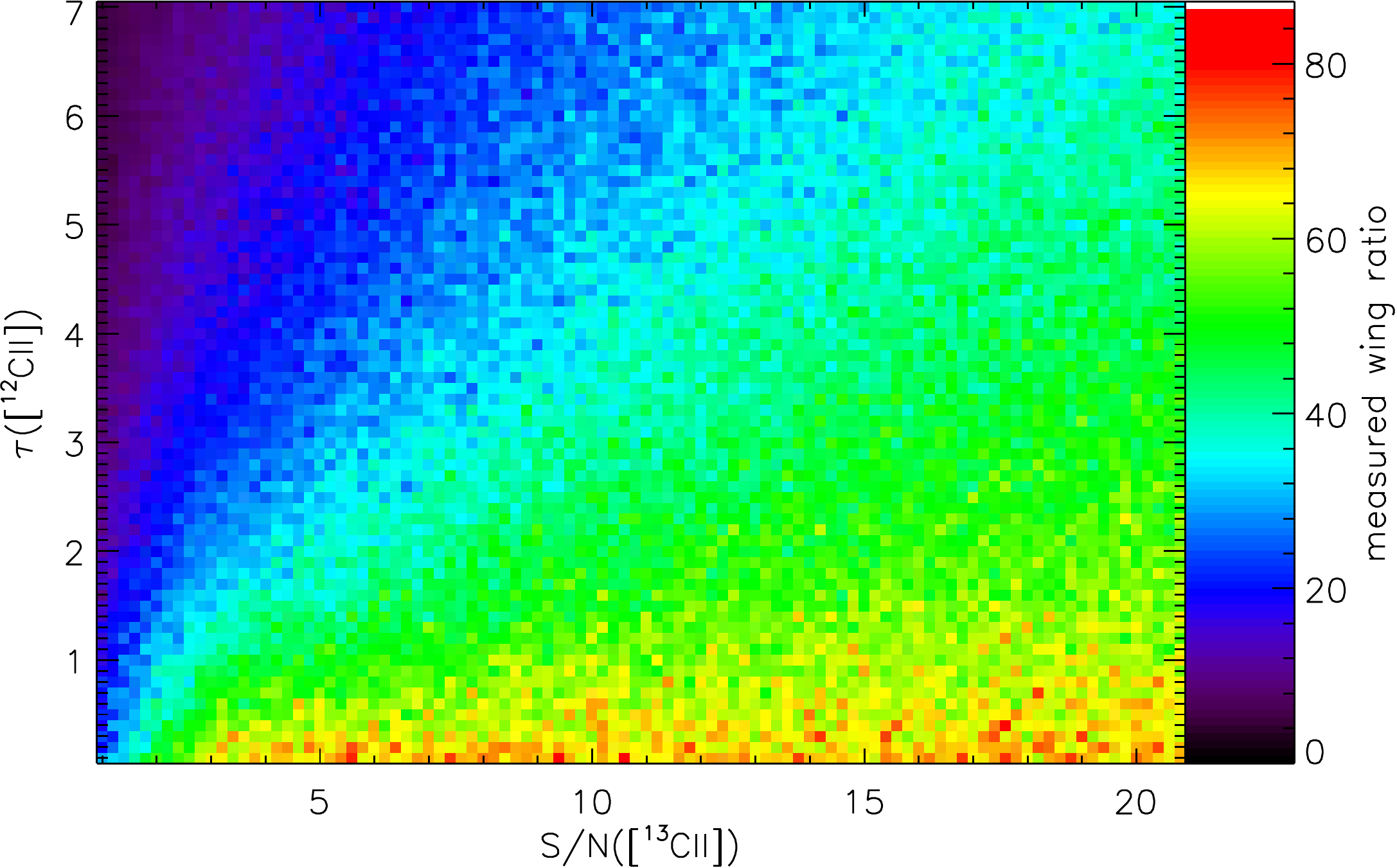}}
\caption{Like Fig.~\ref{fig_parameterscan_default} but for 
an intensity limit of $2.0 \sigma$ of the noise.
\label{fig_parameterscan_limit2_0}}
\end{figure}

\begin{figure}
\centerline{\includegraphics[angle=90,width=0.9\columnwidth]{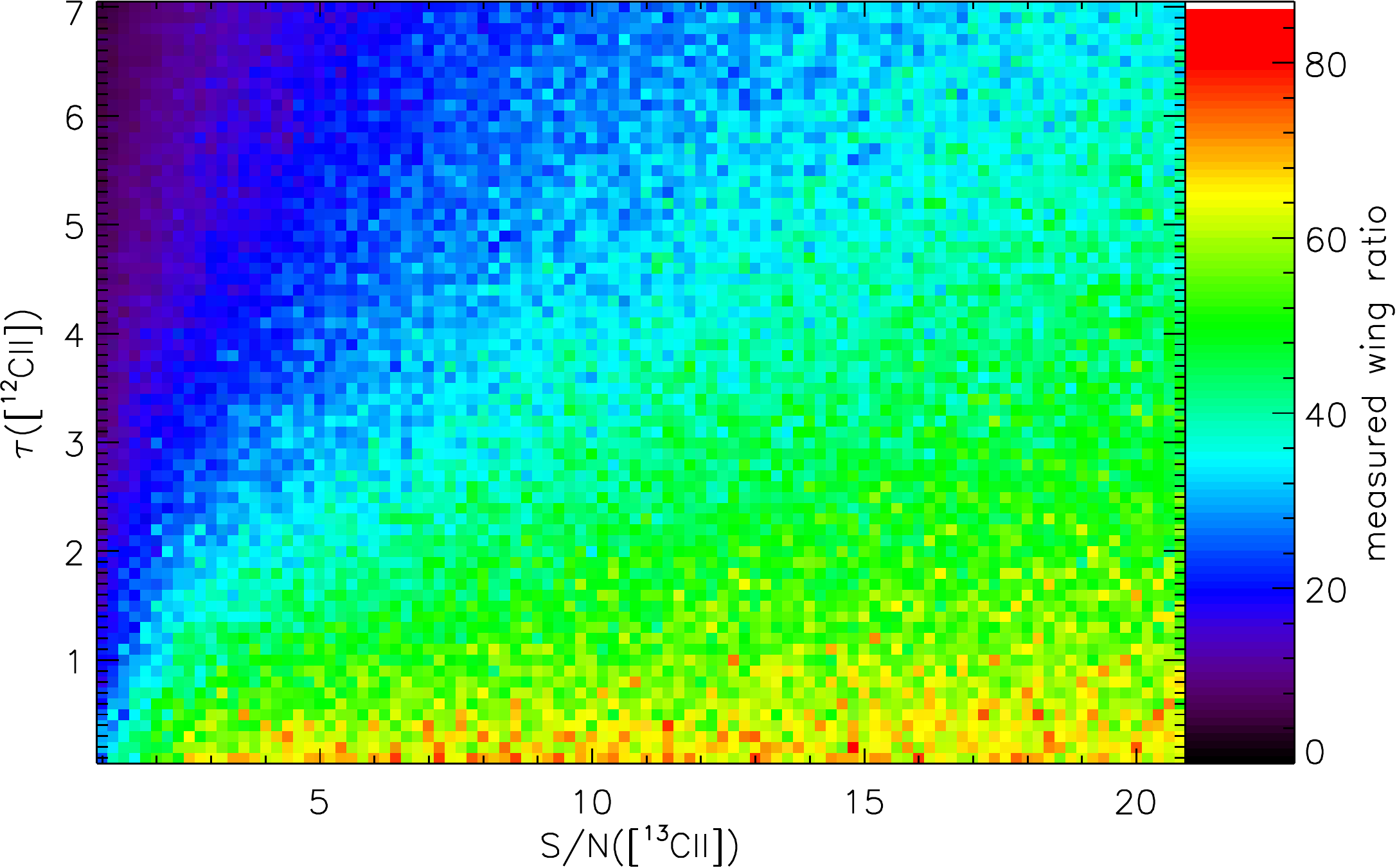}}
\caption{Like Fig.~\ref{fig_parameterscan_default} but for a channel width of 
$0.048 \Delta v$.
\label{fig_parameterscan_width2_5}}
\end{figure}

The results for some parameter combinations are shown in Figs.~\ref{fig_parameterscan_default}-\ref{fig_parameterscan_width2_5}. Figure~\ref{fig_parameterscan_default} shows the dependence on the measured wing intensity ratio from the \Cp{13} signal-to-noise ratio and the \Cp{12} line center optical depth when using an intrinsic
abundance ratio $\alpha^+=67$, a channel width of $0.024 \Delta v$, and an intensity limit of $1.5 \sigma$ of the noise.
In every part of the plot, we find a significant scatter of the measured wing intensity ratio with peak-to-peak deviations of 30-40\,\% of the mean local value. That noisy structure results from the random noise added to the individual spectra. As the noise contribution in the observations is unknown we have to take this scatter as a
fundamental uncertainty of the method even if all other parameters are accurately known. In terms of a $1 \sigma$ uncertainty, this corresponds to about 5-7\,\%. \par

On top of the scatter we see the systematic underestimate of the intrinsic abundance ratio in the case of high optical depths and low signal-to-noise ratios. At a S/N$=5$ even an optical depth around unity leads to an underestimate of $\alpha^+$ by about 10. At a S/N$=20$ the same underestimate is reached at an optical depth of 3.0. Knowing the two parameters, we can use a smooth representation of the plot to estimate the correction that needs to be applied to
the measured wing intensity ratio to obtain the underlying abundance ratio. \par

In Fig.~\ref{fig_parameterscan_alpha50} we show the same plot for an underlying abundance ratio of $\alpha^+=50$. For 
direct comparability, we have fixed the plotting range to values between 0 and 130\,\% of the used intrinsic abundance ratio $\alpha^+$. With this approach, both plots are almost indistinguishable. This indicates that the abundance underestimation in the line wing intensity ratio is a multiplicative effect. When knowing the optical depth and \Cp{13} signal-to-noise, we can use either Fig.~\ref{fig_parameterscan_default} or \ref{fig_parameterscan_alpha50}
to read the ratio between the measured intensity ratio and the underlying abundance ratio to get the correction factor for any observation. \par

Figures~\ref{fig_parameterscan_limit2_0} and \ref{fig_parameterscan_width2_5} show the dependence on the other two parameters. In Fig.~\ref{fig_parameterscan_limit2_0} we have increased the significance threshold below which intensities are ignored in the ratio measurement to $2.0 \sigma$. As more noise is excluded in this way, the local scatter of the ratio drops to well below 30\,\%. However, the bias towards lower ratios becomes stronger
so that in particular for low optical depths the average abundance estimate gets lower, further deviating from the intrinsic ratio. At $\tau($\Cp{12}$)=2$ the systematic effect is about 10\,\% exceeding the random scatter.
Lowering the limit instead increased the scatter over most of the plot significantly so that we conclude that the limit of $1.5 \sigma$ is a good compromise. \par

In Fig.~\ref{fig_parameterscan_width2_5}, we increased the velocity channel width relative to the line width by a factor of two compared to Fig.~\ref{fig_parameterscan_default}. We find a small systematic worsening of the abundance ratio estimate because of the smaller number of channels in the line wings. However, in real observations, one
can only adjust the velocity channel width of the observations simultaneously changing the signal-to-noise ratio that one can obtain in the same observation time. It is therefore necessary to compare every point in Fig.~\ref{fig_parameterscan_width2_5} with the corresponding point in Fig.~\ref{fig_parameterscan_default} that has the same optical depth, but a factor $\sqrt{2}$ lower \Cp{13} S/N. In this comparison, the abundance estimate with the narrower channel width but the high noise is closer to the intrinsic abundance ratio so that we discourage additional binning. \par

Altogether, the numerical experiment shows that even for sources with a moderate \Cp{12} optical depth around unity, the line wings do not allow to directly measure the underlying abundance ratio $\alpha^+$ but that a correction is needed. A direct use of the measured wing intensity ratio as an abundance ratio would need \Cp{13} S/N ratios above 100. The required correction can be read from the plotted parameter study because of the weak dependence on the line
width and the linear dependence on the assumed abundance ratio. It depends on a good estimate of the central optical depth of the \Cp{12} line. We cannot directly use the optical depths measured in Sect.~\ref{subsec:zero} and \ref{Analysis} because they characterize the narrowest component, rather than the broad component that dominates the wings. The relevant optical depth for the wing component should be somewhat lower. \par

One way to measure the relevant optical depth is the use of a variable signal-to-noise ratio. In M17~SW, we measure for one individual \Cp{13} spectrum a S/N$\approx 9$ and an optical depth of 6. There the wing intensity ratio reaches values of only about 20. On the other hand, the averaged M17~SW spectrum, has a S/N$\approx 18$ and the wing intensity ratio grows to about 40. We can use Figs.~\ref{fig_parameterscan_default} or \ref{fig_parameterscan_alpha50} to check for which optical depth the change of the \Cp{13} S/N provides an increase in the wing intensity ratio by 20. Following a horizontal line in the parameter studies, a change by 20 when going from S/N$=9$ to S/N$=18$ occurs at $\tau($\Cp{12}$)\approx 4$ when assuming $\alpha^+=67$ and at $\tau($\Cp{12}$)\approx 5$ for $\alpha^+=50$. This is just somewhat lower than the estimated optical depth from Sect.~\ref{C12od}. Looking up the difference between the measured intensity ratio and the underlying abundance ratio at these optical depths shows that even for S/N$=18,$ we underestimate the abundance ratio by 30\%. This means that the intrinsic abundance ratio for M17~SW is, rather, 60, not 40. \par

\section{Alternative Scenario: Multi-component single layer model} \label{singlelayer}

In the multi-component dual-layer scenario from Section~\ref{Analysis}, the complex line profiles of Mon~R2 and M17~SW have been interpreted as broad emission from the background and narrow absorption notches from cold foreground material. This is not the only possible way of fitting the profiles. An alternative scenario is explored here: a multi-component, but single emission layer model. In order to match the observed profiles, this scenario requires the observed \Cp{13} smooth profiles to be reflected by a similarly broad, but highly optically thick and therefore flat-topped \Cp{12} emission profiles. The corresponding gas has to be relatively cold to not overshoot the observed emission. The complex profile shape is then achieved by adding additional narrow line emission components. We explore this alternative scenario here and we show that it is perfectly feasible in terms of obtaining a reasonable fit solution. However, the follow-up analysis of the resulting column densities in comparison to other tracers, as well as the resulting physical scenario, make this solution physically unrealistic. \par

\begin{figure}
   \centering
   \includegraphics[width=\hsize]{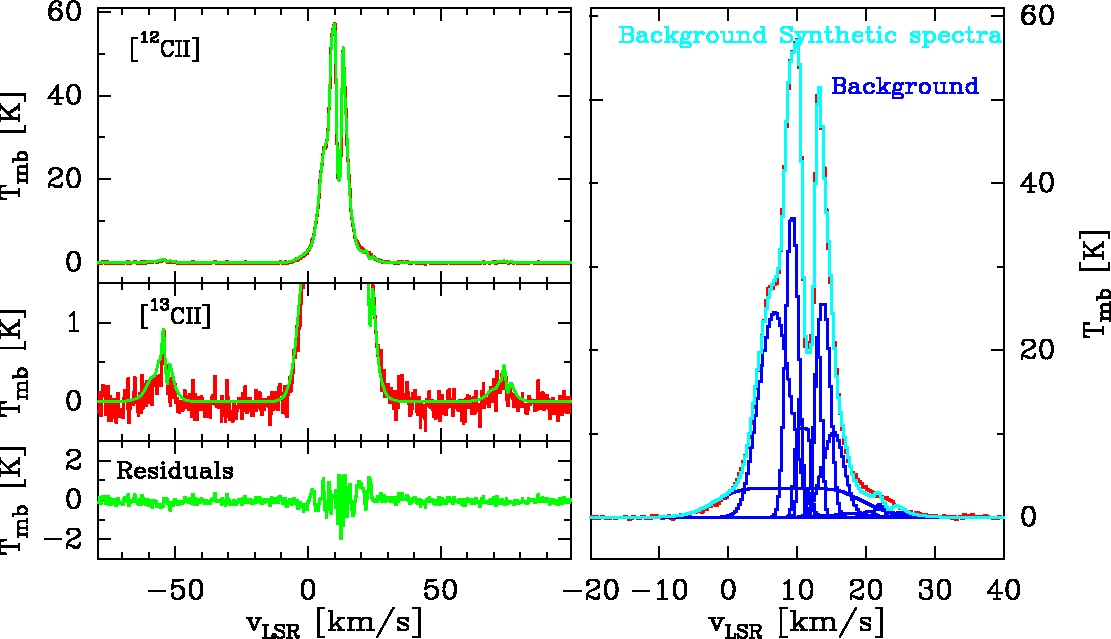}
      \caption{ 
      Same as Fig.~\ref{13CIIsteps}, but for Mon~R2 \Cp{12} spectra of position 2 with no foreground absorption. 
              }
         \label{MonR2alt}
   \end{figure}

\begin{figure}
   \centering
   \includegraphics[width=\hsize]{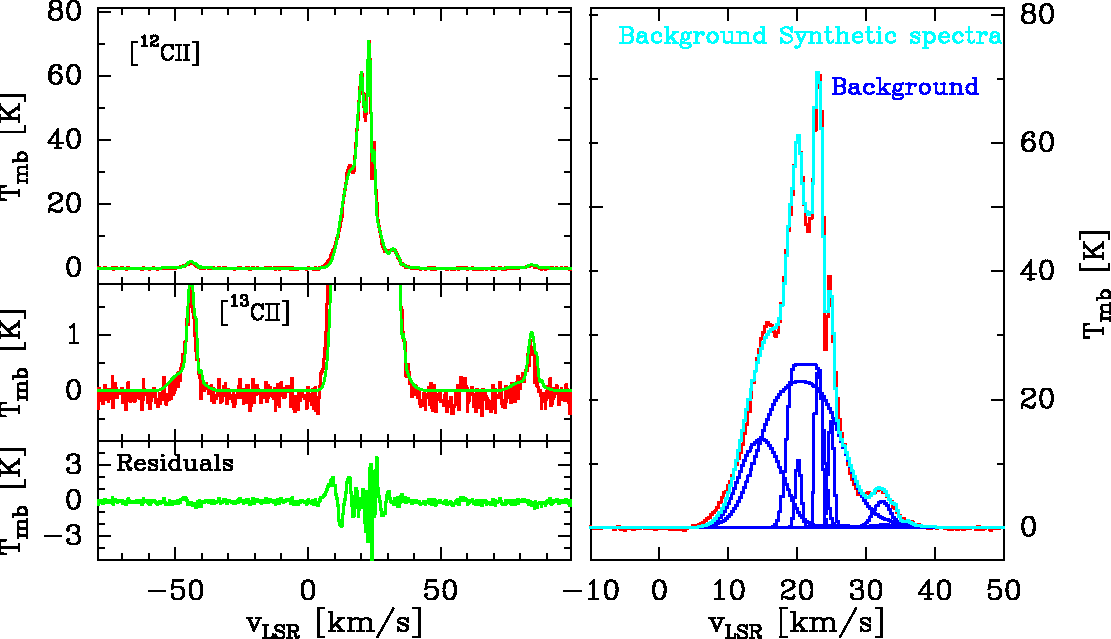}
      \caption{ 
           Same as Fig.~\ref{13CIIsteps}, but for M17~SW \Cp{12} spectra of position 6 with no foreground absorption.
              }
         \label{M17alt}
   \end{figure}

The resulting multi-component fit, assuming this single layer model, is shown in Fig.~\ref{MonR2alt} for Mon~R2 and in Figs.~\ref{M17alt} for M17~SW. In Table~\ref{M17tableback}, we summarize the fitted parameters, with the number of components per position for each source, the $\chi^2$ of the fitting, the \Cp{} column density, the peak optical depth and the equivalent visual extinction. As indicated above, the components can be separated into two types. One is cold high density gas showing a flat-top \Cp{12} profile due to high optical depth that contributes to the \Cp{13} emission. The other is warmer, lower density gas with much narrower \Cp{12} profiles tracing the velocity peaks of the \Cp{12} emission and, due to their low column density, being negligible in \Cp{13}. \par

In this scenario, M17~SW contains two kinds of components. Cold components with $T_{\mathrm{ex}}$ $\sim$ 50~K at extremely high \Cp{} column density and a peak optical depth of 14. This emission is complemented by narrow components with $T_{\mathrm{ex}}$ between 50 and 70~K and lower column density, on the order of 10$^{17}$~cm$^{-2}$ per individual component, fitting the \Cp{12} peaks. The composition is similar for Mon~R2 (See Tables~\ref{Monnofore} and \ref{table:m17single} in the Appendix for a description of each source). \par

We note that in this scenario the warm, narrow line components are only visible when they are located in front of the optically thick low-temperature central emission, as they would otherwise be absorbed away by the central component. \par

 \begin{table*}
  \centering
    \caption{Mon~R2 and M17~SW column density and equivalent extinction for the single layer model}
  \begin{tabular}{l c c c c c c }
      \hline
      \hline
\multicolumn{1}{c}{Positions} & \multicolumn{1}{c}{No.}   & \multicolumn{1}{c}{$\chi^2$} & $T_{\mathrm{ex}}$   & \multicolumn{1}{c}{N$_{12}$(\ion{C}{II})}  & \multicolumn{1}{c}{$\tau^{*}$} & \multicolumn{1}{c}{ A$_{\mathrm V}$}   \\
                              & \multicolumn{1}{c}{Comp.} &                              &                     &                                            &               &   \multicolumn{1}{c}{\Cp{}}    \\
                              &                           &                              &   (K)               & \multicolumn{1}{c}{(cm$^{-2}$)}            &               &  (mag.)            \\
          \hline
MonR2 1 & 5 & 12.8 & 35 & 7.1E18 & 2.62 & 31.7 \\
MonR2 2 & 8 & 3.8  & 40 & 1.7E19 & 10.17 & 75.8  \\
       \hline
M17SW 0 & 11 & 1.3 & 50 & 2.6E19 & 13.88 & 115.9 \\
M17SW 1 & 4  & 1.8 & 50 & 2.5E19 & 14.02 & 111.4 \\
M17SW 2 & 7  & 1.5 & 45 & 1.6E19 & 8.04  & 71.3 \\
M17SW 3 & 12 & 8.0 & 40 & 1.1E19 & 7.26 & 49.0 \\
M17SW 4 & 12 & 1.4 & 55 & 1.8E19 & 7.59 & 80.2 \\
M17SW 5 & 9  & 3.6 & 55 & 7.7E18 & 1.54 & 34.3 \\
M17SW 6 & 7  & 2.9 & 60 & 1.5E19 & 11.26 & 66.8   \\
       \hline
  \end{tabular}
         \label{M17tableback}
\end{table*}

It is important to notice that for both sources with self-absorption effects, the optical depths derived using the multi-component double layer model for both layers, background, and foreground, have values lower than 2, closer to the ones expected by traditional models instead of the extremely large values of the single layer model. \par

\begin{figure}
\centering
\begin{subfigure}{\hsize}
   \centering
   \includegraphics[width=0.95\hsize]{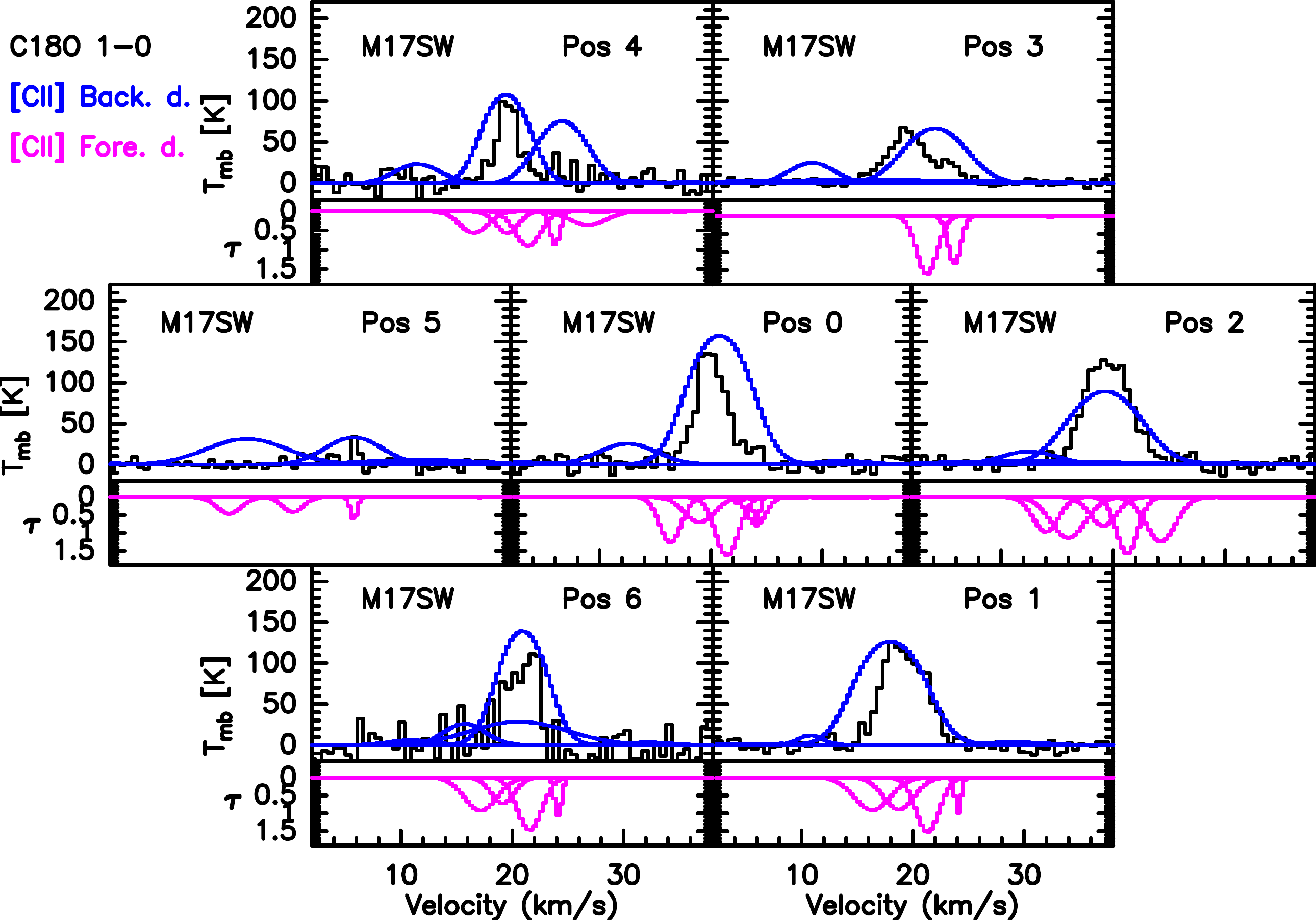}
      \caption{ 
              }
         \label{M17_CO_double}
\end{subfigure}%
\par
\begin{subfigure}{\hsize}
   \centering
   \includegraphics[width=0.95\hsize]{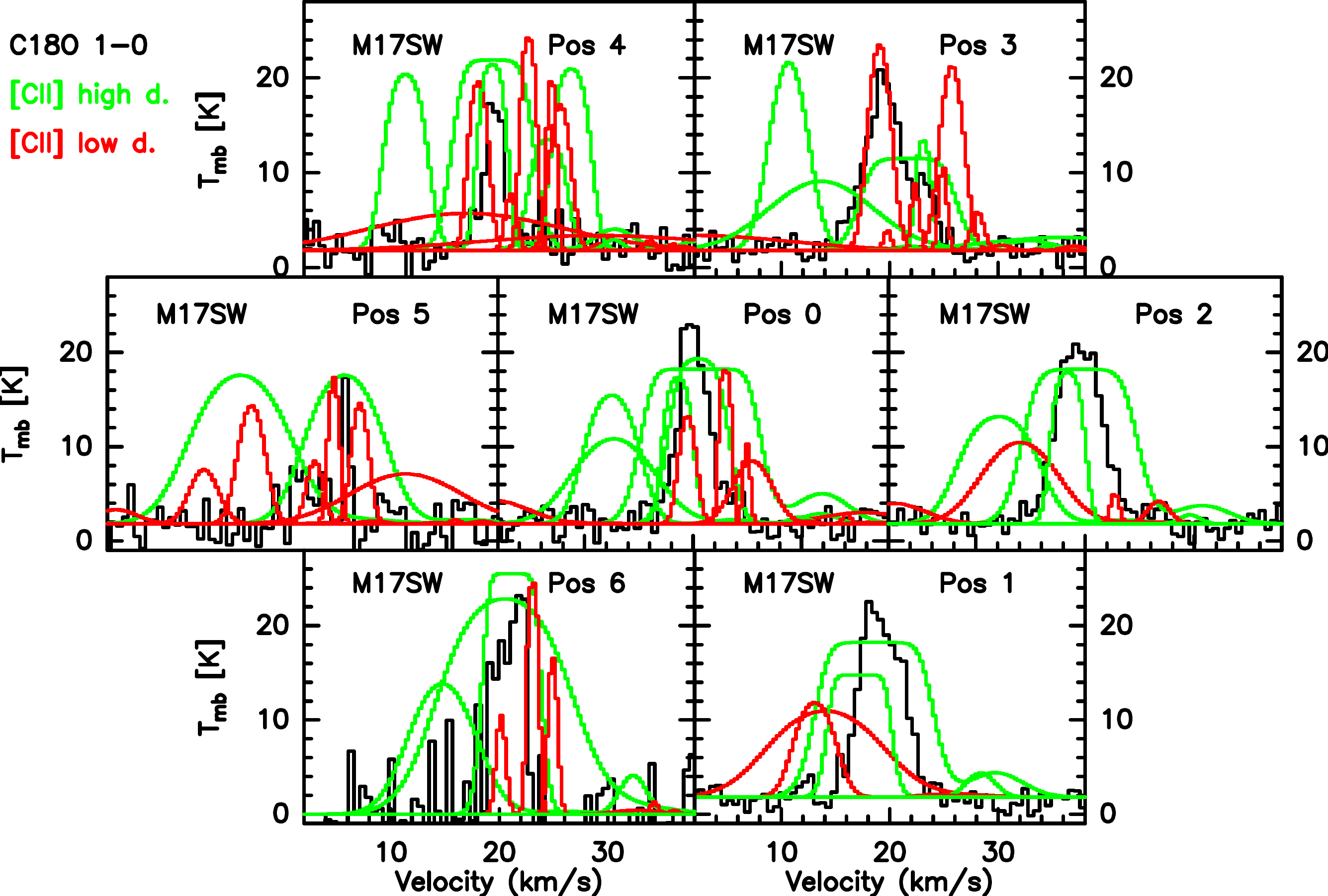}
      \caption{
              }
        \label{M17_CO_single}      
\end{subfigure}
\caption{ M17~SW mosaics of the seven positions observed by upGREAT. \textit{Top Fig. \ref{M17_CO_double}:} comparison between C\element[][18]{O} J = 1-0 and 
      the \Cp{12} Gaussian components from the double layer model. Scaled-up C\element[][18]{O} observations are in black, \Cp{12} background components are in blue and \Cp{12}
      foreground optical depth components are in pink. 
      \textit{Bottom Fig. \ref{M17_CO_single} :} comparison between C\element[][18]{O} J = 1-0 and the \Cp{12} Gaussian components from the single layer model. 
      Scaled-up C\element[][18]{O} observations are in black, \Cp{12} high density low temperature components are in green (\Cp{} high d.) and \Cp{12} low density high 
      temperature components are in red (\Cp{} low d.). 
}
\label{M17COCII}
\end{figure}

Then we compare the multi-component single-layered model with the result from the double layered model from Section~\ref{Analysis} and with the equivalent visual extinctions derived from C\element[][18]{O} molecular emission from Section~\ref{COdis} for Mon~R2 and M17~SW. In Fig.~\ref{M17COCII} , we can see the comparison between the profiles of the double and the single layer \Cp{12} models against the C\element[][18]{O} 1-0 emission profiles (scaled-up to match the \Cp{} components) for M17~SW. We see in Fig.~\ref{M17_CO_double}, that in the double layer model the brightest background emission component in \Cp{12} has a line profile that, though wider in velocity, share a similar line profile with the C\element[][18]{O} emission profile at the central line velocities that correlate with the molecular emission, as was described above. The \Cp{12} absorption dips also show profiles that partially match the C\element[][18]{O} emission. On the other hand, as we can see in Fig.~\ref{M17_CO_single}, the single layer model needs cold and high density \Cp{} emission components without any correlation to the C\element[][18]{O} emission profile. In particular, we would expect that the high density cold components (labeled ``\Cp{} high d.'' in Fig.~\ref{M17_CO_single}) would correlate with the molecular gas, being closer in their physical conditions. Similarly, the bright, narrow, lower density and warmer emission (\Cp{} low d.) notches at various velocities do not correlate with the 
C\element[][18]{O} emission. This comparison suggests that the single layer model, even if it formally provides a good fit, it is physically less plausible.\par

\begin{figure}
   \centering
   \includegraphics[width=\hsize]{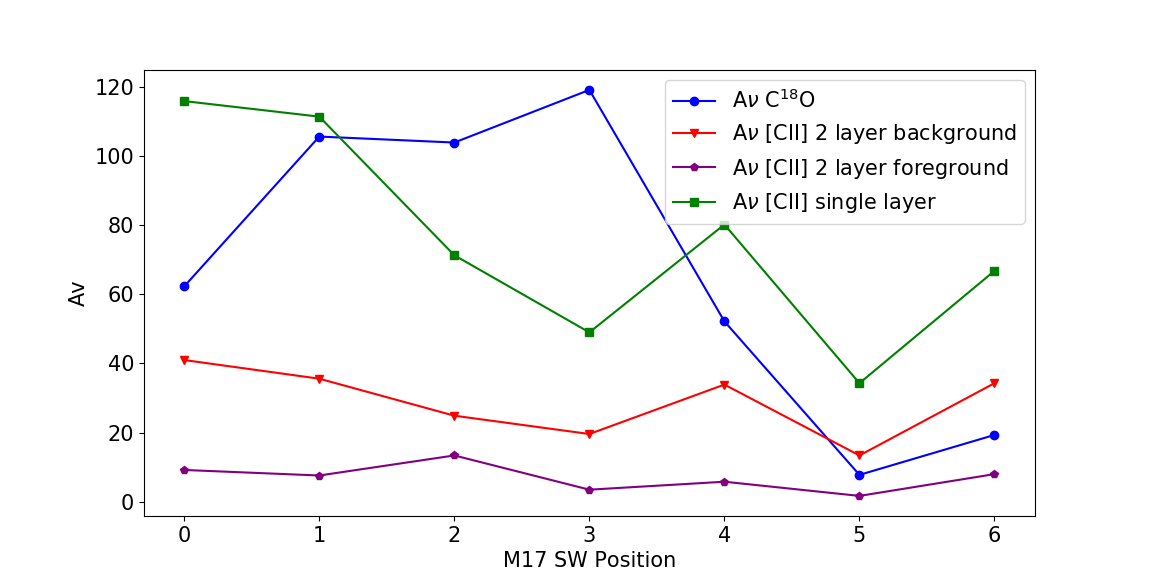}
      \caption{M17~SW Av comparison between the extinction of C\element[][18]{O},the \Cp{} from both the background and foreground layers of the double layer model and 
      the \Cp{} from the single layer model.
              }
         \label{M17Av}
   \end{figure}

We can then compare the equivalent extinctions (or for that matter, the derived H$_2$ column densities) derived from \Cp{} and from the CO isotopologues lines for both models, respectively. These are shown in Figure \ref{M17Av}. For position 5 and 6, the \Cp{} equivalent A$_\mathrm{V}$ is higher for both model scenarios than the one estimated from 
C\element[][18]{O}. This is no surprise because from velocity channel maps between the molecular and ionized line observations \citep{2015A&A...575A...9P}, it is known that these positions are located off the main molecular ridge and, hence, are dominated by PDR material.\par

For the other positions, the equivalent A$_\mathrm{V}$ of the \Cp{} layer estimated for the single layer model is similar or even higher than the one derived from the C\element[][18]{O} emission in some positions, whereas the double-layer \Cp{} emission model gives a  much lower equivalent \Cp{} column density, on average 25\% of the molecular column density. \par

This comparison shows that the single layer model requires extremely high column densities for \Cp{} in comparison to CO. These are unlikely to be present in a dense, high extinction cloud core that is traced by  C\element[][18]{O} and  C\element[][17]{O} to have molecular column densities corresponding alone to visual extinctions of up to 100~A$_\mathrm{V}$. In addition, these high column densities of \Cp{} for the single layer model, are required to have, in the bulk of the emission, low excitation temperature of around 50~K (see above); there is no reasonable physical scenario that can explain these properties. In a PDR, it is expected that the largest amount of hydrogen is located in the molecular core in molecular form. The single layer scenario shows the opposite scenario, with large amounts of material in the form of  CO-dark gas. But, we know that for M17~SW, even if it is affected by strong UV fields, the molecular gas is clumpy and the CO presents high column density and optically thick emission, proof enough that the molecular gas is shielded from the UV field. Hence, we rule out the single layer model for M17~SW. \par
   
For Mon~R2, the situation is similar. The equivalent visual extinction of the molecular gas traced by C\element[][18]{O} is 40.0 for position 1 and 44.5~mag for position 2, whereas the single layer scenario gives an equivalent A$_\mathrm{V}$ of the \Cp{} emission of 31.7 and 75.8~mag, respectively. The double layer scenario has significantly lower, although still relatively high, equivalent visual extinctions: 21~mag for the warm background and 3~mag for the foreground \Cp{} emission. As in the case of M17~SW, the single layer scenarios would thus require to have a  column of \Cp{} emission at relatively low $T_{\mathrm{ex}}$ that is about equally large to that of the molecular gas. \par  
   
In summary, we can rule out the single layer model for several reasons. The physical conditions required to model the \Cp{12} line profile are extremely improbable, with low temperature and high column density \Cp{} components accompanied by warmer and narrow low density bright notches. This would require the presence of extended and cold high density 
ionized gas surrounded by small clumps of bright, low-density ionized material--a scenario that is physically very unlikely. Also, the molecular line emission profile traced in CO does not match the single layer model \Cp{} emission profile, the latter showing high column density velocity components that are not at all matched by the molecular emission. In contrast, the velocity components with the bulk of the column density, match much better between the molecular emission and the \Cp{} emission for the double layer scenario. Moreover, the equivalent visual extinction derived for the \Cp{} emission in the single layer model would be equal or even exceeding the one derived from molecular emission, with the 
implausible scenario of much more hydrogen in the form of CO-dark gas visible in CO. We therefore discard the single layer model, even though it provides a formally fitting scenario, as physically unlikely.

\section{Beam filling and absorption factor effects} \label{factoreffects}

In the following, we discuss how a beam filling factor of\textit{} $\eta_{\phi}$ that is smaller than unity changes the derived 
physical properties. With the multi-component source model we, in principle, have to consider individual beam filling factors for each component. However, this would result in too large a number of free parameters in the fitting. Hence, we restrict the discussion to using one single beam filling factor that is applied to all background emission components in common. For the background emission component, the main effect of a beam filling factor smaller than unity is to raise the source intrinsic brightness, thus requiring higher excitation temperatures or higher optical depth to reach the higher brightness. To the first order, both effects result in a larger column density of the emitting material,  inversely rising proportional to the beam filling factor so that the beam averaged column density stays constant to the first order. \par

To quantify this, we perform, as a first step, a multi-component fit adding as an extra parameter a fixed value $\eta_{\phi}$ for the background layer in emission, decreasing it step by step. We use position 6 of the Horsehead PDR for the beam filling factor analysis. As expected, we find that a decrease in $\eta_{\phi}$ increases the excitation 
temperature, column density and optical depth of the background emission. For example, fixing $\eta_{\phi}$ at 0.5 increases the $T_{\mathrm{ex}}$ from the original fit ($\eta_{\phi}$ = 1) between 15 K and 20 K (from 43~K to 60 to 65~K) and the total column density changes from 1.3$\times$10$^{18}$ to 1.4$\times$10$^{18}$~cm$^{-2}$. An even lower value of $\eta_{\phi}$ =  0.3, results in an increment for the $T_{\mathrm{ex}}$ between 20 and 30~K and a \Cp{} column density of 1.7$\times$10$^{18}$~cm$^{-2}$. A summary of the effect of changing $\eta_{\phi}$ is given in Table \ref{MonR2table}.\par

\begin{figure}
   \centering
   \includegraphics[width=0.95\hsize]{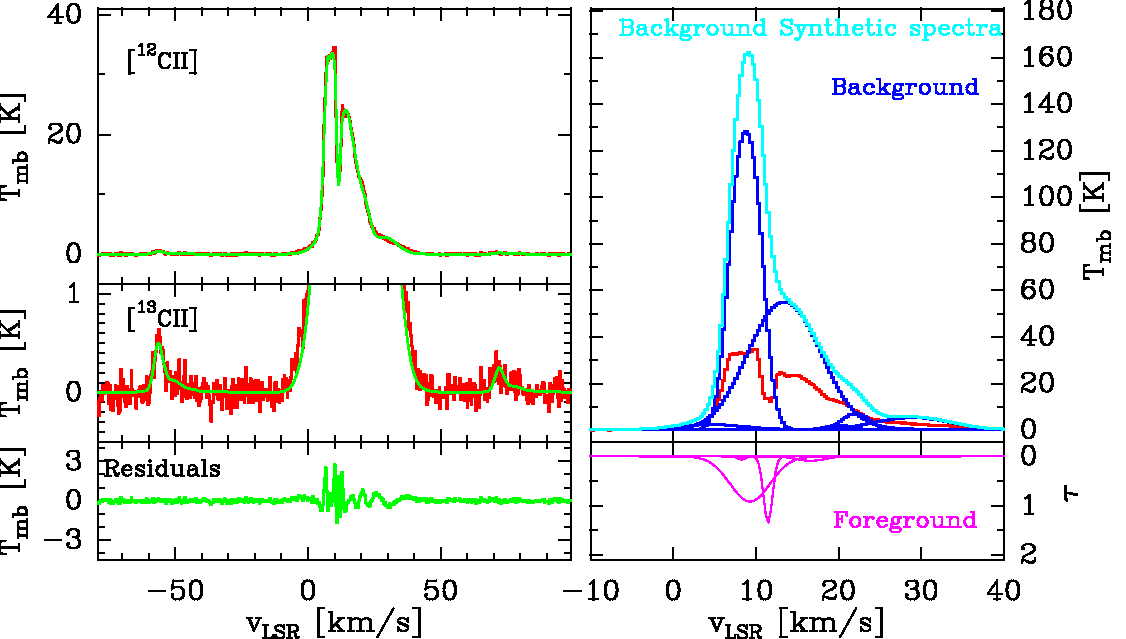}
      \caption{ Mon~R2 \Cp{12} spectra at position 1 fitted with a beam filling factor of 0.5. The observed data is shown in red, the 
      fitted model is in green, each individual fitted background component is in blue, 
      all the background components together in cyan and the optical depth for each foreground component in pink.
              }
         \label{MonR2pos1bff}
   \end{figure}

To consider the case that also foreground absorption is present, we use position 1 of Mon~R2. The resulting fits are shown in Fig.~\ref{MonR2pos1bff} and Table \ref{MonR2table}. Fixing $\eta_{\phi}$ to 0.5, we find for the background layer the same behavior as expected from the Horsehead PDR analysis, namely an increase of 50\% for the excitation 
temperature, and 100\% for the column density and the optical depth. Now, for the foreground layer, there is also an increase of the column density and optical depth similar to the ones of the background layer, of 50\% and 100\%, respectively. This is because the increase in the background requires a corresponding increase in the absorption of the foreground to obtain the same observed T$_{\mathrm{mb}}$. Due to the higher brightness of the background, even a slightly increased excitation temperature of the foreground absorbing layer can be tolerated
to give the same beam averaged brightness in the center of the absorption dip. Therefore, the introduction of a $\eta_{\phi}$ allows us to increase the excitation temperature and in particular, the column density of the background layer, as well as both parameters for the foreground layer, without affecting the observed main beam temperature.\par

In the even more complex case that also the foreground absorption material only partially covers the background emission. We define an absorption
factor\textit{} $\eta_{\mathrm{af}}$. The absorption factor represents the fraction of the background emission layer covered and hence absorbed by the foreground layer. We select, similar to before, position 1 from Mon~R2 as a source with self-absorption to study this effect. \par

We start with a value of $\eta_{\mathrm{af}}$ of 0.9 for the 3 main background components. We find that, naturally, the background remains the same and the foreground column density and optical depth increase. The foreground $N_{i}$(\ion{\element[][12]{C}}{II}) increases from 8.3$\times$10$^{17}$ to 2.0$\times$10$^{18}$~cm$^{-2}$ and the optical depth from 0.99 to 6.26. This is plausible as the smaller fraction of absorption now has to compensate for the fact that the bright background emission shines through, where the foreground does not absorb any. Next, we decrease $\eta_{\mathrm{af}}$ further down to 0.75. This is the lower limit as for a lower value than this one the 25\% of the background shining through cannot be compensated even by complete absorption down to zero brightness in the absorbing part. In this limiting case, the foreground $N_{i}$(\ion{\element[][12]{C}}{II}) increases even more, to 2.9$\times$10$^{18}$~cm$^{-2}$, and similar with the optical depth, which increases to 7.72. In summary, we find that the introduction of an absorption factor $\eta_{\mathrm{af}}$ increases the column density and the optical depth of the foreground material, without affecting  the excitation temperature to a substantial extent, becoming the foreground layer, which is much more massive and thick.\par

   \begin{figure}
   \centering
   \includegraphics[width=0.95\hsize]{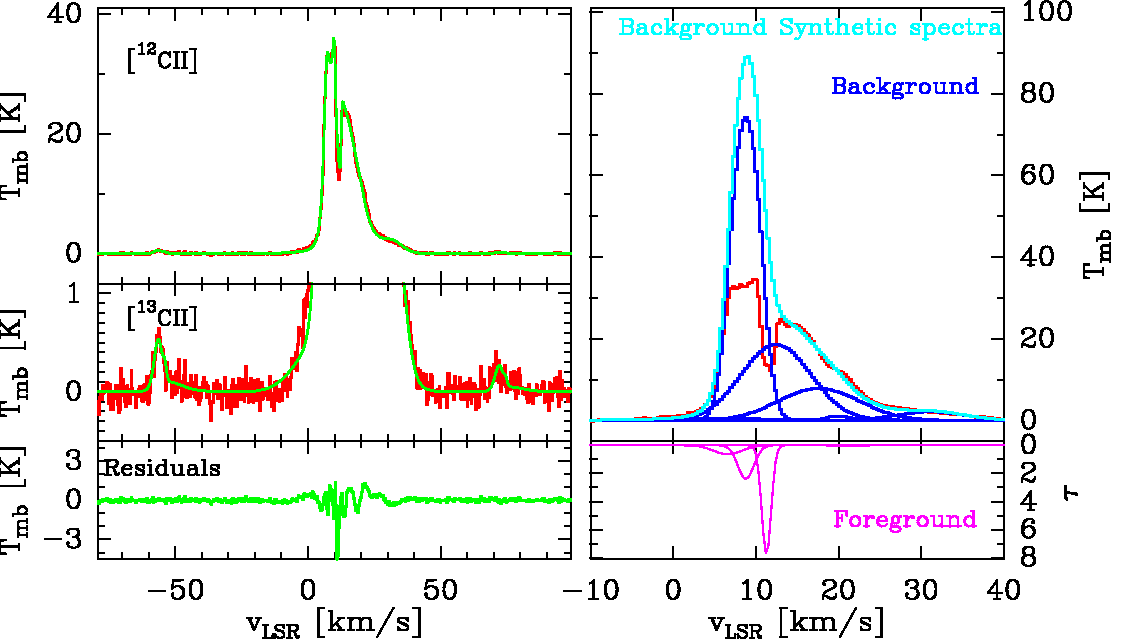}
      \caption{ MonR2 \Cp{12} spectra at position 1, fitted with an foreground absorption factor of 0.75 for the 3 main background components. The observed data is 
                shown in red, the fitted model is in green, each individual fitted background component is in blue, 
                 all the background components together in cyan and the optical depth for each foreground component in pink.
              }
         \label{MonR2pos1af}
   \end{figure}

\begin{table*}
  \centering
  \caption{
Derived fit parameters for the case of background and foreground filling factors as discussed in the text for position 6 in the Horsehead PDR and position 1 of MonR2. 
}
  \begin{tabular}{l  c c c c c  c c c c c c}
      \hline
\hline
\multicolumn{1}{c}{Sources}  & \multicolumn{1}{c}{No.}   & \multicolumn{1}{c}{No.}   & \multicolumn{1}{c}{$\chi^2$} & T$_{\mathrm{ex,bg}}$  & \multicolumn{1}{c}{Background}               & \multicolumn{1}{c}{$\tau_{\mathrm{bg}}^{*}$} & \multicolumn{1}{c}{Back.}                       & T$_{\mathrm{ex,fg}}$ & \multicolumn{1}{c}{Foreground} & \multicolumn{1}{c}{$\tau_{\mathrm{fg}}^{*}$} & \multicolumn{1}{c}{Fore.}      \\
                             & \multicolumn{1}{c}{Back.} & \multicolumn{1}{c}{Fore.} &                              &                       & \multicolumn{1}{c}{$N_{i}$(\ion{\element[][12]{C}}{II})}  &                                              & \multicolumn{1}{c}{A$_{\mathrm V}$ \ion{C}{II}} &                      & \multicolumn{1}{c}{$N_{i}$(\ion{\element[][12]{C}}{II})}  &                 & \multicolumn{1}{c}{A$_{\mathrm V}$ \ion{C}{II}}  \\
                             & \multicolumn{1}{c}{Comp.} & \multicolumn{1}{c}{Comp.} &                              & (K)                   & \multicolumn{1}{c}{(cm$^{-2}$)}              &                                              & \multicolumn{1}{c}{(mag.)}                      & (K)                  & \multicolumn{1}{c}{(cm$^{-2}$)}   &                 &  \multicolumn{1}{c}{(mag.)}          \\
          \hline
    HOR $\eta_{\phi}$ 0.5 & 5 & - & 1.5 & 57.0 & 1.4E18 & 3.07 & 7.1 & - & - & - & -  \\       
    HOR $\eta_{\phi}$ 0.3 & 4 & - & 1.5 & 73.4 & 1.7E18 & 3.19 & 9.0 & - & - & - & -  \\
    \hline      
    MonR2 $\eta_{\phi}$ 0.5 & 5 & 2 & 2.3 & 220.0 & 8.4E18 & 1.28 & 44.9  & 40.0 & 1.2E18 & 1.34 & 6.4  \\      
    MonR2 $\eta_{\mathrm{af}}$ 0.9 & 5 & 4 & 1.6 & 160.0 & 4.2E18 & 0.98 & 22.6   & 35.0 & 2.0E18 & 6.26 & 10.8  \\
    MonR2 $\eta_{\mathrm{af}}$ 0.75 & 5 & 3 & 3.4 & 160.0 & 4.2E18 & 0.98 & 22.6  & 30.0 & 2.9E18 & 7.72 & 15.3  \\
        \hline
  \end{tabular}
         \label{MonR2table}
 \end{table*}   

In conclusion, the introduction of both a foreground and background filling factor less than unity leads to an increase in the excitation temperature of the background component and the column density for both layers, in particular, the foreground absorbing layer. With the current beam resolution, we cannot constrain the beam filling factor for the \Cp{}
emission, nor do we have a way to constrain the absorption factor. The analysis above, however, shows that the excitation temperatures and column densities of the background layers are the lower limits and for the absorbing foreground, the excitation temperatures derived with a unity filling factor are the upper limits, while the absorbing foreground column 
densities are the lower limits. \par

\section{Parameters of the components} \label{Doc}

Here we list the multi-component analysis fit parameters for each position of the different sources. The tables contain the physical parameters of each Gaussian component for all the positions of the sources observed: The excitation temperature $T_{\mathrm{ex}}$, the \Cp{12} column density $N_{i}$(\ion{\element[][12]{C}}{II}), the central velocity of the component, and the velocity width of the line $\Delta$ V. Additionally, we list the optical depth of each component $\tau$ and its equivalent visual extinction A$_{\mathrm V}$.\\

\begin{table*}
  \centering
    \caption{Gaussian components parameters for M43}
    \label{M43gauss}
  \begin{tabular}{l r r r r r r }
      \hline

    \hline
Components & $T_{\mathrm{ex}}$ (K)        & $N_{i}$(\ion{\element[][12]{C}}{II}) 10$^{18}$ (cm$^{-2}$) & Vel (km/s) & $\Delta$ V (km/s) & $\tau$ & A$_{\mathrm V}$   \\
      \hline
0 Background 1 & 110.7 & 1.99 & 10.0 & 2.1 & 2.07 & 8.53 \\
0 Background 2 & 55.4 & 0.63 & 5.0 & 1.9 & 1.42 & 2.68 \\
0 Background 3 & 43.1 & 0.72 & 7.5 & 1.3 & 2.96 & 3.07 \\
0 Background 4 & 33.2 & 0.54 & 3.5 & 3.1 & 1.09 & 2.33 \\
0 Background 5 & 55.0 & 0.21 & 12.1 & 3.4 & 0.27 & 0.91 \\
      \hline
1 Background 1 & 97.3 & 1.81 & 9.3 & 2.6 & 1.75 & 7.73 \\
1 Background 2 & 90.5 & 0.32 & 5.7 & 2.9 & 0.30 & 1.37 \\
1 Background 3 & 54.4 & 0.15 & 3.5 & 7.6 & 0.09 & 0.64 \\
1 Background 4 & 38.5 & 0.47 & 11.6 & 3.3 & 0.80 & 2.02 \\
1 Background 5 & 23.3 & 0.17 & 16.5 & 2.3 & 0.52 & 0.75 \\
      \hline
2 Background 1 & 52.3 & 0.89 & 10.0 & 2.8 & 1.43 & 3.78 \\
2 Background 2 & 60.0 & 1.58 & 8.1 & 5.0 & 1.27 & 6.75 \\
      \hline
3 Background 1 & 76.4 & 0.32 & 10.2 & 2.4 & 0.43 & 1.37 \\
3 Background 2 & 70.0 & 0.77 & 8.0 & 4.5 & 0.60 & 3.31 \\
      \hline
4 Background 1 & 108.4 & 1.44 & 10.4 & 1.9 & 1.68 & 6.17 \\
4 Background 2 & 44.0 & 0.15 & 1.4 & 4.6 & 0.17 & 0.66 \\
4 Background 3 & 80.0 & 0.48 & 4.9 & 2.6 & 0.57 & 2.07 \\
4 Background 4 & 79.6 & 0.20 & 12.1 & 2.6 & 0.24 & 0.87 \\
4 Background 5 & 50.6 & 0.25 & 8.6 & 1.0 & 1.19 & 1.09 \\
4 Background 6 & 99.9 & 0.13 & 7.6 & 1.5 & 0.20 & 0.54 \\
      \hline
5 Background 1 & 108.0 & 1.10 & 10.7 & 2.5 & 0.98 & 4.72 \\
5 Background 2 & 45.0 & 0.12 & 1.3 & 6.8 & 0.09 & 0.51 \\
5 Background 3 & 70.0 & 0.18 & 4.8 & 1.7 & 0.37 & 0.78 \\
5 Background 4 & 66.7 & 0.61 & 8.0 & 6.7 & 0.33 & 2.60 \\
      \hline
6 Background 1 & 101.4 & 1.83 & 9.7 & 2.7 & 1.66 & 7.83 \\
6 Background 2 & 39.7 & 0.20 & 2.3 & 6.1 & 0.18 & 0.87 \\
6 Background 3 & 80.0 & 0.28 & 4.9 & 2.4 & 0.36 & 1.21 \\
6 Background 4 & 30.5 & 0.20 & 12.7 & 1.4 & 0.92 & 0.87 \\
6 Background 5 & 32.4 & 0.45 & 6.9 & 1.2 & 2.41 & 1.93 \\
      \hline
  \end{tabular}
\end{table*}

\clearpage

\begin{table*}
  \centering
    \caption{Gaussian components parameters for Horsehead PDR}
    \label{Horgauss}
  \begin{tabular}{l r r r r r r }
      \hline
      \hline
Components & T$_{\mathrm{ex}}$ (K)        & $N_{i}$(\ion{\element[][12]{C}}{II}) 10$^{18}$ (cm$^{-2}$) & Vel (km/s) & $\Delta$ V (km/s) & $\tau$ & A$_{\mathrm V}$   \\
      \hline
0 Background 1 & 38.0 & 0.27 & 10.6 & 0.7 & 2.16 & 1.17 \\
0 Background 2 & 55.0 & 0.30 & 10.6 & 1.4 & 0.93 & 1.29 \\
0 Background 3 & 31.5 & 0.10 & 17.6 & 14.7 & 0.04 & 0.43 \\
0 Background 4 & 40.0 & 0.11 & 11.6 & 5.3 & 0.12 & 0.48 \\
\hline
1 Background 1 & 26.7 & 0.11 & 10.5 & 1.0 & 0.78 & 0.48 \\  
1 Background 2 & 24.5 & 0.21 & 10.8 & 2.5 & 0.58 & 0.90 \\
1 Background 3 & 36.1 & 0.14 & 14.3 & 18.4 & 0.04 & 0.59 \\
1 Background 4 & 23.6 & 0.11 & 13.8 & 2.5 & 0.31 & 0.48 \\
\hline
2 Background 1 & 37.2 & 0.20 & 10.5 & 0.7 & 1.66 & 0.85 \\
2 Background 2 & 35.5 & 0.40 & 10.4 & 1.3 & 1.80 & 1.71 \\
2 Background 3 & 32.0 & 0.16 & 11.8 & 4.8 & 0.22 & 0.71 \\
2 Background 4 & 39.7 & 0.08 & 14.4 & 17.7 & 0.03 & 0.36 \\
\hline
3 Background 1 & 38.0 & 0.17 & 10.5 & 0.7 & 1.34 & 0.75 \\
3 Background 2 & 39.8 & 0.25 & 10.6 & 1.5 & 0.90 & 1.06 \\
3 Background 3 & 31.9 & 0.17 & 11.9 & 3.9 & 0.26 & 0.71 \\
3 Background 4 & 53.8 & 0.04 & 17.0 & 13.2 & 0.01 & 0.15 \\
\hline
4 Background 1 & 35.5 & 0.14 & 10.8 & 1.1 & 0.75 & 0.60 \\
4 Background 2 & 31.4 & 0.14 & 10.9 & 2.7 & 0.32 & 0.59 \\
4 Background 3 & 27.2 & 0.11 & 20.3 & 9.8 & 0.08 & 0.48 \\
4 Background 4 & 50.1 & 0.04 & 13.3 & 4.4 & 0.04 & 0.16 \\
\hline
5 Background 1 & 48.0 & 0.06 & 10.7 & 1.0 & 0.27 & 0.24 \\
5 Background 2 & 31.6 & 0.17 & 10.9 & 2.0 & 0.52 & 0.71 \\
5 Background 3 & 66.2 & 0.02 & 17.9 & 13.0 & 0.01 & 0.09 \\
5 Background 4 & 31.3 & 0.11 & 12.7 & 4.8 & 0.15 & 0.48 \\
\hline
6 Background 1 & 43.0 & 0.48 & 10.7 & 0.9 & 2.84 & 2.06 \\
6 Background 2 & 40.4 & 0.41 & 10.6 & 1.5 & 1.52 & 1.76 \\
6 Background 3 & 44.6 & 0.07 & 10.8 & 3.3 & 0.11 & 0.30 \\
6 Background 4 & 23.3 & 0.27 & 12.9 & 4.5 & 0.42 & 1.15 \\
6 Background 5 & 29.7 & 0.08 & 18.3 & 8.5 & 0.06 & 0.35 \\
\hline
  \end{tabular}
\end{table*}

\begin{table*}
  \centering
    \caption{Gaussian components parameters for Mon~R2 for the double layer model}
    \label{Mongaus}
  \begin{tabular}{l r r r r r r }
      \hline
      \hline
Components & T$_{\mathrm{ex}}$ (K)        & $N_{i}$(\ion{\element[][12]{C}}{II}) 10$^{18}$ (cm$^{-2}$) & Vel (km/s) & $\Delta$ V (km/s) & $\tau$ & A$_{\mathrm V}$   \\
      \hline
1 Background 1 & 160.0 & 2.37 & 8.7 & 3.3 & 1.07 & 10.56 \\
1 Background 2 & 150.0 & 1.12 & 12.4 & 9.2 & 0.20 & 4.99 \\
1 Background 3 & 150.0 & 0.11 & 31.2 & 11.0 & 0.02 & 0.51 \\
1 Background 4 & 150.0 & 0.55 & 18.2 & 11.6 & 0.08 & 2.43 \\
1 Background 5 & 150.0 & 0.11 & 6.4 & 15.3 & 0.01 & 0.51 \\
1 Foreground 1 & 20.0 & 0.21 & 11.4 & 1.4 & 1.05 & 0.94 \\
1 Foreground 2 & 20.0 & 0.59 & 8.8 & 3.9 & 1.07 & 2.63 \\
\hline
2 Background 1 & 150.0 & 0.33 & 12.0 & 12.3 & 0.04 & 1.41 \\
2 Background 2 & 150.0 & 1.20 & 8.5 & 9.8 & 0.20 & 5.14 \\
2 Background 3 & 150.0 & 2.49 & 10.8 & 5.8 & 0.69 & 10.65 \\
2 Background 4 & 150.0 & 0.64 & 11.0 & 1.0 & 1.03 & 2.74 \\
2 Background 5 & 150.0 & 0.01 & 3.6 & 3.8 & 0.01 & 0.08 \\
2 Background 6 & 150.0 & 0.01 & 20.1 & 2.8 & 0.01 & 0.06 \\
2 Foreground 1 & 20.0 & 0.16 & 12.0 & 1.1 & 1.03 & 0.67 \\
2 Foreground 2 & 20.0 & 0.09 & 7.3 & 1.7 & 0.38 & 0.38 \\
2 Foreground 3 & 20.0 & 0.25 & 11.1 & 1.2 & 1.52 & 1.06 \\
2 Foreground 4 & 20.0 & 0.14 & 9.7 & 4.1 & 0.24 & 0.59 \\
    \hline
  \end{tabular}
\end{table*}

\begin{table*}
  \centering
   \caption{Gaussian components parameters for M17~SW for the double layer model}
    \label{M17gaus}
  \begin{tabular}{l r r r r r r }
      \hline
      \hline
Components & T$_{\mathrm{ex}}$ (K)        & $N_{i}$(\ion{\element[][12]{C}}{II}) 10$^{18}$ (cm$^{-2}$) & Vel (km/s) & $\Delta$ V (km/s) & $\tau$ & A$_{\mathrm V}$   \\
      \hline
0 Background 1 & 250.0 & 8.13 & 20.8 & 5.4 & 1.43 & 34.78 \\
0 Background 2 & 200.6 & 0.14 & 2.5 & 9.4 & 0.02 & 0.60 \\
0 Background 3 & 200.0 & 0.06 & 33.0 & 4.5 & 0.02 & 0.25 \\
0 Background 4 & 250.0 & 0.82 & 12.5 & 5.9 & 0.13 & 3.53 \\
0 Foreground 1 & 40.6 & 0.74 & 21.4 & 2.5 & 1.63 & 3.15 \\
0 Foreground 2 & 40.7 & 0.09 & 24.1 & 0.6 & 0.81 & 0.37 \\
0 Foreground 3 & 40.7 & 0.03 & 22.4 & 0.8 & 0.17 & 0.11 \\
0 Foreground 4 & 53.7 & 0.65 & 16.4 & 2.2 & 1.28 & 2.78 \\
0 Foreground 5 & 30.6 & 0.35 & 19.0 & 3.2 & 0.71 & 1.51 \\
0 Foreground 6 & 30.6 & 0.24 & 24.1 & 2.0 & 0.77 & 1.03 \\
\hline
1 Background 1 & 200.0 & 7.76 & 18.0 & 5.7 & 1.61 & 33.18 \\
1 Background 2 & 200.0 & 0.17 & 10.8 & 2.7 & 0.08 & 0.74 \\
1 Background 3 & 200.0 & 0.05 & 5.1 & 6.5 & 0.01 & 0.21 \\
1 Background 4 & 200.0 & 0.02 & 25.9 & 1.6 & 0.01 & 0.09 \\
1 Background 5 & 200.0 & 0.01 & 8.2 & 1.4 & 0.01 & 0.06 \\
1 Foreground 1 & 30.0 & 0.09 & 24.1 & 0.6 & 0.98 & 0.40 \\
1 Foreground 2 & 30.0 & 0.58 & 21.3 & 2.5 & 1.52 & 2.49 \\
1 Foreground 3 & 30.0 & 0.53 & 16.4 & 3.7 & 0.91 & 2.29 \\
1 Foreground 4 & 30.9 & 0.45 & 18.7 & 3.2 & 0.89 & 1.93 \\
\hline
2 Background 1 & 200.0 & 4.66 & 19.3 & 6.6 & 0.84 & 19.94 \\
2 Background 2 & 200.0 & 0.51 & 12.5 & 5.8 & 0.11 & 2.19 \\
2 Background 3 & 200.0 & 0.43 & 12.5 & 19.5 & 0.03 & 1.86 \\
2 Background 4 & 200.0 & 0.04 & 34.8 & 6.2 & 0.01 & 0.16 \\
2 Foreground 1 & 30.0 & 0.61 & 24.3 & 3.1 & 1.24 & 2.60 \\
2 Foreground 2 & 30.0 & 0.55 & 21.2 & 2.2 & 1.58 & 2.33 \\
2 Foreground 3 & 52.2 & 0.91 & 16.0 & 3.6 & 1.14 & 3.90 \\
2 Foreground 4 & 50.0 & 0.55 & 14.0 & 2.7 & 0.97 & 2.37 \\
2 Foreground 5 & 30.0 & 0.35 & 19.1 & 2.8 & 0.80 & 1.48 \\
\hline
3 Background 1 & 180.0 & 2.92 & 22.0 & 5.9 & 0.66 & 12.49 \\
3 Background 2 & 150.0 & 0.69 & 15.4 & 29.4 & 0.04 & 2.95 \\
3 Background 3 & 150.0 & 0.72 & 10.9 & 4.5 & 0.26 & 3.08 \\
3 Background 4 & 150.0 & 0.04 & 46.2 & 12.3 & 0.01 & 0.19 \\
3 Foreground 1 & 25.0 & 0.51 & 21.3 & 2.2 & 1.61 & 2.19 \\
3 Foreground 2 & 25.0 & 0.26 & 23.7 & 1.4 & 1.32 & 1.12 \\
   \hline
4 Background 1 & 200.0 & 4.00 & 19.4 & 4.2 & 1.14 & 17.11 \\
4 Background 2 & 200.0 & 2.63 & 24.5 & 4.8 & 0.65 & 11.22 \\
4 Background 3 & 200.0 & 0.62 & 11.5 & 4.6 & 0.16 & 2.64 \\
4 Background 4 & 200.0 & 0.27 & 20.6 & 40.5 & 0.01 & 1.17 \\
4 Background 5 & 200.0 & 0.04 & 31.3 & 3.4 & 0.01 & 0.17 \\
4 Foreground 1 & 30.0 & 0.38 & 21.4 & 2.7 & 0.90 & 1.64 \\
4 Foreground 2 & 30.0 & 0.14 & 23.8 & 1.0 & 0.86 & 0.59 \\
4 Foreground 3 & 41.0 & 0.26 & 26.8 & 3.8 & 0.36 & 1.10 \\
4 Foreground 4 & 30.9 & 0.27 & 16.6 & 3.0 & 0.56 & 1.14 \\
4 Foreground 5 & 30.0 & 0.21 & 19.6 & 2.3 & 0.56 & 0.88 \\    
\hline
5 Background 1 & 200.0 & 1.14 & 23.9 & 5.7 & 0.24 & 4.87 \\
5 Background 2 & 200.0 & 0.38 & 29.8 & 12.4 & 0.04 & 1.61 \\
5 Background 3 & 200.0 & 1.52 & 14.3 & 8.2 & 0.22 & 6.49 \\
5 Background 4 & 200.0 & 0.03 & 2.5 & 3.8 & 0.01 & 0.13 \\
5 Foreground 1 & 30.0 & 0.14 & 18.4 & 2.2 & 0.42 & 0.62 \\
5 Foreground 2 & 30.0 & 0.07 & 23.9 & 0.7 & 0.63 & 0.29 \\
5 Foreground 3 & 30.0 & 0.18 & 12.7 & 2.5 & 0.46 & 0.79 \\
\hline    
6 Background 1 & 250.0 & 5.20 & 20.9 & 4.4 & 1.11 & 22.24 \\
6 Background 2 & 200.0 & 1.65 & 20.6 & 9.8 & 0.20 & 7.05 \\
6 Background 3 & 200.0 & 0.15 & 10.9 & 4.5 & 0.04 & 0.65 \\
6 Background 4 & 200.0 & 0.64 & 15.7 & 4.2 & 0.18 & 2.73 \\
6 Background 5 & 200.0 & 0.04 & 31.7 & 5.7 & 0.01 & 0.19 \\
6 Foreground 1 & 40.5 & 0.32 & 19.1 & 2.4 & 0.72 & 1.38 \\
6 Foreground 2 & 45.3 & 0.76 & 21.6 & 2.6 & 1.46 & 3.23 \\
6 Foreground 3 & 39.9 & 0.57 & 17.2 & 3.4 & 0.92 & 2.44 \\
6 Foreground 4 & 41.5 & 0.14 & 24.1 & 0.7 & 1.06 & 0.61 \\  
    \hline
  \end{tabular}
\end{table*}

\begin{table*}
  \centering
    \caption{Gaussian components parameters for Mon~R2 considering a single layer model}
    \label{Monnofore}
  \begin{tabular}{l r r r r r r }
      \hline
      \hline
Components & T$_{\mathrm{ex}}$ (K)        & $N_{i}$(\ion{\element[][12]{C}}{II}) 10$^{18}$ (cm$^{-2}$) & Vel (km/s) & $\Delta$ V (km/s) & $\tau$ & A$_{\mathrm V}$   \\
\hline
1 Background 1 & 70.0 & 2.51 & 8.3 & 3.3 & 2.62 & 10.32 \\
1 Background 2 & 35.0 & 2.85 & 14.7 & 25.1 & 0.68 & 12.17  \\
1 Background 3 & 50.0 & 0.56 & 20.1 & 4.8 & 0.55 & 2.39 \\
1 Background 4 & 50.0 & 0.03 & 5.4 & 2.9 & 0.04 & 0.15 \\
1 Background 5 & 70.0 & 1.19 & 14.7 & 4.3 & 0.96 & 5.08 \\
\hline
2 Background 1 & 40.4 & 2.44 & 10.8 & 1.3 & 10.17 & 10.42 \\
2 Background 2 & 27.7 & 10.89 & 8.7 & 12.5 & 5.75 & 46.59 \\
2 Background 3 & 95.6 & 1.08 & 6.7 & 4.9 & 0.56 & 4.62 \\
2 Background 4 & 120.0 & 0.07 & 10.4 & 0.7 & 0.20 & 0.31 \\
2 Background 5 & 85.0 & 0.22 & 13.0 & 1.0 & 0.61 & 0.92 \\
2 Background 6 & 85.0 & 0.31 & 15.3 & 3.7 & 0.24 & 1.31 \\
2 Background 7 & 85.0 & 0.58 & 13.8 & 2.1 & 0.78 & 2.48 \\
2 Background 8 & 120.0 & 0.57 & 9.3 & 1.9 & 0.60 & 2.44 \\
    \hline
  \end{tabular}
\end{table*}

 \longtab[6]{
 \begin{longtable}{l r r r r r r }
 \caption{Gaussian components parameters for M17~SW considering a single layer model}\\
  \label{table:m17single} \\
 \hline
 \hline
 Components & T$_{\mathrm{ex}}$ (K)        & $N_{i}$(\ion{\element[][12]{C}}{II}) 10$^{18}$ (cm$^{-2}$) & Vel (km/s) & $\Delta$ V (km/s) & $\tau$ & A$_{\mathrm V}$   \\
 \hline
 \endfirsthead
 \caption{Continued.} \\
 \hline
 Components & T$_{\mathrm{ex}}$ (K)        & $N_{i}$(\ion{\element[][12]{C}}{II}) 10$^{18}$ (cm$^{-2}$) & Vel (km/s) & $\Delta$ V (km/s) & $\tau$ & A$_{\mathrm V}$   \\
 \hline
 \endhead
 \hline
 \endfoot
 \hline
 \endlastfoot
0 Background 1 & 52.0 & 3.21 & 20.5 & 3.6 & 4.03 & 13.74\\
0 Background 2 & 50.0 & 16.22 & 20.7 & 5.5 & 13.88 & 69.40 \\
0 Background 3 & 50.0 & 0.14 & 35.7 & 8.8 & 0.08 & 0.61 \\
0 Background 4 & 50.0 & 1.30 & 12.7 & 7.6 & 0.80 & 5.54 \\
0 Background 5 & 50.0 & 0.27 & 1.5 & 7.5 & 0.17 & 1.14 \\
0 Background 6 & 50.0 & 1.64 & 12.5 & 4.3 & 1.78 & 7.00 \\
0 Background 7 & 50.0 & 1.29 & 18.6 & 2.1 & 2.92 & 6.88 \\
0 Background 8 & 50.0 & 0.40 & 19.5 & 1.6 & 1.20 & 5.51 \\
0 Background 9 & 50.0 & 0.11 & 24.9 & 0.7 & 0.74 & 0.46 \\
0 Background 10 & 50.0 & 0.82 & 23.0 & 0.8 & 4.67 & 3.51 \\
0 Background 11 & 50.0 & 0.43 & 25.5 & 3.9 & 0.52 & 1.86 \\
    \hline
1 Background 1 & 50.0 & 13.15 & 18.6 & 5.6 & 10.96 & 56.24 \\
1 Background 2 & 45.0 & 2.33 & 14.1 & 9.7 & 1.22 & 10.00 \\
1 Background 3 & 45.0 & 8.14 & 17.2 & 2.9 & 14.02 & 34.81 \\
1 Background 4 & 45.0 & 1.01 & 13.1 & 3.4 & 1.49 & 4.31 \\    
\hline
2 Background 1 & 50.0 & 9.86 & 19.5 & 5.7 & 8.04 & 42.17 \\
2 Background 2 & 45.0 & 2.44 & 12.1 & 5.9 & 2.12 & 10.46 \\
2 Background 3 & 50.0 & 1.13 & 14.0 & 7.1 & 0.75 & 4.85 \\
2 Background 4 & 50.0 & 1.73 & 18.4 & 2.1 & 3.87 & 7.40 \\
2 Background 5 & 50.0 & 0.20 & 2.6 & 6.5 & 0.14 & 0.84 \\
2 Background 6 & 50.0 & 0.08 & 26.7 & 2.3 & 0.16 & 0.33 \\
2 Background 7 & 50.0 & 0.04 & 22.7 & 0.9 & 0.22 & 0.18 \\
\hline
3 Background 1 & 40.0 & 6.57 & 21.7 & 5.0 & 7.26 & 29.13 \\
3 Background 2 & 45.0 & 0.53 & 23.1 & 1.2 & 2.28 & 2.29 \\
3 Background 3 & 75.3 & 0.89 & 10.7 & 3.6 & 0.79 & 3.79 \\ 
3 Background 4 & 70.0 & 0.29 & 2.1 & 19.7 & 0.05 & 1.24 \\
3 Background 5 & 70.0 & 0.80 & 13.7 & 10.7 & 0.26 & 3.43 \\
3 Background 6 & 70.0 & 0.69 & 19.1 & 2.1 & 1.14 & 2.95 \\
3 Background 7 & 70.0 & 0.61 & 25.7 & 2.2 & 0.94 & 2.60\\
3 Background 8 & 70.0 & 0.02 & 19.8 & 0.8 & 0.07 & 0.07 \\
3 Background 9 & 70.0 & 0.20 & 35.9 & 15.9 & 0.04 & 0.87 \\
3 Background 10 & 70.0 & 0.06 & 22.3 & 0.8 & 0.25 & 0.25 \\
3 Background 11 & 70.0 & 0.05 & 28.1 & 1.4 & 0.14 & 0.23 \\
3 Background 12 & 70.0 & 0.08 & 24.9 & 0.8 & 0.33 & 0.33 \\
    \hline
4 Background 1 & 55.0 & 7.35 & 19.4 & 4.2 & 7.59 & 31.46 \\
4 Background 2 & 44.2 & 1.44 & 24.5 & 2.6 & 2.89 & 6.18 \\
4 Background 3 & 55.0 & 0.79 & 18.1 & 1.6 & 2.16 & 3.40 \\
4 Background 4 & 55.0 & 0.97 & 16.9 & 19.6 & 0.22 & 4.17 \\
4 Background 5 & 55.0 & 1.81 & 11.4 & 3.0 & 2.58 & 7.76 \\
4 Background 6 & 55.0 & 1.85 & 26.6 & 2.6 & 3.07 & 7.93 \\
4 Background 7 & 55.0 & 0.62 & 25.6 & 1.8 & 1.44 & 2.64 \\
4 Background 8 & 55.0 & 1.64 & 19.4 & 1.9 & 3.76 & 7.02 \\
4 Background 9 & 55.0 & 0.10 & 21.1 & 1.2 & 0.35 & 0.40 \\
4 Background 10 & 60.0 & 0.84 & 22.7 & 1.2 & 2.80 & 3.58 \\
4 Background 11 & 55.0 & 0.46 & 28.9 & 24.2 & 0.08 & 1.94 \\
4 Background 12 & 55.0 & 0.42 & 24.8 & 0.8 & 2.33 & 1.78 \\    
    \hline
5 Background 1 & 55.0 & 2.25 & 23.9 & 6.3 & 1.54 & 9.64 \\
5 Background 2 & 55.0 & 2.87 & 14.3 & 8.0 & 1.54 & 12.30 \\
5 Background 3 & 55.0 & 0.57 & 15.3 & 2.5 & 0.98 & 2.42 \\
5 Background 4 & 55.0 & 0.80 & 29.5 & 11.2 & 0.31 & 3.41 \\
5 Background 5 & 55.0 & 0.23 & 10.9 & 3.0 & 0.34 & 1.00 \\
5 Background 6 & 55.0 & 0.16 & 21.1 & 1.7 & 0.40 & 0.67 \\
5 Background 7 & 55.0 & 0.38 & 25.3 & 1.6 & 1.01 & 1.63 \\
5 Background 8 & 55.0 & 0.34 & 22.9 & 1.0 & 1.49 & 1.47 \\
5 Background 9 & 55.0 & 0.08 & 2.8 & 4.5 & 0.08 & 0.35 \\    
    \hline
6 Background 1 & 60.0 & 5.28 & 20.5 & 9.4 & 2.25 & 23.52  \\
6 Background 2 & 60.0 & 7.84 & 21.1 & 2.8 & 11.26 & 33.50 \\
6 Background 3 & 60.0 & 1.22 & 14.8 & 6.3 & 0.78 & 5.24 \\
6 Background 4 & 60.0 & 0.13 & 20.2 & 1.0 & 0.53 & 0.54 \\
6 Background 5 & 60.0 & 0.02 & 32.6 & 4.5 & 0.02 & 0.09 \\
6 Background 6 & 60.0 & 0.23 & 24.9 & 0.9 & 1.06 & 0.98 \\
6 Background 7 & 60.0 & 0.68 & 23.1 & 0.8 & 3.35 & 2.91 \\     
 \hline
 \end{longtable}
}

\longtab[7]{
 \begin{longtable}{l r r r r r r }
 \caption{Gaussian components parameters for M17~SW with an $\alpha^+$=60.} \\
  \label{table:m1760}  \\
 \hline
 \hline
 Components & T$_{\mathrm{ex}}$ (K)        & $N_{i}$(\ion{\element[][12]{C}}{II}) 10$^{18}$ (cm$^{-2}$) & Vel (km/s) & $\Delta$ V (km/s) & $\tau$ & A$_{\mathrm V}$   \\
 \hline
 \endfirsthead
 \caption{Continued.} \\
 \hline
 Components & T$_{\mathrm{ex}}$ (K)        & $N_{i}$(\ion{\element[][12]{C}}{II}) 10$^{18}$ (cm$^{-2}$) & Vel (km/s) & $\Delta$ V (km/s) & $\tau$ & A$_{\mathrm V}$   \\
 \hline
 \endhead
 \hline
 \endfoot
 \hline
 \endlastfoot
0 Background 1 & 250.0 & 10.03 & 20.8 & 5.2 & 1.81 & 44.51 \\
0 Background 2 & 200.6 & 0.10 & 1.2 & 6.5 & 0.02 & 0.43 \\ 
0 Background 3 & 200.0 & 0.06 & 33.0 & 4.5 & 0.01 & 0.24 \\
0 Background 4 & 200.0 & 2.22 & 15.1 & 8.4 & 0.31 & 9.87 \\
0 Background 5 & 200.0 & 0.03 & 38.5 & 4.1 & 0.01 & 0.13 \\
0 Background 6 & 200.0 & 0.04 & 28.2 & 2.0 & 0.02 & 0.16 \\
0 Foreground 1 & 40.0 & 0.45 & 24.3 & 3.6 & 0.69 & 1.99 \\
0 Foreground 2 & 40.0 & 0.74 & 21.4 & 3.0 & 1.35 & 3.29 \\
0 Foreground 3 & 40.0 & 1.61 & 16.6 & 6.3 & 1.41 & 7.14 \\
0 Foreground 4 & 40.0 & 0.18 & 24.1 & 0.7 & 1.31 & 0.79 \\
0 Foreground 5 & 40.0 & 0.10 & 21.4 & 1.2 & 0.44 & 0.43 \\
    \hline
1 Background 1 & 200.0 & 11.62 & 18.0 & 5.7 & 2.41 & 51.57 \\ 
1 Background 2 & 200.6 & 0.08 & 10.8 & 1.7 & 0.06 & 0.37 \\
1 Background 3 & 200.0 & 0.04 & 5.1 & 5.8 & 0.01 & 0.15 \\
1 Background 4 & 200.0 & 0.01 & 25.9 & 1.1 & 0.01 & 0.04 \\
1 Background 5 & 200.0 & 0.05 & 9.0 & 2.3 & 0.03 & 0.23 \\
1 Background 6 & 200.0 & 0.01 & 22.2 & 0.0 & 0.39 & 0.04 \\
1 Background 7 & 200.0 & 0.01 & 0.0 & 5.2 & 0.00 & 0.07 \\
1 Foreground 1 & 30.0 & 0.16 & 24.0 & 0.8 & 1.26 & 0.72 \\
1 Foreground 2 & 30.0 & 0.55 & 21.6 & 2.5 & 1.41 & 2.45 \\
1 Foreground 3 & 30.0 & 0.62 & 17.2 & 6.4 & 0.62 & 2.73 \\
1 Foreground 4 & 30.0 & 0.91 & 18.1 & 6.9 & 0.85 & 4.04 \\  
\hline
2 Background 1 & 200.0 & 6.95 & 19.3 & 6.6 & 1.25 & 30.85 \\ 
2 Background 2 & 200.0 & 0.21 & 11.6 & 2.4 & 0.11 & 0.94 \\
2 Background 3 & 200.0 & 0.18 & 8.7 & 4.3 & 0.05 & 0.79 \\
2 Background 4 & 200.0 & 0.05 & 34.3 & 6.2 & 0.01 & 0.21 \\
2 Background 5 & 200.0 & 0.11 & 3.4 & 12.6 & 0.01 & 0.49 \\ 
2 Foreground 1 & 30.0 & 0.88 & 24.5 & 3.5 & 1.61 & 3.91 \\
2 Foreground 2 & 30.0 & 0.53 & 21.4 & 2.2 & 1.51 & 2.33 \\
2 Foreground 3 & 30.0 & 0.27 & 15.9 & 2.5 & 0.69 & 1.20 \\
2 Foreground 4 & 30.0 & 0.72 & 19.0 & 4.2 & 1.10 & 3.20 \\
\hline
3 Background 1 & 180.0 & 4.56 & 21.9 & 5.0 & 1.21 & 20.22 \\ 
3 Background 2 & 150.0 & 0.04 & 15.4 & 2.3 & 0.03 & 0.17 \\
3 Background 3 & 150.0 & 0.92 & 11.0 & 5.2 & 0.29 & 4.08 \\
3 Background 4 & 150.0 & 0.08 & 37.5 & 9.4 & 0.01 & 0.37 \\
3 Background 5 & 150.0 & 0.14 & -0.2 & 10.8 & 0.02 & 0.62 \\
3 Background 6 & 150.0 & 0.09 & 27.7 & 2.3 & 0.06 & 0.39 \\
3 Background 7 & 150.0 & 0.03 & 30.5 & 2.2 & 0.02 & 0.12 \\
3 Background 8 & 150.0 & 0.02 & 11.0 & 2.4 & 0.01 & 0.09 \\
3 Background 9 & 150.0 & 0.03 & 48.4 & 8.5 & 0.01 & 0.15 \\
3 Foreground 1 & 20.0 & 0.08 & 23.9 & 0.8 & 0.68 & 0.36 \\
3 Foreground 2 & 20.0 & 0.25 & 20.3 & 2.4 & 0.74 & 1.13 \\
3 Foreground 3 & 20.0 & 0.30 & 21.3 & 1.6 & 1.39 & 1.35 \\
3 Foreground 4 & 20.0 & 0.34 & 23.4 & 2.3 & 1.09 & 1.53 \\
3 Foreground 5 & 20.0 & 0.01 & 10.4 & 0.9 & 0.05 & 0.03 \\
    \hline
4 Background 1 & 200.0 & 6.75 & 19.7 & 4.1 & 1.97 & 29.95 \\ 
4 Background 2 & 200.0 & 4.11 & 24.4 & 4.7 & 1.04 & 18.22 \\
4 Background 3 & 200.0 & 0.56 & 11.4 & 4.0 & 0.17 & 2.47 \\
4 Background 4 & 200.0 & 0.07 & 38.1 & 8.2 & 0.01 & 0.29 \\
4 Background 5 & 200.0 & 0.06 & 4.0 & 8.4 & 0.01 & 0.27 \\
4 Background 6 & 200.0 & 0.04 & 31.7 & 2.4 & 0.02 & 0.16 \\
4 Background 7 & 200.0 & 0.02 & 13.9 & 1.4 & 0.02 & 0.10 \\
4 Background 8 & 200.0 & 0.03 & 7.0 & 3.0 & 0.01 & 0.12 \\
4 Background 9 & 200.0 & 0.02 & 47.3 & 6.6 & 0.00 & 0.11 \\  
4 Foreground 1 & 20.0 & 0.06 & 24.0 & 0.7 & 0.63 & 0.27 \\
4 Foreground 2 & 20.0 & 0.52 & 20.2 & 3.3 & 1.12 & 2.29 \\
4 Foreground 3 & 20.0 & 0.28 & 26.7 & 3.4 & 0.59 & 1.24 \\
4 Foreground 4 & 20.0 & 0.14 & 17.3 & 1.9 & 0.52 & 0.61 \\
4 Foreground 5 & 20.0 & 0.05 & 21.4 & 1.0 & 0.38 & 0.24 \\
4 Foreground 6 & 20.0 & 0.31 & 23.0 & 2.8 & 0.78 & 1.36 \\
4 Foreground 7 & 20.0 & 0.06 & 16.2 & 1.3 & 0.31 & 0.25 \\ 
    \hline
5 Background 1 & 200.0 & 1.64 & 23.9 & 5.7 & 0.34 & 7.29 \\ 
5 Background 2 & 200.0 & 1.82 & 14.1 & 6.2 & 0.35 & 8.08 \\
5 Background 3 & 200.0 & 0.06 & 4.6 & 5.4 & 0.01 & 0.25 \\
5 Background 4 & 200.0 & 0.12 & 30.8 & 3.6 & 0.04 & 0.52 \\
5 Background 5 & 200.0 & 0.11 & 35.6 & 7.1 & 0.02 & 0.47 \\
5 Foreground 1 & 20.0 & 0.06 & 24.0 & 0.7 & 0.58 & 0.26 \\
5 Foreground 2 & 20.0 & 0.13 & 24.5 & 4.4 & 0.21 & 0.59 \\
5 Foreground 3 & 20.0 & 0.19 & 12.3 & 2.4 & 0.57 & 0.86 \\
5 Foreground 4 & 20.0 & 0.20 & 14.1 & 2.8 & 0.49 & 0.87 \\
5 Foreground 5 & 20.0 & 0.03 & 21.8 & 0.9 & 0.22 & 0.13 \\
5 Foreground 6 & 20.0 & 0.10 & 17.3 & 2.2 & 0.31 & 0.42 \\
    \hline
6 Background 1 & 250.0 & 4.49 & 21.1 & 3.0 & 1.42 & 19.93 \\ 
6 Background 2 & 200.0 & 5.82 & 19.7 & 7.0 & 0.99 & 25.84 \\
6 Background 3 & 200.0 & 0.12 & 11.7 & 3.0 & 0.05 & 0.55 \\
6 Background 4 & 200.0 & 0.06 & 30.4 & 4.2 & 0.02 & 0.27 \\
6 Background 5 & 200.0 & 0.04 & 27.1 & 2.1 & 0.02 & 0.17 \\
6 Background 6 & 200.0 & 0.02 & 8.6 & 2.3 & 0.01 & 0.09 \\
6 Background 7 & 200.0 & 0.01 & 13.4 & 0.9 & 0.02 & 0.06 \\
6 Foreground 1 & 35.0 & 0.20 & 19.4 & 1.8 & 0.66 & 0.88 \\
6 Foreground 2 & 35.0 & 0.83 & 21.5 & 2.8 & 1.77 & 3.70 \\
6 Foreground 3 & 35.0 & 0.63 & 17.3 & 3.3 & 1.14 & 2.79 \\
6 Foreground 4 & 35.0 & 0.08 & 24.1 & 0.6 & 0.86 & 0.36 \\
 \hline
 \end{longtable}
}

\begin{table*}
  \centering
    \caption{Gaussian components parameters for the Horsehead PDR for Position 6 considering a beam filling factor of 0.5}
    \label{Horbff05}
  \begin{tabular}{l r r r r r r }
      \hline
      \hline
Components & T$_{\mathrm{ex}}$ (K)        & $N_{i}$(\ion{\element[][12]{C}}{II}) 10$^{18}$ (cm$^{-2}$) & Vel (km/s) & $\Delta$ V (km/s) & $\tau$ & A$_{\mathrm V}$   \\
      \hline
 Background 1 & 57.0 & 0.63 & 10.7 & 0.9 & 3.07 & 2.70 \\
 Background 2 & 57.3 & 0.57 & 10.6 & 1.5 & 1.63 & 2.44 \\
 Background 3 & 60.8 & 0.12 & 11.4 & 4.2 & 0.11 & 0.51 \\
 Background 4 & 47.2 & 0.08 & 16.1 & 7.7 & 0.05 & 0.35 \\
 Background 5 & 24.6 & 0.04 & 22.8 & 1.5 & 0.19 & 0.17 \\
    \hline
  \end{tabular}
\end{table*}

\begin{table*}
  \centering
    \caption{Gaussian components parameters for the Horsehead PDR for Position 6 considering a beam filling factor of 0.3}
    \label{Horbff03}
  \begin{tabular}{l r r r r r r }
      \hline
      \hline
Components & T$_{\mathrm{ex}}$ (K)        & $N_{i}$(\ion{\element[][12]{C}}{II}) 10$^{18}$ (cm$^{-2}$) & Vel (km/s) & $\Delta$ V (km/s) & $\tau$ & A$_{\mathrm V}$   \\
      \hline
 Background 1 & 73.4 & 0.81 & 10.7 & 0.8 & 3.19 & 3.47 \\
 Background 2 & 79.3 & 0.72 & 10.6 & 1.5 & 1.49 & 2.98 \\
 Background 3 & 104.1 & 0.13 & 11.5 & 4.4 & 0.07 & 0.56 \\
 Background 4 & 50.0 & 0.13 & 16.8 & 9.5 & 0.06 & 0.56 \\
    \hline
  \end{tabular}
\end{table*}

\begin{table*}
  \centering
    \caption{Gaussian components parameters for Mon~R2 for Position 1 considering a beam filling factor of 0.5}
    \label{Horbff11}
  \begin{tabular}{l r r r r r r }
      \hline
      \hline
Components & T$_{\mathrm{ex}}$ (K)        & $N_{i}$(\ion{\element[][12]{C}}{II}) 10$^{18}$ (cm$^{-2}$) & Vel (km/s) & $\Delta$ V (km/s) & $\tau$ & A$_{\mathrm V}$   \\
      \hline
 Background 1 & 220.0 & 4.33 & 8.8 & 3.6 & 1.28 & 18.52 \\
 Background 2 & 200.0 & 3.44 & 13.4 & 9.6 & 0.43 & 14.72 \\
 Background 3 & 170.0 & 0.33 & 29.0 & 10.9 & 0.04 & 1.41 \\
 Background 4 & 170.0 & 0.15 & 22.1 & 3.8 & 0.06 & 0.64 \\
 Background 5 & 160.0 & 0.14 & 4.4 & 10.4 & 0.02 & 0.60 \\
 Foreground 1 & 40.0 & 0.31 & 11.5 & 1.3 & 1.34 & 1.33 \\
 Foreground 2 & 30.0 & 0.81 & 9.3 & 5.7 & 0.92 & 3.47 \\
 Foreground 3 & 30.0 & 0.06 & 16.6 & 4.0 & 0.09 & 0.26 \\
 Foreground 4 & 30.0 & 0.01 & 8.3 & 1.1 & 0.06 & 0.04 \\
    \hline
  \end{tabular}
\end{table*}

\begin{table*}  
 \centering
    \caption{Gaussian components parameters for MonR2 for Position 1 considering an absorbing factor of 0.9}
    \label{Monr2bff21}
  \begin{tabular}{l r r r r r r }
      \hline
      \hline
Components & T$_{\mathrm{ex}}$ (K)        & $N_{i}$(\ion{\element[][12]{C}}{II}) 10$^{18}$ (cm$^{-2}$) & Vel (km/s) & $\Delta$ V (km/s) & $\tau$ & A$_{\mathrm V}$   \\
      \hline
 Background 1 & 160.0 & 2.37 & 8.8 & 3.6 & 0.98 & 10.14 \\
 Background 2 & 150.0 & 1.12 & 12.4 & 9.6 & 0.19 & 4.79\\
 Background 3 & 150.0 & 0.53 & 17.6 & 11.4 & 0.08 & 2.27 \\
 Background 4 & 150.0 & 0.12 & 31.2 & 9.9 & 0.02 & 0.51 \\
 Background 5 & 150.0 & 0.06 & 4.4 & 20.2 & 0.00 & 0.26 \\
 Foreground 1 & 35.0 & 1.14 & 11.5 & 1.1 & 6.26 & 4.88\\
 Foreground 2 & 30.0 & 0.88 & 8.9 & 4.2 & 1.34 & 3.76 \\
    \hline
  \end{tabular}
\end{table*}

\begin{table*}
\centering
\caption{Gaussian components parameters for Mon~R2 for Position 1 considering an absorbing factor of 0.75}
\label{Monr2bff22}
  \begin{tabular}{l r r r r r r }
      \hline
      \hline
Components & T$_{\mathrm{ex}}$ (K)        & $N_{i}$(\ion{\element[][12]{C}}{II}) 10$^{18}$ (cm$^{-2}$) & Vel (km/s) & $\Delta$ V (km/s) & $\tau$ & A$_{\mathrm V}$   \\
      \hline
 Background 1 & 160.0 & 2.37 & 8.8 & 3.6 & 0.98 & 10.14 \\
 Background 2 & 150.0 & 1.12 & 12.4 & 9.6 & 0.19 & 4.79 \\
 Background 3 & 150.0 & 0.53 & 17.6 & 11.4 & 0.08 & 2.27 \\
 Background 4 & 150.0 & 0.12 & 31.2 & 9.9 & 0.02 & 0.51 \\
 Background 5 & 150.0 & 0.06 & 4.4 & 20.2 & 0.00 & 0.26 \\
 Foreground 1 & 30.0 & 1.62 & 11.3 & 1.3 & 7.72 & 6.93 \\
 Foreground 2 & 30.0 & 0.85 & 8.8 & 2.3 & 2.39 & 3.64 \\
 Foreground 3 & 30.0 & 0.40 & 6.6 & 4.0 & 0.64 & 1.71 \\
    \hline
\end{tabular}
\end{table*}

\end{appendix}

\end{document}